\documentclass[journal]{IEEEtran}

\ifCLASSINFOpdf
\else
\fi

\usepackage{framed,multirow}
\usepackage{graphicx}
\usepackage{amssymb}
\usepackage{amsmath}
\usepackage{multirow}
\usepackage{booktabs}
\usepackage{array}
\usepackage{todonotes}
\newcolumntype{L}[1]{>{\raggedright\let\newline\\\arraybackslash\hspace{0pt}}m{#1}}
\newcolumntype{C}[1]{>{\centering\let\newline\\\arraybackslash\hspace{0pt}}m{#1}}
\newcolumntype{R}[1]{>{\raggedleft\let\newline\\\arraybackslash\hspace{0pt}}m{#1}}

\usepackage{bm}

\usepackage{lipsum}
\usepackage{changepage}
\usepackage{color,soul}
\usepackage{amssymb}
\usepackage{amsmath}
\usepackage{enumitem}   

\usepackage{float}

\usepackage{booktabs,xcolor}
\definecolor{lightgray}{gray}{0.9}

\usepackage{tikz}

\usepackage{lipsum}

\let\OLDthebibliography\thebibliography
\renewcommand\thebibliography[1]{
  \OLDthebibliography{#1}
  \setlength{\parskip}{0pt}
  \setlength{\itemsep}{0pt plus 0.3ex}
}

\usepackage{array}

\makeatletter
\newcommand{\thickhline}{%
    \noalign {\ifnum 0=`}\fi \hrule height 1pt
    \futurelet \reserved@a \@xhline
}
\newcolumntype{"}{@{\hskip\tabcolsep\vrule width 1pt\hskip\tabcolsep}}
\makeatother

\usepackage{amssymb}
\usepackage{latexsym}

\usepackage{url}
\usepackage{xcolor}

\usepackage{hyperref}

\begin{document}

\title{Diffusion MRI with Machine Learning}

\author{Davood Karimi$^*$, Simon K. Warfield \\ Harvard Medical School and Boston Children's Hospital, Boston, Massachusetts, USA \\
$^*$davood.karimi@childrens.harvard.edu}


\maketitle

\begin{abstract}

\hspace{2mm} Diffusion-weighted magnetic resonance imaging (dMRI) of the brain offers unique capabilities including noninvasive probing of tissue microstructure and structural connectivity. It is widely used for clinical assessment of disease and injury, and for neuroscience research. Analyzing the dMRI data to extract useful information for medical and scientific purposes can be challenging. The dMRI measurements may suffer from strong noise and artifacts, and may exhibit high inter-session and inter-scanner variability in the data, as well as inter-subject heterogeneity in brain structure. Moreover, the relationship between measurements and the phenomena of interest can be highly complex. Recent years have witnessed increasing use of machine learning methods for dMRI analysis. This manuscript aims to assess these efforts, with a focus on methods that have addressed data preprocessing and harmonization, microstructure mapping, tractography, and white matter tract analysis. We study the main findings, strengths, and weaknesses of the existing methods and suggest topics for future research. We find that machine learning may be exceptionally suited to tackle some of the difficult tasks in dMRI analysis. However, for this to happen, several shortcomings of existing methods and critical unresolved issues need to be addressed. There is a pressing need to improve evaluation practices, to increase the availability of rich training datasets and validation benchmarks, as well as model generalizability, reliability, and explainability concerns.

\end{abstract}

\begin{IEEEkeywords}
Diffusion MRI, Machine Learning, Artificial Intelligence, Deep Learning
\end{IEEEkeywords}

\section{Introduction}

\subsection{Background and motivation}

Diffusion-weighted Magnetic Resonance Imaging (dMRI) is a widely used medical imaging modality \cite{johansen2013diffusion, le2014diffusion}. It has a unique role in neuroimaging, where it stands as the only noninvasive method for probing the tissue microstructural makeup and structural connectivity of the brain \cite{lerch2017studying, tournier2019diffusion}. It has facilitated the study of normal brain development and quantitative characterization of the impact of diseases and disorders \cite{bodini2009diffusion, salat2014diffusion}. As a result, dMRI has become an indispensable tool in medicine and neuroscience, and it has been a major component of large neuroimaging initiatives \cite{miller2016multimodal, van2013wu}.

Raw dMRI data can suffer from a host of imperfections and artifacts \cite{tax2022s, jones2010challenges}. Yet, these data need to be analyzed to uncover subtle differences or minute changes that reflect the underlying normal or abnormal variations in brain development. These neurodevelopmental processes, in turn, are highly complex, heterogeneous, and multi-factorial. Consequently, development and validation of computational methods for dMRI analysis are difficult. Accurate and reproducible processing of dMRI data has been a long-standing challenge, and thousands of research papers have been devoted to addressing its various aspects. Several computational pipelines and software projects have aimed at standardizing and streamlining some of the more routine dMRI computations \cite{tournier2019mrtrix3, garyfallidis2014dipy, theaud2020tractoflow}. However, there are many persistent challenges that have not been fully addressed and there is an urgent need for new methods to enable higher accuracy, reproducibility, reliability, and computational speed \cite{tax2022s, bach2014methodological, jones2010twenty}.

Classical dMRI analysis methods have appropriately been based on conventional signal processing, biophysical modeling, and numerical optimization techniques. Meanwhile, many studies have advocated for machine learning and data-driven techniques \cite{golkov2016q, wasserthal2018tractseg, poulin2019tractography} motivated by the opportunities to exploit advances in hardware, new software libraries, and models learned from data. These techniques have become more popular in recent years. This trend has been driven by more powerful machine learning models (mostly based on deep neural networks) and a greater appreciation of the power and flexibility of these methods. Overall, these methods have been shown to possess the potential to improve the speed, accuracy, and reproducibility of various computations in dMRI analysis such as data pre-processing \cite{hu2020distortion} and harmonization \cite{moyer2020scanner}, tissue microstructure mapping \cite{golkov2016q, de2021neural, karimi2021deep}, tractography \cite{poulin2019tractography}, tract-specific analysis \cite{wasserthal2018tractseg}, and population studies \cite{li2021longitudinal}. It appears that recent studies herald a new generation of dMRI analysis methods that may significantly complement, if not at least in some cases entirely replace, the more conventional techniques. Therefore, it is good time to review, summarize, and critically assess the achievements of these works, highlight their shortcomings and limitations, and point out future work that may contribute to this new field in dMRI.

\subsection{Scope and organization of this manuscript}

This manuscript focuses almost entirely on applications in neuroimaging. It deals with the processing steps after the reconstruction of the so-called q-space data. It is concerned with data pre-processing steps such as denoising and artifact correction as well as downstream computations such as white matter microstructure mapping and tract-specific analysis.

The search to find the relevant works to be included in this paper was conducted in PubMed and Google Scholar. The initial search was performed in June 2023 and then repeated in June 2024. The search terms included ``diffusion MRI" and ``machine learning" or ``deep learning". More detailed searches were performed in reputable journals (e.g., Medical Image Analysis and IEEE Transactions on Medical Imaging) and conferences (e.g., MICCAI) that have a focus on this topic. Given the wide scope of work in this domain, we sought to focus on papers with interesting methodological advances or extensive experimental results.

This paper starts with a brief description of challenges in dMRI data analysis (Section \ref{sec:challenges}) and general overview of the reasons why machine learning may be effective in addressing these challenges (Section \ref{sec:Why_ML}). Section \ref{sec:prior_works} reviews prior works that have used machine learning in dMRI. The main applications considered include data preprocessing and quality enhancement, data harmonization, estimation of tissue microstructure and fiber orientation, tractography, tract-specific analysis, registration and segmentation. Figure \ref{fig:outline_methods} shows the outline of that section. Following that, Section \ref{sec:discussion} discusses the main technical considerations, challenges, and open questions. Some of the main topics discussed in that section include validation approaches, inherent limitations of machine learning, data, ground truth, and modeling considerations, as well as model explainability, uncertainty, and reliability concerns. Figure \ref{fig:outline_discussion} shows the outline of that section. A short Conclusion section will present the closing remarks.

\section{The challenging nature of dMRI analysis}
\label{sec:challenges}

The challenges and pitfalls of analyzing the dMRI data have been discussed in dedicated publications \cite{tax2022s, jones2010twenty, jones2013white}. A primary source of difficulty in dMRI analysis is measurement noise and artifacts \cite{tax2022s, pierpaoli2010artifacts}. Measurement noise can significantly alter the analysis results even in relatively simple computations such as diffusion tensor estimation \cite{jones2004squashing, roalf2016impact}. It can be difficult to suppress the noise and artifacts due to their complex distributions. The echo planar imaging method that is used to acquire dMRI measurements can give rise to eddy-current- and susceptibility-induced distortions. Subject motion during data acquisition is another persistent challenge in MRI. These factors can significantly impact the accuracy of quantitative tissue microstructure mapping and structural connectivity analysis \cite{yendiki2014spurious, oldham2020efficacy, baum2018impact}. Furthermore, the relationship between the dMRI measurements and the underlying microstructure of neuronal tissue is complex \cite{novikov2018modeling, novikov2019quantifying}. Another source of difficulty is the measurement requirements of advanced dMRI models. Estimation of multi-compartment models of tissue microstructure and of complex white matter fiber configurations requires densely-sampled q-space data. Models often seek to represent properties of brain tissue with parameters corresponding to aspects of microstructure. Classical signal processing approaches are concerned with characterizing the identifiability of model parameters, and with characterizing any potential bias or variance in estimated values of parameters. For some model parameters to be identifiable, imaging strategies may need to be carefully chosen to obtain suitable measurements.  Obtaining such measurements may require long scan times, leading to a tradeoff between the acquisition duration and the phenomena captured by a model. Even with a reasonably large number of measurements, standard computational methods can produce erroneous results \cite{kerkela2022improved}. As only one example, conventional linear and non-linear least-squares techniques for diffusion kurtosis imaging (DKI \cite{jensen2005diffusional}) tend to produce physically implausible results \cite{tabesh2011estimation, neto2018advanced}. Streamline tractography and tract-specific analysis face inherent challenges and ambiguities such as fiber crossings and bottlenecks \cite{jones2013white, maier2017challenge, thomas2014anatomical, calixto2024white}, which can give rise to high false positive rates and erroneous results. Cross-subject comparisons and population studies can also be significantly hampered by high inter-scanner data heterogeneity  \cite{cai2021masivar, ning2020cross}. Furthermore, most applications are also challenged by paucity of ground truth and lack of universally accepted performance metrics.

\section{Why machine learning may help}
\label{sec:Why_ML}

Machine learning encompasses a rich set of flexible and powerful methods that have the potential to improve upon the more conventional techniques for dMRI analysis. The advantages of machine learning methods over classical techniques depend on the specific tasks, which are discussed in more detail in the following sections. Here, we list some of the main reasons why machine learning may be useful for dMRI analysis.

\begin{itemize}

\item Machine learning methods can learn the complex relationships between the input and output from observing a large number of examples. This may be a fundamentally more plausible and more powerful approach than conventional methods that are based on approximate biophysical models \cite{fick2017assessing} or ad-hoc and simplistic rules such as those used in tractography \cite{poulin2019tractography}.

\item Machine learning models can represent highly complex functions. For neural networks, universal approximation theorems state that they can represent all functions of practical interest \cite{hornik1989multilayer}. This means that machine learning models are far less restricted than conventional methods in terms of the complexity of the phenomena that they can represent. This can be a significant advantage because the true mapping between the input and output may be outside the scope of the mathematical models imposed by conventional methods \cite{nedjati2014machine, nedjati2017machine}.

\item As a result of the above properties, machine learning methods can simplify the complex multi-stage pipelines into simpler end-to-end methods. Many dMRI analysis tasks such as estimation of tissue microstructure or segmentation of white matter tracts rely on a sequence of computations that are optimized separately and independently \cite{golkov2016q, wasserthal2018tractseg}. With machine learning methods, it is often possible to combine these into a single model that can be optimized jointly \cite{wasserthal2018tractseg}.

\item Many machine learning models can effectively and seamlessly integrate various inputs, constraints, and sources of prior knowledge, such as other MRI contrasts or spatial information. As an example, recent works have shown that deep learning methods can easily leverage anatomical MRI data (on top of dMRI data) to improve microstructure estimation \cite{tian2020deepdti}, super-resolution \cite{qin2021multimodal}, tissue segmentation \cite{golkov2016q}, and distortion correction \cite{hu2020distortion}. Given the rich spatial regularity in many neuroimaging problems, incorporation of spatial information may significantly improve the model performance. Spatial information can be too complex to mathematically formulate with conventional methods and, if not done properly, can have a negative impact on performance \cite{gruen2023spatially}. Machine learning models, on the other hand, can effortlessly learn this information directly from data. Modern neural networks such as Convolutional Neural Networks (CNNs), graph CNNs, and transformers have been shown to be especially well-suited for this purpose \cite{hong2019longitudinal, weine2022synthetically, tian2020deepdti, karimi2022diffusion}.

\item Machine learning methods may offer much faster computation \cite{kaandorp2021improved}. This is especially the case for neural networks, where the models consist of a large number of basic operations that can be parallelized on graphical processing units (GPUs). Although the training of these models may require much time, prediction/inference on a new data sample can be orders of magnitude faster than conventional methods \cite{kerkela2022improved, barbieri2020deep, de2021neural, wasserthal2018tract}. The speed advantage has become increasingly more important as image resolution and dataset size continue to grow.

\item Machine learning models may be trained on data with various types and different levels of imperfections such as noise, motion, sub-optimal acquisition protocols, and imaging artifacts. This way, they may gain a degree of robustness with respect to these data imperfections by learning to factor them out in their computations \cite{li2021superdti, de2021neural, neher2015machine, gong2019robust, gong2021deep, weine2022synthetically}. This can be a unique advantage since mathematically modeling these data imperfections can be difficult or impossible. It has been reported, for example by Wegmayr et al. for tractography \cite{wegmayr2018data}, that machine learning models trained on data with noisy labels may produce more accurate and less noisy predictions on test data. Weine et al. developed a pipeline for generating synthetic cardiac diffusion tensor imaging (DTI) data that incorporated simulated heart motion and showed that a deep learning model trained on such data outperformed standard DTI estimation methods \cite{weine2022synthetically}.

\item Because machine learning methods are model-free, they avoid the simplifying approximations, such as Gaussianity of the diffusion process or Ricianity of the measurement noise, which are bound to impact the performance of conventional methods. Moreover, existing mathematical models may be intractable and standard computational methods may generate unreliable results \cite{hill2021machine, nedjati2017machine}. If adequate training data is available, machine learning models can side-step these stumbling blocks and learn the underlying mapping from data.

\item By offering new ways of analyzing the dMRI data, machine learning methods provide new insights into the potential and limitations of this imaging modality. A good case in point is the recent work of Cai et al. on tractography \cite{cai2023convolutional}, where the authors have investigated the possibility of performing tractography based purely on anatomical MRI data without any diffusion encoding. Their results suggest that tractography and structural connectivity analysis based on anatomical MRI may be comparable with those based on dMRI measurements, raising questions about whether standard streamline tractography is driven by tissue microstructure. Other works have used autoencoders and similar bottleneck networks to characterize the information content of dMRI signal and discover common features between different biophysical models \cite{zucchelli2021brain, nath2021dw}.

\end{itemize}

\section{Prior works on dMRI analysis with machine learning}
\label{sec:prior_works}

\begin{figure*}[!ht]
\centering
\includegraphics[width=\textwidth]{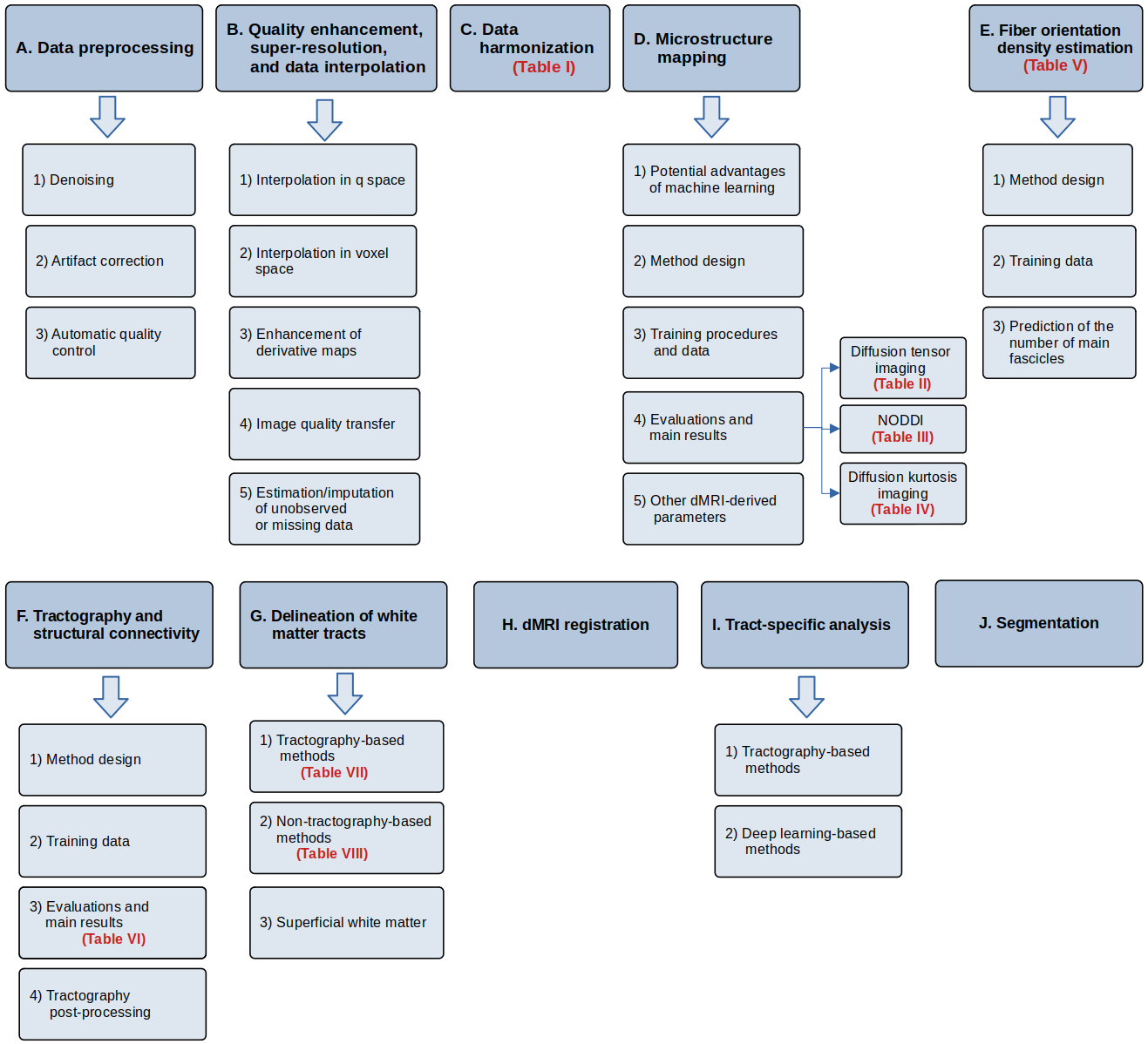}
\caption{Outline of the 10 classes of methods that have been surveyed in Section \ref{sec:prior_works}.}
\label{fig:outline_methods}
\end{figure*}

Due to its unique advantages mentioned above, machine learning has the potential to improve the accuracy, robustness, generalizability, and computational speed for many dMRI analysis tasks. This section describes some of the recent studies that have investigated this potential. The diagram in Figure \ref{fig:outline_methods} shows the outline of the methods covered in this section.

\subsection{Data preprocessing}

Diffusion-weighted MRI measurements suffer from artifacts and imperfections that can influence the results of downstream analysis steps and the conclusions drawn from the data. Therefore, proper data preprocessing is an essential first step for dMRI analysis \cite{tax2022s}. This section describes the recent progress in using machine learning methods to improve pre-processing of dMRI data.

\noindent
\subsubsection{Denoising}

\hspace{3mm}

Measurement noise in dMRI has a spatially-varying distribution and depends on factors such as acquisition and reconstruction methods. It is typically approximated using Rician or non--central chi--square distributions \cite{dietrich2008influence, canales2015spherical}. The noise is stronger at high b-values that are essential for estimating fiber orientation and multi-compartment models, and it is of high practical importance because it can influence the analysis results in subtle but significant and complex ways \cite{jones2004effect, jones2004squashing, roalf2016impact}. Denoising methods based on the principal components analysis and non-local means have been popular in dMRI \cite{veraart2016diffusion, manjon2010adaptive}. These methods were often adapted to the dMRI setting in innovative ways. For example St-Jean et al. \cite{st2016non} combined the non-local matching with dictionary learning. In order to adapt the non-local means method to curved non-Euclidean domain of dMRI signals, Chen et al. \cite{chen2017neighborhood} developed a method based on convolutions on graphs that enabled computing multi-scale features from dMRI data.

In recent years, several neural network-based image denoising algorithms ave been proposed by the computer vision society. These methods, such as Noise2Noise \cite{lehtinen2018noise2noise}, Noise2Void \cite{krull2019noise2void}, and Noise2Self \cite{batson2019noise2self} have often surpassed the performance of classical denoising algorithms. They offer several advantages that are especially useful for medical imaging applications including dMRI: (1) They do not need to model the image prior; (2) They do not need to know the noise distribution; (3) They do not require clean images or paired noisy-clean images for training. The ideas presented in these methods have been adopted to develop dMRI denoising techniques in a few recent studies \cite{fadnavis2020patch2self, tian2022sdndti}. Fadnavis et al. proposed Patch2Self, which is based on learning locally-linear relations between different dMRI volumes \cite{fadnavis2020patch2self}. Patch2Self predicts the signal in one of the q-space elements (i.e., one channel in a 4D representation of the dMRI data) based on the other elements. In order to exploit spatial correlations, the model uses 3D spatial patches as input and target. Effectiveness of Patch2Self was demonstrated in a study on spinal cord \cite{schilling2021patch2self}, where Patch2Self improved FA repeatability and multiple sclerosis (MS) lesion conspicuity. SDnDTI \cite{tian2022sdndti} is another self-supervised method that uses a CNN model. It is specially tailored to single-shell data for DTI estimation. Given a set of N dMRI volumes (each acquired with a different gradient direction but the same gradient strength), SDnDTI selects subsets of six volumes and denoises them. This is done by training the CNN to take these six volumes as input and predict a low-noise target that is computed from all N volumes. This is repeated until all N volumes are denoised. Experiments show that SDnDTI is superior to standard methods such as those based on PCA and non-local means.

\noindent
\subsubsection{Artifact correction}

\hspace{3mm}

Diffusion MRI measurements can also suffer from various artifacts such as those caused by off-resonance fields, subject motion, and Gibbs ringing \cite{pierpaoli2010artifacts, andersson2021diffusion}. In the past two decades, much research has been devoted to computational methods for retrospective correction of these artifacts. A recent review of these methods has been written by Tax et al. \cite{tax2022s}. Machine learning methods have also been proposed to address these artifacts.

Much of the distortion correction algorithms for dMRI, such as those targeting susceptibility-induced distortions, are based on image registration. Image registration for dMRI data is especially challenging because of several factors: (1) Low SNR, especially at higher b-values; (2) Different appearance of dMRI volumes of the same brain acquired with different gradient strengths and directions, which is again exacerbated for higher b values; (3) Difficulty of finding a reference image for registration since all dMRI volumes are affected by distortions; (4) Complex interaction of various artifacts such as signal loss that can further challenge the registration process. Standard methods for distortion correction, such as topup \cite{andersson2003correct}, use iterative optimization to compute the displacement field. Deep learning-based registration methods have the potential to address these challenges because they can be trained in an unsupervised manner on distorted data.

Unsupervised \cite{duong2020unsupervised, zahneisen2020deep, alkilani2023fd}, semi-supervised \cite{legouhy2022correction}, and supervised \cite{hu2020distortion} deep learning-based registration methods have been applied on reversed phase-encoding dMRI data to estimate a displacement field for distortion correction. One study \cite{legouhy2022correction} combined unsupervised training with supervised training using the displacement fields computed by topup as the target. Experiments showed that this semi-supervised method led to better results than purely supervised and purely unsupervised techniques. Most prior works have shown that deep learning techniques match the accuracy of the conventional methods such as topup while being orders of magnitude faster \cite{duong2020unsupervised, alkilani2023fd, legouhy2022correction, zahneisen2020deep, schilling2020distortion}. One study used distortion-free volumes obtained with point-spread-function encoded EPI as registration target and achieved more accurate distortion correction than topup \cite{hu2020distortion}. The input to the network, which was a 2D Fully Convolutional Network (FCN), consisted of distorted dMRI volumes as well as an anatomical T2-weighted image, and the network output was undistorted dMRI volumes. Machine learning methods also have good generalizability. Studies have shown that deep learning models trained on normal brains work well on abnormal brains \cite{hu2020distortion, zahneisen2020deep} and on different (left-right) phase encoding directions \cite{zahneisen2020deep}.

Most deep learning-based methods have used the dMRI volumes as input for computing the displacement fields for distortion correction. Qiao et al., on the other hand, propose to use fiber orientation distribution (FOD) as input \cite{qiao2019fod, qiao2021unsupervised}. They argue that the information in the individual dMRI data volumes is not sufficient to resolve detailed white matter structures that are essential for distortion correction in areas such as the brainstem \cite{qiao2019fod}. Instead, they compute FOD maps from dMRI volumes acquired with opposite phase encodings and train an FCN to compute the displacement fields. This method achieved better results than topup and two alternative deep learning methods in brainstem and white matter. Another interesting work is that of Schilling et al. \cite{schilling2020distortion, schilling2019synthesized}, which enables susceptibility-induced distortion correction when scans with reversed phase-encodings are not available. They have developed deep learning models to generate an undistorted b0 (non-diffusion-weighted) volume given (i) a distorted b0 volume and (ii) an anatomical (T1-weighted) image. The undistorted and distorted images can then be used to compute the displacement field with topup. Their methods are based on FCNs and generative adversarial networks (GANs, \cite{goodfellow2020generative}).

Deep learning methods have been developed to address other types of artifacts in dMRI. Ayub et al. have proposed a method based on variational auto-encoders for in-painting cropped dMRI volumes \cite{ayub2020inpainting}. The model, trained on artificially-cropped volumes and tested on real-world cropped data, was able to correctly inpaint the cropped data. Deep learning methods have been successfully applied to remove Gibbs artifacts in dMRI data. One study trained a CNN on T2-weighted images to estimate Gibbs ringing artifacts \cite{zhang2019mri}. The trained model removed Gibbs artifacts in dMRI volumes as well. Another work trained a CNN on natural images with simulated artifacts and showed that the trained model effectively suppressed Gibbs artifacts in dMRI data and in derived parameter maps such as fractional anisotropy (FA) images \cite{muckley2021training}.

\subsubsection{Automatic quality control}

\hspace{3mm}

Quality control in dMRI data processing pipelines can be challenging because of the diversity of artifacts and large data sizes. A number of deep learning methods have been proposed for automatic quality control in large dMRI studies. Ahmad et al. \cite{ahmad20233d} trained a CNN to classify dMRI volumes as normal or containing artifacts. The artifacts considered in their study included motion-induced signal dropout, inter-slice instability, ghosting, chemical shift and susceptibility-induced distortions. Their method had an accuracy of 92\% on a pool of seven datasets with different acquisition protocols. Another work used a mix of real and simulated data to train a quality control method, based on CNNs, for detecting intra-volume motion and signal dropout \cite{graham2018supervised}. To detect motion-corrupted volumes in neonatal dMRI volumes, Kelly et al. trained a random forest on the outputs of an ensemble of CNNs and achieved classification accuracies of well above 90\% \cite{kelly2017transfer}. Samani et al. also trained a CNN to detect a range of artifacts such as motion artifacts, ghosting, susceptibility-induced artifacts, and chemical shift \cite{samani2020qc}. They reported a detection accuracy of 98\% on multi-scanner dMRI data. Both \cite{graham2018supervised} and \cite{samani2020qc} used transfer learning to train their networks.

\subsection{Quality enhancement, super-resolution, and data interpolation}

Increasing the spatial resolution or q-space sampling density during image acquisition would require increasing the scan time. As an alternative, recent studies have proposed to use computational methods. Much of the proposed methodology have been inspired by deep learning-based super-resolution techniques. However, often these standard deep learning techniques have to be significantly modified to handle the dMRI data. A summary of these works is provided below.

\subsubsection{Interpolation in q space (i.e., angular space)}

\hspace{3mm}

Several works have used neural networks to enhance the angular resolution of q-space data \cite{chen2023deep, yin2019fast, koppers2021enhancing, lyon2023spatio}. Yin et al. used a method based on sparse representation of the dMRI signal and a neural network built on 1D convolutions to map low angular resolution data to high angular resolution data. This method improved the FOD reconstruction accuracy for complex fiber configurations \cite{yin2019fast}. In another study, Ren et al. have developed a method for predicting dMRI data for arbitrary gradient directions \cite{ren2021q}. The input to this model includes only T1-weighted, T2-weighted, and b0 (non-diffusion-weighted) images. Methodologically, the model is a GAN, where both the generator and the discriminator are conditioned on the gradient direction and strength. Experiments are performed with training and test data from the Human Connectome Project (HCP). The results show that the synthesized data can be used to estimate DKI, Neurite Orientation Dispersion and Density Imaging model (NODDI, \cite{zhang2012noddi}), and FOD.

Lyon et al. proposed a parametric continuous convolutional network for angular super-resolution of dMRI data \cite{lyon2023spatio}. Their method improved the results of fixel-based analysis and estimation of NODDI. Another work has proposed a flexible model, based on a recurrent convolutional neural network (RCNN), that can predict the unobserved data for arbitrary target b-vectors \cite{lyon2022angular}. To achieve this flexibility, the model follows an encoder-decoder design where the encoder uses the gradient table information of the measured data as additional input. The decoder, then, takes the latent representation generated by the encoder and the gradient table of the target (unobserved) data to predict the unobserved data. This model accurately upsampled the dMRI data by a factor of 8, and it was more accurate than predictions based on spherical harmonic interpolation and a 1D variant of the model architecture. At higher sub-sampling rates, spatial information study improved the accuracy of angular interpolation \cite{lyon2022angular}. A more recent study proposed a geometric deep learning method to predict unobserved q-space data for arbitrary target acquisition schemes such as single/multi-shell and grid-based schemes \cite{ewert2024geometric}. Experiments showed that this method was superior to model-based (i.e., non-machine learning) techniques and it improved the estimation accuracy for tissue microstructure and FOD.

\subsubsection{Interpolation in voxel space}

\hspace{3mm}

Different from the methods above, a number of works have attempted to increase the spatial resolution of the dMRI data in voxel space \cite{elsaid2019super, tian2021srdti}. It is also possible to jointly improve the spatial resolution and q-space resolution, as proposed in super-resolved q-space deep learning (SR-q-DL) framework \cite{qin2021super}. In this framework, patches from the source low-resolution and the target high-resolution dMRI volumes are represented in dictionaries and a neural network is trained to map the representation of the source signal to that of the target signal. This method significantly improved the accuracy of estimating the NODDI parameters from low-quality dMRI data. In a more recent work, Spears and Fletcher \cite{spears2023learning} have argued that prior voxel-space super-resolution methods such as \cite{qin2021multimodal} are not ideal for tractography, where a continuous FOD field is desired. Instead, they have proposed a method based on continuous and differentiable signal representations with neural networks. They show that their new method achieves state of the art FOD estimation and tractography on HCP data.

\subsubsection{Enhancement of derivative maps}

\hspace{3mm}

A number of works have proposed to enhance the quality/resolution of the derived parameter maps. Zeng et al. developed a method to enhance the resolution of FOD maps \cite{zeng2022fod}. Their method used the FOD estimated from low angular resolution single-shell data to predict the FOD estimated from multi-shell high angular resolution data. Experiments on clinical quality data showed that this method significantly improved the tractography and structural connectivity assessment. A similar study has been reported by Lucena et al., \cite{lucena2021enhancing}, where the authors trained CNNs to map the FODs computed from single-shell data to those computed from multi-shell data. The method worked well and showed good generalizability to different scanners and data acquisition protocols. Ye et al. \cite{ye2019super} proposed to combine a learning-based parameter mapping method similar to \cite{ye2017estimation} with a super-resolution method similar to \cite{tanno2017bayesian} in a joint framework to estimate high-resolution NODDI maps from dMRI measurements with low resolution in both voxel space and q-space.

\subsubsection{Image quality transfer}

\hspace{3mm}

In a series of works, Alexander et al. have developed and promoted the notion of image quality transfer (IQT), which uses machine learning methods such as random forests and neural networks to improve the quality of dMRI parameter maps \cite{alexander2017image, blumberg2018deeper}. The goal of IQT is to use high-quality data (such as the HCP data) to improve the quality of regular dMRI data that are typically obtained in clinical applications. The method works by learning a regression function to map the low-quality dMRI patches to high-quality patches. This idea has been successfully applied to increase the spatial resolution of diffusion tensor images and to estimate NODDI parameters from low-quality single-shell measurements. Furthermore, it has been shown that on low-quality data, IQT enables tractography-based reconstruction of thin fiber tracts that can typically only be reconstructed using specialized data.

\subsubsection{Estimation/imputation of unobserved or missing data}

\hspace{3mm}

Estimating unobserved q-space shells from observed ones has been reported by several works \cite{murray2023neural, koppers2017diffusion, chen2023deep}. Murray et al. used neural networks to predict multi-shell data from single-shell data and used the predicted data to estimate the NODDI model \cite{murray2023neural}. They achieved satisfactory results on healthy brains as well as brain of MS patients. In a similar work, Chen et al. \cite{chen2023deep} successfully trained an multilayer perceptron (MLP) to predict six-shell data from two-shell data. A closely related application is imputation of missing data, which is of particular importance because some dMRI measurements often have to be discarded due to strong noise, artifacts, or excessive motion, such as in young or noncooperative subjects. To address this problem, Hong et al. developed a method based on graph CNNs \cite{hong2019multifold}. They represented the relation between the measurements in the voxel space and q-space using a graph, and trained a residual graph CNN to learn this relation. They trained the model using adversarial techniques and successfully applied it on data from healthy infants between 0 and 12 months old. This method was able to reconstruct accurate maps of generalized FA (GFA) and FOD from five-fold-reduced slice acquisitions.

\subsection{Data harmonization}
\label{sec:harmonization}

Data harmonization is crucial for reliable analysis and comparison of inter-scanner and inter-site dMRI data. It has become an increasingly more relevant problem as larger and geographically/demographically more diverse datasets are used in multi-center neuroimaging studies. Statistical methods such as ComBat \cite{fortin2017harmonization} and RISH \cite{mirzaalian2016inter} have achieved considerable success in addressing this problem. However, they have important limitations. For example, they depend on data from matched subjects in the reference and target datasets, and their performance deteriorates when the differences between reference and target datasets are nonlinear \cite{cetin2020exploring}.

\begin{table*}[!htb]
\centering
\caption{A listing of some of the machine learning-based methods for data harmonization in dMRI.}
 \label{table:harmonization_table}
\begin{tabular}{C{30mm} | L{135mm} }
\thickhline
Method & Summary of methodology and results \\ \thickhline
Semi-supervised contrastive learning using CNNs \cite{hansen2022contrastive} & This method uses paired data to learn subject-specific and acquisition-specific representations. A decoder is trained to map the subject-specific representations to the target contrast. The method is shown to be superior to a method based on CycleGAN \cite{zhu2017unpaired} as well as interpolation based on SHORE \cite{ozarslan2009simple}. It can handle heterogeneous data from multiple sites, different acquisitions protocols, and demographics.  \vspace{2mm} \\
Null space learning with an MLP architecture \cite{nath2019inter} & The model is trained using paired dMRI scans of human subjects as well as dMRI-histology data from squirrel monkeys. The loss function is designed to encourage accurate FOD estimation and consistent scan-rescan GFA. Experiments show that this method has high accuracy in FOD estimation and high reproducibility on test data from unseen scanners. Experiments also show that this approach results in higher FOD estimation accuracy across 1.5T and 3T scanners. \vspace{2mm}  \\
Variational auto-encoder \cite{moyer2020scanner} & The method is based on learning a latent representation that is invariant to scanner, scanning protocol, or similar confounders. The latent representation can then be used to reconstruct the image content that is stripped of those factors. The method can be trained without paired scans. Training involves an adversarial loss that attempts to predict the source (scanner, etc.) of the acquisition. Results show that, compared with a well-established conventional technique \cite{mirzaalian2018multi}, this new method achieves superior results in terms of several parameters including FA, MD, and fiber orientation. \vspace{2mm}  \\
Residual CNN \cite{koppers2019spherical} & This method is based on predicting the spherical harmonics representation of the dMRI signal in the target domain from that of the source domain. It works on 3D patches of size 3. A final non-learned projection in the spherical harmonics space is needed to make sure fiber orientations are not changed due to the intensity harmonization. This method requires that the same subjects be scanned in the source and target scanner/acquisition protocol. Evaluations have shown that this method can achieve effective harmonization of the dMRI signal, FA, and MD.  \vspace{2mm} \\
Hierarchical CNNs \cite{tong2020deep} & The focus of this work is on harmonization of DKI measures. The method requires a set of subjects to be scanned with both target and source scanners/acquisition protocols. It computes the DKI metrics using the data in the target domain. The network is trained to map the dMRI data in the source domain directly to the DKI measures in the target domain, after nonlinear intra-subject registration. This method reduced the inter-scanner variation in DKI measures by 51-66\%.  \vspace{2mm} \\
CNN with a scan-rescan consistency loss \cite{yao2023deep} & This method is proposed and evaluated specifically for FOD estimation, although it seems to be directly applicable for estimating any other parameter. A consistency loss penalizes the model for divergent FOD predictions from different scans of the same subject. It achieves better FOD estimation accuracy and inter-scanner consistency than standard techniques on external test data. \vspace{2mm}   \\
Ensemble of different neural networks \cite{blumberg2019multi} & The authors present their method as a multi-task learning approach. Essentially, a set of neural networks, which can have different architectures and trained for different prediction tasks, are combined via training a set of additional neural networks that utilize the features learned by these networks to predict the parameter of interest. Compared with state of the art deep learning techniques, this method achieved better dMRI signal prediction. \vspace{2mm}  \\
Adaptive dictionary learning \cite{st2020harmonization} & A dictionary is learned on the set of reference datasets/scanner(s). It is assumed that representing a test scanner's data in this dictionary automatically harmonizes the data towards the reference dictionary by suppressing the features that are specific to the test data. The method does not require paired subjects in the source and reference datasets. Successful harmonization results are reported in terms of FA, MD, and rotationally invariant spherical harmonics representations.  \\
\thickhline
\end{tabular}
\end{table*}

Recently, various machine learning-based methods have been proposed, but they still have not achieved the popularity of methods such as ComBat and RISH. A recent study compared a range of interpolation methods, statistical regression techniques, and deep learning methods for cross-scanner and cross-protocol multi-shell dMRI data harmonization \cite{ning2020cross}. A regression method based on \cite{karayumak2019retrospective} performed better than all deep learning methods. However, some of the deep learning methods were among the best performing techniques. The authors of \cite{ning2020cross} hypothesized that the performance of the deep learning methods may significantly improve with larger training datasets than the 10 training subjects used in that work. Tax et al. evaluated five different learning-based harmonization methods (including four neural networks and a dictionary-based technique) on data from different scanners and with different maximum gradient strengths \cite{tax2019cross}. Their results showed that overall these learning-based methods were successful in harmonizing cross-scanner and cross-protocol data, although no comparison with the state of the art non-learning methods was performed. Their analysis also showed that the learning-based methods were more effective on isotropic measures such as MD than on anisotropic measures such as FA. They attribute this behavior to the possibility that, because the spatial variations in isotropic measures are less abrupt, imperfect spatial alignments may be less harmful when applying machine learning methods on imperfectly-registered pairs of dMRI volumes.

Table \ref{table:harmonization_table} lists some of the existing machine learning methods for dMRI data harmonization. In general, these methods offer higher flexibility in modeling the sources of heterogeneity and in handling data from unpaired subjects. For example, one study developed a method based on representing the data in a disentangled latent space that allowed for separating the effects of anatomy and acquisition \cite{hansen2022contrastive}. Similarly, variational auto-encoders have been proposed to harmonize dMRI data by learning a latent representation that is invariant to site/protocol-specific factors \cite{moyer2020scanner}. Focusing on the specific task of FOD estimation, another study developed a method, named null space deep network, that seamlessly integrated a small dMRI dataset with histology ground truth and a large scan-rescan dMRI dataset without histology ground truth \cite{nath2019inter}. The method showed superior FOD estimation accuracy on data from scanners that had not been included in the training set. A similar method was applied to harmonize structural connectivity metrics on a two-center dataset \cite{newlin2023comparing}. The authors found that optimal harmonization was achieved when the deep learning method was applied to harmonize the dMRI data at a voxel level followed by applying ComBat \cite{fortin2017harmonization} to harmonize the structural connectivity metrics. Another work from the same research team, again focusing on FOD prediction, proposed an extra loss term to encourage consistency of FOD estimations computed from scan-rescan data of the same subject \cite{yao2023robust}. This new loss function improved the generalizability of FOD estimation as well as downstream tasks such as structural connectivity assessment. Blumberg et al. have proposed a multi-task learning strategy, where a neural network is trained to combine the information learned by an ensemble of different neural networks \cite{blumberg2019multi}. The individual neural networks in the ensemble may have been trained on separate datasets, possibly for completely different tasks. The added neural network model uses the high-level features learned by the ensemble to perform data harmonization. The authors argue that this approach can be especially effective in scenarios where one or some of the datasets are very small.

\subsection{Microstructure mapping}
\label{sec:microstructure_mapping}

The brain white matter consists of a network of neuronal fibers that are supported by other cellular components \cite{edgar2009white, novikov2019quantifying}. Microstructural characteristics of the white matter tissue such as myelination influence its physical properties such as viscosity, density, and permeability \cite{kuroiwa2006ex, mellergaard1989time}. These physical properties, in turn, can influence the diffusion of water molecules. Although the size of these microstructural elements are in the micrometer range, dMRI can probe the tissue microstructure because dMRI signal is sensitive to the displacement of water molecules at the micrometer scale \cite{le2014diffusion, assaf2014inferring}. Therefore, the dMRI signal can be an indicator of changes/variations in brain tissue microstructure due to normal/abnormal development or diseases \cite{bozzali2007diffusion, lakhani2020advanced, calixto2024detailed, bodini2009diffusion, salat2014diffusion}.

There have been many efforts and much progress in developing bio-physical models that relate the measured dMRI signal to the microstructure of brain tissue \cite{novikov2019quantifying, novikov2018modeling}. Advanced models rely on specialized measurement protocols and complex numerical optimization methods. The underlying estimation problem is typically nonlinear, sensitive to measurement noise and initialization, and may be unstable and computationally intensive \cite{harms2017robust, jelescu2016degeneracy}. Accuracy and precision of model fitting is hard to verify and can significantly depend on the optimization algorithm used to fit the model to the measurements \cite{harms2017robust}. To avoid local minima, some works have resorted to computationally intensive approaches such as grid search, multi-start methods, cascade optimization, and stochastic optimization techniques \cite{alexander2010orientationally, alexander2008general, jelescu2016degeneracy, harms2017robust}. There is a great interest in developing methods, such as those based on dictionary matching \cite{rensonnet2019towards, daducci2015accelerated}, to reduce the computational time without compromising the estimation accuracy. An example of such methods is AMICO \cite{daducci2015accelerated}, which reformulates the microstructure estimation equations as convex optimization problems that can be solved much faster.

\subsubsection{Potential advantages of machine learning methods}

\hspace{3mm}

In recent years, machine learning has increasingly been applied to these estimation tasks. Overall, five main justifications have been cited for preferring a machine learning-based approach for this application.

\begin{enumerate}

\item Unlike the conventional methods that presume a known fixed relationship between the dMRI measurements and the target parameter, machine learning methods can learn this relationship from data \cite{fick2017assessing, golkov2016q}. For certain microstructural parameters, such as residence time of water inside axons, it is believed that existing mathematical models that express the relationship with the dMRI signal are either too simplistic or intractable, and that the existing numerical forward models are computationally too expensive to be used in estimating the parameters from data \cite{hill2021machine, nedjati2017machine}. Fick et al. show that, for axonal diameter estimation, existing signal models fail outside a narrow range of diameters while a machine learning method can achieve accurate estimation for the whole range of diameters in the data \cite{fick2017assessing}.

\item Conventional methods often involve several steps with no feedback from the later steps to the earlier steps \cite{golkov2016q, harms2017robust}. As a result, these methods are hard to design and optimize. Machine learning methods such as deep neural networks, on the other hand, may be optimized end-to-end as a single processing step.

\item Machine learning methods can be much faster than numerical optimization routines. Again, such is the case for neural networks that run on GPUs \cite{kaandorp2021improved, barbieri2020deep, de2021neural}.

\item Standard methods perform the model fitting in a voxel-wise manner, which fails to exploit the spatial correlations to improve the estimation accuracy. Machine learning models can effectively learn complex spatial correlations directly from data and leverage this knowledge to improve the estimation accuracy \cite{gong2023machine}.

\item With machine learning methods it is typically much easier to incorporate prior knowledge or additional information. For example, it is generally easy to impose constraints on the parameter values to be estimated and to include other MRI contrasts as input to machine learning models.
    
\end{enumerate}

Tables \ref{table:DTI_table}, \ref{table:NODDI_table}, and \ref{table:DKI_table} list some of the recent works that have employed machine learning to estimate, respectively, DTI, DKI, and NODDI parameters. A summary of the methods developed in these works and their experimental results is presented below.

\begin{table*}[!ht]
\centering
\caption{Machine learning methods for estimating the diffusion tensor or its derived parameters such as FA and MD.}
\label{table:DTI_table}
\begin{tabular}{C{30mm} | L{140mm}  }
\thickhline
Method & Summary of methodology and results  \\ \thickhline
SuperDTI \cite{li2021superdti} & This work reports accurate estimation of FA, MD, and the main diffusion tensor eigenvector from six measurements. The model trained on healthy HCP brains works well on pathological brains. Evaluations included qualitative and quantitative tractography assessment.  \vspace{1mm}  \\
DeepDTI \cite{tian2020deepdti} & This method trains a CNN to compute \emph{dMRI data residuals}. Specifically, it computes the residuals between under-sampled ($n=6$) data and high-quality targets computed from densely-sampled data. The diffusion tensor is then computed with a standard method. Compared with conventional estimation techniques, this method reduced the number of measurements by a factor of 3.3-4.6. It improved DTI estimation, DTI-based tractography, and tract-specific analysis on twenty prominent tracts. Only HCP data is used for validation.  \vspace{1mm}   \\
Aliotta et al. \cite{aliotta2019highly} & This work reported higher FA and MD reconstruction accuracy and precision than standard estimation methods from as few as 3 diffusion-weighted measurements. With a model trained on data from 10 healthy subjects, FA-based delineation of brain tumors was more accurate than with a standard method.   \vspace{1mm} \\
DIFFnet \cite{park2021diffnet} & The main contribution of this work is a method to handle the measurements acquired with different schemes. The method simply projects and bins the the q-space data in standard orthogonal planes. Compared with standard methods, they report faster computation and lower error.   \vspace{1mm} \\
Fetal DTI estimation \cite{karimi2021deep} & This work has reported reconstruction of fetal DTI with unprecedented accuracy. The method was trained using synthetic data generated with a novel pipeline that used both fetal in-utero data and scans of premature neonates. Evaluations included quantitative comparisons with conventional methods as well as detailed assessment by human experts. \vspace{1mm}  \\
Patch-CNN \cite{goodwin2023patch} & The proposed method, a patch-wise CNN, reduces the required number of measurements by a factor of two compared with standard estimation methods. The estimated DTI maps were used to trace major white matter tracts with high accuracy. \vspace{1mm}  \\
Transformer-based DTI estimation \cite{karimi2022diffusion} & This work used transformer networks to learn the spatial correlations in dMRI signal and in diffusion tensor. The method reconstructed the diffusion tensor with superior accuracy while reducing the required number of measurements by factors of 5-15. Evaluations include tractography and structural connectivity.  \vspace{1mm} \\
Cardiac DTI estimation (FG-Net) \cite{liu2023accelerated} & This study has achieved accurate cardiac DTI estimation from six dMRI measurements. It includes a basic FCN that estimates the dMRI data for additional directions than the six measured. The method predicts DTI metrics more accurately than conventional methods on ex-vivo data. \vspace{1mm}  \\
Cardiac DTI \cite{weine2022synthetically} & The model (a residual CNN) trained with purely synthetic data performs well on synthetic as well as in-vivo test data for cardiac DTI estimation. It is less prone to predicting implausible values and enables more accurate detection of tissue lesion. \vspace{1mm}  \\
SwinMR (cardiac DTI) \cite{huang2024deep} & Deep learning methods can reconstruct cardiac DTI with k-space under-sampling rates of up to 4 without any significant quality reduction compared to reference. A transformer network is shown to achieve superior results than a CNN. \vspace{1mm}  \\
\thickhline
\end{tabular}
\end{table*}

\subsubsection{Method design}

\hspace{3mm}

Methodologically, these works have been dominated by neural networks, although other models such as random forests \cite{nedjati2017machine, fick2017assessing, hill2021machine} and Bayesian estimation methods \cite{mozumder2019population, reisert2017disentangling} have also been used. The first work to propose deep neural networks for this purpose was the ``q-space deep learning'' method \cite{golkov2016q}. It reported accurate estimation of DKI parameters with 12 measurements and NODDI parameters with 8 measurements, resulting in a twelve-fold reduction in scan time. Many studies have followed \cite{golkov2016q} to develop neural network models for tissue microstructure mapping. This should not be surprising given that deep neural networks have emerged as the best-performing regression models in many applications \cite{lathuiliere2019comprehensive}. Most non-neural-network methods either predate the recent surge of deep learning or do not include a rigorous comparison with neural networks.

Incorporation of population-estimated priors has been advocated by several studies \cite{mozumder2019population, karimi2022atlas}. For estimation of NODDIDA (NODDI with Diffusivity Assessment), it was shown that estimating a multivariate Gaussian prior from 35 subjects significantly improved the prediction accuracy and robustness \cite{mozumder2019population}. The authors concluded that incorporation of the prior reduced the ill-posedness of the estimation problem and made it possible to estimate this complex model from clinically-feasible measurements. Some works have suggested utilizing other inputs in addition to the dMRI measurements. For instance, one work has used T1-weighted and T2-weighted images (registered to the dMRI data) to improve diffusion tensor estimation \cite{tian2020deepdti}.

\subsubsection{Training procedures and data}

\hspace{3mm}

Most studies have adopted an end-to-end training approach, where the input dMRI measurements are mapped to the target parameter of interest using a single (albeit deep) neural network. However, there are notable exceptions that are instructive. For DKI, Kerkela et al. \cite{kerkela2022improved} used the neural network predictions as input to a regularized non-linear least-squares method to compute the final values. This approach improved the estimation robustness of the standard method and reduced the probability of predicting implausible values. Other studies have also proposed to combine neural network estimators with conventional methods. For example for computing the NODDI parameters under very low signal to noise ratio (SNR), Gong et al. used neural networks to obtain good initial estimates, which were then used to obtain more accurate estimates using maximum likelihood estimation (MLE) \cite{gong2023machine}. The justification for this approach is that when the SNR is low, MLE is prone to erroneous predictions because of its vulnerability to local minima. The predictions of the neural networks, on the other hand, although very fast, can suffer from relatively small but significant biases. Hence, one can use the neural network to compute a good initial estimate, which can then be refined using an unbiased MLE estimator. A similar approach was proposed in \cite{faiyaz2022single}, where a shallow neural network was applied to estimate the isotropic volume fraction in the NODDI model, which was subsequently used in an MLE formulation to compute the complete NODDI model from single-shell data.

For diffusion tensor estimation from six measurements, Liu et al. argue that directly mapping the measurements to the diffusion tensor is not optimal \cite{liu2023accelerated}. Given the measurements along six gradient directions, they employ a neural network to estimate the signal along additional directions. These measurements are subsequently used by a second network to estimate the diffusion tensor. Ye et al. \cite{ye2019deep, ye2017tissue} have proposed methods based on sparse representations and deep neural networks. In one implementation, their method is a two-stage pipeline: the first stage uses an LSTM model to compute a sparse decomposition of the signal in a dictionary. An MLP computes the desired microstructure indices based on the sparse representation coefficients. This work has been in part inspired by the widely-used AMICO method \cite{daducci2015accelerated}, which decouples the estimation of NODDI and ActiveX \cite{alexander2010orientationally} models into linear problems and solves them using sparsity-based optimization. Ye et al.'s experiments show that this method can estimate NODDI parameters with higher accuracy than standard estimation techniques. This method was further improved with the help of a separable spatial-angular dictionary for signal representation \cite{ye2020improved}.

\begin{table*}[!ht]
\centering
\caption{A summary of recent machine learning techniques for estimating the parameters of the NODDI model.}
\label{table:NODDI_table}
\begin{tabular}{C{30mm} | L{140mm}  }
\thickhline
Method & Summary of methodology and results  \\ \thickhline
DLpN \cite{faiyaz2022single} & A dictionary learning method is first used to estimate the isotropic volume fraction. Subsequently, a neural network computes the remaining NODDI parameters. This method estimates the NODDI parameters from a single-shell (b=2000) acquisition as accurately as conventional methods with multi-shell data. The method is also validated on clinical data. \vspace{1mm} \\
Deep sparse representation methods \cite{ye2019deep} \cite{ye2020improved} & These related works compute sparse representation of the signal in dictionaries. Sparse representation coefficients are then used by neural networks to predict the microstructure indices. Using only HCP data, more than twice reduction in estimation error compared with optimization methods are reported.  \vspace{1mm}  \\
MEDN \cite{ye2017tissue} & This method is based on representing the dMRI signal in a dictionary-matching framework inspired by AMICO \cite{daducci2015accelerated} and related to the above two methods \cite{ye2019deep, ye2020improved}. An MLP-type network is used to compute the NODDI parameters from the signal representations. The method achieves more than twice lower estimation error and faster computation compared with standard techniques.  \vspace{1mm} \\
Gibbons et al. \cite{gibbons2019simultaneous} & This work simultaneously estimated NODDI and generalized FA. The trained model worked well on data from healthy individuals and stroke patients. The method is a basic 2D FCN. \vspace{1mm}  \\ 
Chen at al. \cite{chen2023deep} & This work used MLP models to predict unmeasured dMRI signal and microstructure indices including NODDI parameters. New loss functions are introduced to encourage accurate prediction of the dMRI signal and microstructure indices. Using only HCP-style data, the effectiveness of the new loss functions are demonstrated. No comparisons with standard estimation methods are reported.  \vspace{1mm} \\
METSC \cite{zheng2023microstructure} & This work makes use of sparsity-based representation of the dMRI signal, computed using an unfolded iterative hard thresholding method, and a transformer network. It achieves an eleven-fold reduction in the required number of measurements compared with conventional methods and reduces the estimation error compared with other learning-based methods.  \vspace{1mm} \\
q-space deep learning \cite{golkov2016q} & Using a voxel-wise estimation with an MLP, this work reported a twelve-fold reduction in the required number of measurements to achieve the same level of accuracy as standard methods. As few as 8 measurements were sufficient for accurate estimation. Four different datasets including one from MS patients are used. \vspace{1mm} \\
Machine learning-informed estimation \cite{gong2023machine} & The focus of this work is on very low SNR scenarios such as spinal cord imaging. An MLP is used to estimate good initial values for the NODDI parameters. Refined estimates are subsequently computed using a maximum likelihood estimation technique.  \vspace{1mm} \\
SR-q-DL \cite{qin2021super} & This method follows a super-resolution framework. Specifically, the low-resolution dMRI signal volumes are used to compute high-resolution tissue microstructure maps. The method itself has two stages. The first stage learns a sparse representation of the dMRI signal with 1D convolutions. The second stage is a CNN that performs the computation. Evaluations have shown that this method outperforms conventional and machine learning methods. However, only HCP data is used to develop and validate the method. \vspace{1mm}  \\
DIFFnet \cite{park2021diffnet} & This is a CNN-based method that achieves low reconstruction error while reducing the computation time by three orders of magnitude.  \vspace{1mm}  \\
Population‐based Bayesian regularization \cite{mozumder2019population} & This is a Bayesian estimation method, where a population-informed prior is computed from a cohort of healthy subjects and included in the estimation for test subjects. Results show that the use of the population-based prior helps alleviate the ill-posedness of the estimation problem and significantly improves the estimation accuracy and robustness. The model was developed using HCP data and validated on an independent dataset.  \vspace{1mm}  \\
\thickhline
\end{tabular}
\end{table*}

The most common training approach has been supervised training, where the loss function to be minimized is computed based on the difference between the predicted and ground truth parameter values. However, there have been important and instructive exceptions. Kaandorp propose to train a neural network model in an unsupervised manner by using the predicted tissue parameters to predict the corresponding dMRI signal and optimizing the network weights to minimize the difference between the predicted and measured signal \cite{kaandorp2021improved, kaandorp2023deep}. They found that predictions of the unsupervised methods had higher variability compared with the supervised methods. Predictions of the supervised methods, on the other hand, displayed a strong bias towards the mean of the training data and were deceptively smooth \cite{kaandorp2023deep}. Epstein et al. have analyzed the bias-variance trade-off in supervised and self-supervised microstructure estimation methods \cite{epstein2022choice}. They argue that the reason for the lower bias of self-supervised methods is because they are based on the same optimization objective as MLE, while the higher prediction bias of supervised methods is because they deviate from this objective and are based on a poor choice of training target. They show that using an MLE-computed parameter value as the estimation target can reduce the estimation bias of supervised methods.

Most commonly, target microstructure values for the training data are computed by applying a standard estimation method on dMRI measurements. This is often justified by using densely-acquired high-SNR measurements to improve the accuracy of this computation. Then, the machine learning model is trained with downsampled measurements to match the estimation accuracy of the standard method. However, this approach inevitably inherits some of the limitations of the standard method that is used to compute the prediction target. Very little attention has been paid to this issue \cite{zucchelli2021brain}. One study proposed to inspect the results computed by the standard fitting method and remove the voxels that contain implausible values \cite{kerkela2022improved}.

In some applications, obtaining reliable in-vivo training data may be hard or impossible. Two examples are cardiac DTI \cite{weine2022synthetically} and fetal imaging \cite{karimi2021deep}. For cardiac DTI, Weine et al. \cite{weine2022synthetically} developed a parameterized pipeline to synthesize training data with realistic spatial correlations and cardiac motion. For fetal brain DTI, another work proposed a pipeline that synergistically combined data from preterm neonates and fetuses to synthesize realistic fetal dMRI data \cite{karimi2021deep}. Both studies reported good results on independent real test data. Using synthetic data has the added advantage that it can incorporate a much wider range of parameters and factors such as noise and motion than can be available in any in-vivo dataset.

\begin{table*}[!ht]
\centering
\caption{A summary of recent machine learning methods for DKI estimation.}
\label{table:DKI_table}
\begin{tabular}{ C{30mm} | L{140mm}  }
\thickhline
Method & Summary of methodology and results  \\ \thickhline
Separable dictionaries and deep learning \cite{hashemizadehkolowri2022jointly} & A comparison of different deep learning methods showed that a technique that utilized separable dictionaries \cite{ye2020improved} achieved the best results for jointly estimating DTI, DKI, and NODDI parameters. Only HCP data were used. Moreover, the deep learning methods show some systematic biases, such as underestimation of radial kurtosis. \\
q-space deep learning \cite{golkov2016q}  & This work has reported accurate estimation of DKI parameters from 12 measurements. The model is a voxel-wise MLP, which is tested in multiple external datasets.  \vspace{1mm} \\ 
Li et al. \cite{li2019fast} & The method estimates DTI and DKI metrics jointly. Results show that DKI parameters can be estimated with 8 measurements.   \vspace{1mm} \\
Masutani \cite{masutani2019noise} & This work trained an MLP using synthetic data. Only four diffusion-weighted measurements were used for DKI estimation. Tests on synthetic and real data showed that high estimation accuracy was achieved. This work also demonstrated the importance of noise level matching between training and test datasets.  \vspace{1mm} \\
Hierarchical CNNs \cite{gong2021deep} & The deep learning model is a hierarchical CNN that is combined with a motion detection and rejection technique. Experiments, including data from children with attention deficit hyperactivity disorder, show that the new method is robust to head motion and can estimate the DKI parameters from eight measurements. It works well in the presence of severe motion, when up to 90\% of the data are motion-corrupted.  \vspace{1mm} \\
MLP + least squares \cite{kerkela2022improved} & This work uses an MLP to estimate initial DKI parameter values, which are then used within a regularized least-squares procedure to compute the final values. This approach avoids implausible predictions and improves model robustness compared with existing linear and non-linear optimization methods. The data includes scans of 10 healthy volunteers.   \vspace{1mm} \\
\thickhline
\end{tabular}
\end{table*}

\subsubsection{Evaluations and main results}

\hspace{3mm}

Overall, the results of the published studies suggest that machine learning methods may be capable of achieving higher estimation accuracy than conventional methods (Tables \ref{table:DTI_table}, \ref{table:NODDI_table} and \ref{table:DKI_table}). They may also be able to reduce the required number of measurements. Some works have found that estimation accuracy of deep learning methods may be less affected by sub-optimal acquisition protocols than numerical optimization methods \cite{de2021neural}. There are, however, studies that have reported contrary results that deserve careful consideration. For example, one study has reported that neural network and random forest models trained on simulated data may produce less accurate and more biased results compared with nonlinear least-squares estimation methods \cite{gyori2022training}. Another study has shown that deep learning methods can achieve good estimation accuracy when the optimization landscape is well-behaved but that they have poor performance when the optimization landscape is degenerate such as for multi-compartment models with even two compartments \cite{de2021use, mozumder2019population}.

It has also been reported that machine learning methods are less sensitive to measurement noise and other imperfections \cite{jung2021artificial, gong2019robust}. As an example, Gong et al. trained a deep learning model to compute DTI and DKI parameters using data from healthy subjects with voluntary head motion as well as attention deficit hyperactivity disorder (ADHD) patients with non-voluntary motion \cite{gong2019robust}. They found that predictions of the deep learning model were less sensitive to motion and comparable to predictions with motion-free data.

Furthermore, it has been shown that models that leverage spatial correlations in a neighborhood (e.g., an image patch) around the voxel of interest can lead to more accurate estimation \cite{gibbons2019simultaneous, li2021superdti, hashemizadehkolowri2022jointly, aliotta2021extracting}. Early works used CNNs to learn the spatial correlations. More recent studies have relied on attention-based neural networks that are considered to be more effective in learning correlations \cite{zheng2023microstructure, karimi2021deep, karimi2020robust}. Zheng et al. argue for superiority of transformer architectures for tissue microstructure mapping in dMRI \cite{zheng2023microstructure}. In order to enable effective implementation of these architectures with small datasets, they introduce additional computational modules. Specifically, they use a sparse representation stage to compute the signal embeddings in a dictionary using unfolded iterative hard thresholding.

Although most studies have evaluated the new methods only in terms of estimation accuracy metrics such as root mean square of the error (RMSE), several works have investigated downstream use of the estimated tissue microstructure parameters. For example, Gibbons et al. showed that post‐stroke outcome prediction was the same for microstructure maps estimated with standard methods and those estimated with a neural network with 10-fold fewer measurements \cite{gibbons2019simultaneous}. Aliotta et al. observed that deep learning-estimated FA maps resulted in more accurate delineation of the brain tumor boundaries in glioblastoma multiforme patients than with conventional methods \cite{aliotta2019highly}. Kaandorp et al. showed that intravoxel incoherent motion (IVIM) parameters computed by a deep learning method better predicted the chemoradiotherapy response of pancreatic cancer patients than a standard method \cite{kaandorp2021improved}. Another work assessed the microstructure maps estimated by several deep learning methods in terms of their effectiveness in studying the impact of migraine on the brain white matter \cite{aja2023validation}. Specifically, FA and mean diffusivity (MD) values computed by different methods were compared using Tract-Based Spatial Statistics (TBSS) \cite{smith2006tract}. It was observed that, compared with a standard estimation technique, deep learning methods improved the true positive rate but also increased the false positive rate.

For DTI estimation, several works have reported good reconstructions from the theoretical minimum of six measurements (Table \ref{table:DTI_table}). DeepDTI \cite{tian2020deepdti} uses six diffusion-weighted measurements and a non-weighted (b0) measurement, T1-weighted, and T2-weighted images as input and computes the residuals with respect to high-quality measurements. The estimated high-quality measurements are then used to estimate the diffusion tensor with a standard estimation technique. Detailed analysis, including local assessment of DTI-derived parameters, tract-based analysis, and tractography show that DeepDTI achieves accurate estimation while reducing the number of measurements by a factor of at least three. DeepDTI is based on a 3D FCN and requires that the six diffusion-weighted measurements be acquired along the optimal directions proposed in \cite{skare2000condition}. SuperDTI, proposed by Li et al. \cite{li2021superdti}, also used an FCN for direct estimation of the diffusion tensor. It achieved accurate FA computation and tractography with six measurements and accurate MD computation with three measurements. Similar results were reported in \cite{aliotta2019highly, aliotta2021extracting}. Using transformer networks, another study made a similar claim and further evaluated the method using tractography and quantitative structural connectivity assessment \cite{karimi2022diffusion}. DiffNet \cite{park2021diffnet} estimated FA and MD with 3-20 diffusion-weighted measurements and showed that the estimation accuracy was higher than standard methods. DiffNet-estimated FA was more accurate for segmenting brain tumor volume, while also reducing the scan time.

\subsubsection{Other dMRI-derived parameters}

\hspace{3mm}

In addition to the DTI, DKI, and NODDI parameters, machine learning methods have also been applied to compute other parameters. Examples include myelin water fraction \cite{jung2021artificial, fick2017assessing}, water residence time in white matter (related to axonal permeability) \cite{nedjati2017machine, hill2018deep, hill2021machine}, axonal radius \cite{nedjati2017machine, fick2017assessing}, spherical mean technique (SMT) \cite{ye2019deep, gyori2022training, hashemizadehkolowri2022jointly}, ensemble average propagator (EAP) models \cite{ye2017learning, ye2019deep}, and relaxation-diffusion models of white matter microstructure \cite{de2021neural, grussu2021deep, de2021use}. Parik et al. developed a novel magnetic resonance fingerprinting sequence to simultaneously estimate T1-weighted, T2-weighted, and diffusion tensor parameters. Instead of the standard dictionary matching approach, they used a deep neural network to estimate the target parameters and achieved good results in healthy brains and brains of MS patients \cite{pirk2020deep}. 

A number of works have attempted to use deep neural networks to estimate the IVIM model \cite{kaandorp2023deep, barbieri2020deep, bertleff2017diffusion, kaandorp2021improved, parker2023rician, zhang2019implicit, zheng2023microstructure, epstein2022choice}. One study showed that an MLP could achieve results that were comparable with or more accurate than least-squares estimation and Bayesian techniques \cite{barbieri2020deep}. Another study found that a neural network could outperform the state of the art estimation methods for computing the parameters of a combined IVIM-kurtosis imaging model \cite{bertleff2017diffusion}. It also showed that the neural network method made fewer predictions that were outside the range of plausible parameter values. One study showed that a neural network, trained using an unsupervised approach, produced more consistent and more accurate estimations than standard methods and it better predicted the chemoradiotherapy response of pancreatic cancer patients \cite{kaandorp2021improved}.

\subsection{Fiber orientation density estimation}

Estimation of fiber orientation distribution (FOD), especially in regions with complex fiber configurations, requires high angular resolution diffusion imaging (HARDI) measurements. However, even when such measurements are available, this is a challenging estimation task. Recent studies have highlighted the inherent limitations of FOD estimation based on dMRI data \cite{schilling2018histological, schilling2016comparison}. Nonetheless, the advantages of estimating crossing fiber configurations for applications such as tractography and connectivity analysis has been shown time and time again \cite{prvckovska2016reproducibility, bucci2013quantifying}. As a result, there have been ongoing efforts to improve the accuracy of FOD estimation. The simplest approach to characterizing crossing fibers may be the multi-tensor model \cite{peled2006geometrically, tuch2002high}. However, these methods require determination of the number of tensors in each voxel, which is a difficult model selection problem \cite{scherrer2013reliable, schultz2010multi}. At present, spherical deconvolution methods such as constrained spherical deconvolution (CSD) are the most widely used techniques for assessing crossing tracts \cite{tournier2007robust, jeurissen2014multi}. These methods consider the dMRI signal in q-space to be the result of the convolution between a spherical point-spread function representing the fiber response function and the FOD. Naturally, they estimate the FOD via deconvolution of the dMRI signal with this response function.

Many machine learning methods have been proposed for FOD estimation. Table \ref{table:fod_table} provides a brief listing of some of these methods. We summarize our main observations and conclusions from our study of these works below.

\begin{table*}[!htb]
\centering
\caption{A summary of some of the machine learning methods for FOD estimation.}
\label{table:fod_table}
\begin{tabular}{C{30mm} | L{140mm}  }
\thickhline
Method & Summary of methodology and results  \\ \thickhline
Autoencoder-based FOD regularization \cite{patel2018better} & This work uses autoencoders to learn FOD priors from high-quality dMRI data. The prior is used to constrain CSD for FOD estimation on test data. Compared with standard estimation techniques, the new method reduces the required number of measurements by a factor of two and can work with a lower diffusion strength (b-value). The method is tested on only one subject. \vspace{1mm}  \\
Equivariant networks \cite{elaldi2021equivariant}, \cite{elaldi20233} & These works are based on rotation- and translation-equivariant convolutional networks. Training is performed in an unsupervised manner by convolving the computed FOD with the tissue response function to predict the dMRI signal and using the error in the predicted signal as optimization loss. The new methods results in lower FOD estimation error and more accurate tractography on in-vivo human data and phantom data.  \vspace{1mm} \\
MLP trained with histology ground truth \cite{nath2019deep} & These works have developed and validated FOD estimation models using training data with histology-derived FOD ground truth. The models are MLPs with residual connections. Results show that the deep learning methods lead to higher estimation accuracy (quantified in terms of angular correlation coefficient) compared with standard methods such as CSD. On in-vivo human data, the new method shows higher reproducibility.  \vspace{1mm}  \\
CNN method for fetal and neonatal brains \cite{kebiri2024deep} & A CNN has been used to predict FOD from six diffusion-weighted measurements for neonatal brains. Estimation accuracy is on par with the state of the art on neonatal scans. The method also performs well qualitatively on out-of-distribution clinical datasets of fetuses and newborns. \vspace{1mm}  \\
CNN classifiers \cite{koppers2016direct} & The method, which is based on CNN classifiers, shows better results than the state of the art, especially when the number of measurements is small. It can reconstruct voxels with three crossing fascicles from 10 dMRI measurements, but the method is only tested on synthetic data.  \vspace{1mm} \\
Lightweight CNN \cite{lin2019fast} & The model is a CNN that works on cubic patches of size three voxels. It achieves accurate estimation of FOD with 25 measurements. It estimates crossing fibers better than multi-tissue multi-shell constrained spherical deconvolution (MSMT-CSD \cite{jeurissen2014multi}). \vspace{1mm}  \\
FORDN \cite{ye2017fiber} & This method, named Fiber Orientation Reconstruction guided by a Deep Network (FORDN), is based on overcomplete dictionaries and MLP networks. A coarse dictionary is first used to represent the signal. In the second stage, a larger dictionary is used to compute a finer FOD that is close (in an $\ell_1$ sense) to the coarse FOD computed in the first step. Compared with methods that are based on sparse reconstruction or deep learning methods, this method is more accurate especially in voxels with two and three crossing fibers. However, the method is tested on phantom data and in-vivo scan of one human subject. \vspace{1mm} \\
Voxel-wise MLP \cite{karimi2021learning} & This work uses a voxel-wise MLP that is trained on simulated or real data. Extensive experiment show that this method is superior to a range of conventional FOD estimation methods when the number of measurements is small.  \vspace{1mm} \\
Voxel-wise or small-patch CNNs \cite{koppers2017reliable,koppers2017reconstruction} & These works apply 2D and 3D CNNs on data from individual voxels or small patches to compute the number of fascicles and the complete FOD. Experiments on HCP data show the deep learning methods can estimate the number of major fascicles and the FOD in voxels with complex fiber configurations with as few as 15 measurements and they are more accurate than CSD.  \vspace{1mm} \\
Method for heterogeneous multi-shell data \cite{yao2023unified} & To enable the method to work with different q-space shells, the model has three input heads for three common b-values (1000, 2000, and 3000). Either one of the shells or any combination thereof can be supplied at test time. The architecture itself is a CNN. Only HCP data is used to validate the method.   \vspace{1mm}  \\
MLP applied on dMRI signal decay features \cite{karimi2021machine} & A novel feature vector is proposed based on the decay of the diffusion signal as a function of orientation. Using this hand-crafted feature vector as input, an MLP is trained to estimate the ``angle to the closest fascicle'' for a large set of directions on the sphere. This information is used to infer the number and orientation of the major fascicles or to approximate the complete FOD.  \vspace{1mm} \\
Spherical deconvolution network \cite{bartlett2023recovering} & The network is inspired by a reformulation of CSD, hence it is presented as a ``model-driven deep learning method''. The reformulation is turned into an iterative optimization method that is unfolded and implemented as a deep neural network. The loss function has an $\ell_2$ term for the predicted FOD and a cross entropy term for the predicted number of fixels (i.e., major peaks). The study has used HCP data only. \\ 
\thickhline
\end{tabular}
\end{table*}

\subsubsection{Method design}

\hspace{3mm}

Prior to the surge of deep learning, a few studies used classical machine learning methods for FOD estimation \cite{schultz2012learning, taquet2013estimation}. One study computed priors from a population of training subjects and employed the learned prior in a maximum a posteriori framework to improve the estimation accuracy on test data \cite{taquet2013estimation}.

In more recent years, a number of studies have trained neural networks to estimate the FOD \cite{nath2020deep, koppers2017reconstruction}. These studies are different in terms of the neural network architecture, training strategy, and evaluation metrics. However, most share a similar claim that a deep neural network can estimate FODs more accurately than spherical deconvolution methods. Methodologically, most deep learning studies have followed a straight estimation approach, where the input dMRI measurements are mapped to the target FOD. The input signal as well as the target FOD are usually expressed either as functions on a discrete spherical grid or using spherical harmonics. However, there have been many exceptions, three of which we briefly describe below.

One study has proposed to use an auto-encoder to learn a model of plausible FOD shapes from a high-quality training dataset \cite{patel2018better}. This prior is then used to regularize the CSD method to obtain more accurate predictions than the standard CSD. Results show that this technique outperforms conventional methods when measurements are few or the diffusion strength (i.e., the b-value) is low. Koppers and Merhof \cite{koppers2016direct} use a succession of two CNN classifiers, where the first CNN estimates the number (either 1, 2, or 3) of fibers in the voxel and the second CNN estimates the fiber orientations. Ye and Prince \cite{ye2017fiber} used deep neural networks to estimate the FOD by solving sparse estimation problems in dictionaries. A coarse dictionary was used in the first stage to compute an initial estimate, which was then refined using a finer dictionary. Bartlett et al. start by reformulating the CSD equation to derive an iterative optimization algorithm for FOD estimation \cite{bartlett2023recovering}. They propose to solve this problem using a deep neural network. In addition to an $\ell_2$ loss for FOD prediction error, they introduce a cross entropy loss to encourage correct prediction of the number of major fascicles. Their experiments show that their method performs better than MSMT-CSD in estimating FOD and its peaks.

Unlike the scalar microstructural indices, FOD is a function of angle and, hence, it is rotation-sensitive. Elaldi et al. argue that the standard convolutional layers (which provide equivariance to planar translation) are insufficient for spherical signals in dMRI, where rotation equivariance is additionally needed \cite{elaldi2021equivariant}. Instead, they employ rotation-equivariant graph convolutions proposed in \cite{perraudin2019deepsphere}. Furthermore, they opt for spherical harmonics of degree 20 (as opposed to degree 6 or 8 in standard methods \cite{tournier2007robust}) to enable reconstruction of nearby FOD peaks. They show that, compared with the CSD baseline, their neural network achieves superior FOD estimation and better performance on downstream tasks such as tractography on phantom data and human brain scans. They have further extended this work by developing neural network layers that are equivariant to translations and grid reflections (as are standard FCNs) as well as to voxel and grid rotations \cite{elaldi20233}. They argue that these extra equivariance properties are needed to give the network the necessary spatio-spherical inductive biases to effectively learn to compute the FOD from data. Their results show more accurate FOD estimation, tractography, and brain tissue segmentation in dMRI compared with a range of state of the art methods. Related neural network architectures are reviewed in more detail in Section \ref{sec:modeling_considerations} of this paper.

\subsubsection{Training data}

\hspace{3mm}

The majority of studies have used standard estimation methods such as CSD to generate target FODs for training and validation data \cite{koppers2017reconstruction, yao2023unified}. Typically, these works have aimed at matching the standard technique while using fewer measurements as input. It has been claimed that neural networks can compute the FOD using only 6-20 measurements from a single shell to match the FOD computed with standard methods using multi-shell HARDI data. Therefore, the validation has been based on comparison with standard methods. This is unlike the common validation approach for FOD estimation techniques, which has been primarily based on simulation \cite{jelescu2020challenges}. However, with this approach, the trained model will inevitably inherit some of the shortcomings of the standard method that is used to generate the target FOD \cite{elaldi2021equivariant, schilling2018histological, schilling2016comparison}. Moreover, the assessments can only tell us how close the new machine learning method is to the standard technique. 

A few important studies have used histology data for training and/or validation. Nath et al. used ground truth histology FOD data from squirrel monkey brains, registered to dMRI data, to assess and compare a deep learning technique and CSD \cite{nath2019deep}. Their experimental results showed that there was additional information in the dMRI measurements that CSD failed to utilize, and that a deep learning method was capable of using that extra information for FOD estimation. The deep learning model, which was a voxel-wise MLP, outperformed CSD in terms of estimation accuracy using histology as ground truth. Further experiments with in-vivo human scans from the HCP dataset showed better scan-rescan reproducibility of the deep learning method, which the authors interpreted as evidence that the method could be used in clinical applications. Other studies \cite{elaldi2021equivariant, elaldi20233} have proposed novel methods for sidestepping the need for an FOD ground truth. They compute the convolution of the estimated FOD with the fiber response function to obtain the corresponding dMRI signal and use the difference between this predicted signal and actual measurements as the optimization loss for model training.

\subsubsection{Prediction of the number of main fascicles}

\hspace{3mm}

Instead of estimating the complete FOD, several studies have addressed the less ambitious but still challenging problem of determining the number of major fascicles in each voxel. Schultz trained a support vector regression model to estimate the number of major fascicles using simulated training data \cite{schultz2012learning}. Evaluations on simulated and real brain data showed that this method determined the number of major fascicles more accurately than CSD. Another study \cite{karimi2021machine} devised novel feature vectors to characterize the decay of the diffusion signal as a function of orientation. The feature vectors were then used by an MLP to compute the angle to the closest fascicle. This information was further processed via smoothing and local minimum extraction to determine the number and orientation of major fascicles in the voxel. Comparisons with several classical methods showed that this machine learning technique was more accurate.

\subsection{Tractography and structural connectivity}
\label{sec:Tractography}

Tractography algorithms build on local fiber orientations to compute virtual streamlines that connect different brain regions \cite{behrens2014mr}. Tractography has important applications, most prominently delineation of specific white matter tracts and quantitative structural connectivity \cite{sotiropoulos2019building, liu2019deepbundle}. It is one of the most common, challenging, and controversial computations enabled by dMRI \cite{yeh2021mapping, zhang2022quantitative, lemkaddem2014global}. Early tractography methods relied on the diffusion tensor model for computing the local fiber orientations \cite{basser2000vivo}. Over the past two decades, more advanced streamline tracing methods have been developed including anatomically constrained tractography \cite{smith2012anatomically}, global tractography algorithms \cite{mangin2013toward}, ensemble tractography \cite{takemura2016ensemble}, augmented tractography \cite{yeh2020shape}, and microstructure-informed tractography filtering \cite{smith2020quantitative}. These methods aim to utilize anatomical context information or the correspondence between tractography streamlines and the local dMRI signal to overcome the limitations of conventional tractography techniques.

However, tractography is intrinsically ill-posed and suffers from high false positive and false negative rates \cite{maier2017challenge, thomas2014anatomical}. Some of the main sources of error, unreliability, and ambiguity in tractography include \cite{zhang2022quantitative, sotiropoulos2019building, yang2021diffusion, rheault2020common}: (1) Difficulty of modeling crossing fibers, (2) Ambiguity of streamline tracing in voxels with crossing or kissing fibers, (3) Presence of bottleneck regions, where several tract bundles trace the same voxels in the same direction, and (4) Anatomical biases such as the gyral bias and termination bias. Standard tractography methods often produce inaccurate results that can influence tract-specific analysis and structural connectivity assessment \cite{schilling2019challenges, zhang2022quantitative, sotiropoulos2019building}.

Machine learning may offer a framework for developing better tractography algorithms. They are model-free and allow for a seamless integration of anatomical priors and other sources of information that may be useful for tractography. Some of these information, such as the direction of previous tractography steps, have been used in conventional tractography methods in the past \cite{malcolm2010filtered, lazar2003white}. Nonetheless, machine learning methods offer higher flexibility in integrating various inputs in a unified model and to optimize the model to reduce the tractography error with respect to all inputs jointly. As a result, they have the potential to have lower false positive rates, to be more robust to measurement noise, and to overcome the inherent biases of classical tractography algorithms \cite{poulin2019tractography, neher2017fiber, sarwar2020towards}. Although various methods such as self-organizing maps \cite{duru2013self} and random forests \cite{neher2017fiber} have been successfully used for tractography, machine learning-based tractography techniques have increasingly relied on neural networks \cite{poulin2019tractography, neher2017fiber, dedeep2018, poulin2018bundle}. Table \ref{table:tractography_table} lists some of the works that have developed machine learning methods for streamline tractography. We summarize the main technical aspects of these methods and their reported results below.

\begin{table*}[!htb]
\centering
\caption{A summary of selected machine learning methods for streamline tractography.}
\label{table:tractography_table}
\begin{tabular}{C{30mm} | C{20mm} L{55mm} L{55mm} }
\thickhline
Reference & Model (D: deterministic, P: probabilistic) & Model input & Main findings \\ \thickhline
Fiber tracking with a random forest classifier \cite{neher2017fiber}  & Random forests (D \& P) & Raw diffusion signal in a small neighborhood of the current voxel \& direction of the three preceding streamline steps. & Higher sensitivity and specificity and lower angular error than 12 state of the art methods.  \vspace{1mm}  \\
Learn to Track \cite{poulin2017learn} & GRU (D) & Raw diffusion signal in the current voxel. & Better tract coverage and lower false positive rate than classical methods.   \vspace{1mm} \\
DeepTract \cite{benou2019deeptract}  & RNN (D \& P) & Raw diffusion signal in the current voxel. & Lower percentage of invalid connections and non-connections, higher rate of valid bundles, and higher bundle coverage than a range of learning and classical methods. \vspace{1mm}  \\
Jörgens et al. \cite{jorgens2018learning} & MLP (D) & Raw diffusion signal \& directions of the two preceding streamline steps. & This work reports extensive experiments to determine the optimal input and network architectures.  \vspace{1mm} \\ 
Wegmayr et al. \cite{wegmayr2018data} & MLP (D)  & Raw diffusion signal in a $3^3$-voxel neighborhood \& directions of the two preceding streamline steps. & Better results than standard tractography methods when trained on small datasets. The method is robust to noisy training data.  \vspace{1mm} \\
Cai et al. \cite{cai2023convolutional} & Convolutional-recurrent neural network & Anatomical MRI and streamline memory. & This work has reported that the variability in tractography and structural connectivity metrics computed based on T1-weighted MRI is similar to that for standard dMRI-based results. \vspace{1mm}  \\
Liu et al. \cite{liu2024streamline} & Convolutional, attention, and MLP modules & FOD, brain parcellation and tissue segmentation maps, prior fixel map \& directions of six preceding steps. & This method was designed for \emph{fetal} brain tractography. Experiments showed superior ability to reconstruct various white matter tracts compared with existing methods. \\
Track-to-Learn \cite{theberge2021track} & Deep reinforcement learning (D \& P) & FOD at the current voxel and its six neighbors \& directions of the past four streamline steps. & A reinforcement learning approach can be competitive with supervised learning methods while also being more generalizable to unseen or out-of-distribution data.  \vspace{1mm} \\
Bundle-Wise Deep Tracker \cite{poulin2018bundle} & GRU (D) & dMRI signal. & Superior to classical methods in terms of true positive rate and better bundle volume coverage than existing probabilistic techniques.  \vspace{1mm} \\
Entrack \cite{wegmayr2021entrack} & MLP (P) & Directions of the three last streamline steps \& FOD, represented as SH of order 4, in a voxel neighborhood of size 3 voxels. & The method achieves competitive results compared with a range of conventional and machine learning methods on independent synthetic test data.  \\
\thickhline
\end{tabular}
\end{table*}

\subsubsection{Method design}

\hspace{3mm}

An increasingly popular class of models in tractography is recurrent neural networks (RNNs) \cite{poulin2017learn, poulin2018bundle, benou2019deeptract, jorgens2018learning, cai2023implementation}. Poulin et al. \cite{poulin2017learn} trained MLPs and Gated Recurrent Units (GRUs) \cite{cho2014learning} to map the dMRI signal to the streamline directions. The input to their model was the b0-normalized signal, resampled to a fixed spherical grid with 100 directions. The information from the preceding tracing steps was passed to the current step via the hidden state vector. Experiments showed that the GRU-based method outperformed 96 competing techniques on a simulated dataset and achieved expert-level results on a scan from the HCP dataset. Benou et al. \cite{benou2019deeptract} used an RNN to estimate a probabilistic FOD, which could then be used for deterministic or probabilistic tractography. The streamline tracing was stopped when the model predicted a termination label or when the entropy of the predicted FOD was below a threshold. This method showed competitive performance compared with classical methods and machine learning techniques, especially in terms of false positive rate and ability to reconstruct different fiber bundles.

The majority of the proposed methods have followed a supervised learning strategy. The main difficulty with these approaches is obtaining rich data with reliable ground truth, discussed further in the next sub-section. Given the difficulty of obtaining training data, there is a growing interest in reinforcement learning techniques \cite{sinzinger2022reinforcement, theberge2021track, wanyan2018tractography}. Theberge et al. developed the first deep reinforcement learning-based tractography method \cite{theberge2021track}. In their method, the state is derived from the diffusion signal while the reward is computed from the FOD peaks and the direction of the streamline. The model input consists of SH-represented FOD at the current voxel and its six neighbors. Additionally, the past four streamline directions are included as extra inputs to the model. The reward function is based on local information. Specifically, it promotes the closeness of the predicted streamline tracing step with the major FOD peaks and encourages streamline smoothness. Experiments show that the reinforcement learning method is competitive with supervised learning methods while also having better generalizability to unseen datasets. However, a true reinforcement learning method requires the knowledge of streamline start and end points, which is unavailable for in-vivo data.

Designing a machine learning-based tractography method involves several important choices including: (1) model input, (2) whether to adopt a probabilistic or a deterministic streamline propagation approach, (3) classification \cite{neher2015machine, jorgens2018learning} or regression \cite{poulin2017learn, wegmayr2021entrack} strategy, and (4) bundle-specific \cite{poulin2018bundle, reisert2018hamlet} or whole-brain tractography \cite{poulin2017learn, benou2019deeptract}. Most works have not justified their choices and there is a lack of experimental results to inform these decisions.


\subsubsection{Training data}

\hspace{3mm}

One approach to obtaining ground truth tractography data is via elaborate manual editing of automatically generated tractograms by experts \cite{maier2016tractography}. However, even for experts it is often impossible to resolve ambiguous situations and this practice suffers from high inter-observer variability \cite{rheault2020tractostorm, schilling2021tractography}. Physical phantoms represent an alternative \cite{wilkins2012development, neher2014fiberfox}. Poupon et al. have developed the FiberCup phantom to simulate a coronal slice of the brain at the level of corticospinal tracts \cite{poupon2010diffusion, fillard2011quantitative}. This phantom includes a rich set of crossing, fanning, and kissing tracts as well as U-fibers. The main problem with this approach is that no phantom can represent the complexity and inter-subject variability of the human brain. Lack of reliable data for model development and validation has remained one of the main stumbling blocks for advancing machine learning-based tractography, similar to other applications discussed in this paper. A notable recent database that has been made publicly available is TractoInferno \cite{poulin2022tractoinferno}. It includes multi-scanner data acquired with different protocols. In addition to the processed dMRI data, TractoInferno includes 30 white matter bundles reconstructed using an ensemble of 4 different tractography approaches followed by automatic and manual quality control.

\subsubsection{Evaluations and main results}

\hspace{3mm}

Machine learning methods have been effective in reconstructing various tracts with low false positive rates, reducing the systematic errors and biases that are prevalent with classical tractography methods, and improving generalizability to unseen data \cite{neher2015machine, poulin2019tractography}. Several reasons have been put forward to explain the success of machine learning methods. First, they avoid making rigid, ad-hoc, and possibly sub-optimal modeling assumptions or streamline propagation and stopping rules that are common in standard methods \cite{neher2015machine, poulin2019tractography, wegmayr2021entrack}. Some of the proposed machine learning methods use the dMRI data (rather than diffusion tensor or FOD estimates) as input, thereby sidestepping the unavoidable errors in fiber orientation estimation \cite{neher2015machine, benou2019deeptract, wegmayr2018data}. Moreover, they have the potential to learn tissue probabilities from dMRI data, rather than relying on the information provided by a registered anatomical MRI or ambiguous parameters such as FA. Machine learning methods can also learn the noise and artifacts from data, instead of imposing simplified noise models and assuming artifact-free scans \cite{neher2015machine}. Furthermore, these methods can incorporate estimation uncertainty into streamline tractography in a systematic way \cite{wegmayr2021entrack}. Machine learning techniques are also by far more flexible in incorporating non-local information, such as anatomical context, which can greatly improve tractography results \cite{poulin2019tractography}. An important piece of information that can be easily incorporated into most machine learning models is the directions of preceding streamline tracing steps. The direction of the last step indicates the streamline slope and the directions of the last two steps indicate its curvature. This information has been useful in improving tractography accuracy \cite{poulin2017learn}. Incorporation of the direction of preceding steps can also improve the tractography in voxels with crossing fibers and bottleneck regions \cite{theberge2021track, poulin2017learn}.

\subsubsection{Tractography post-processing}

\hspace{3mm}

Machine learning methods have also been used to enhance the generated tractograms via post-processing operations such as identification of false positives or anatomically implausible streamlines. Deep learning has shown great success in this task as well. An example of a supervised method is the work of Astolfi \cite{astolfi2020tractogram}, where an existing method \cite{petit2019half} is used to generate labels on training data based on anatomical priors. A limitation of supervised approaches is that it is hard to obtain training data that can adequately represent the complete range of invalid streamlines \cite{ugurlu2019supervised}. Unsupervised methods sidestep this limitation. An example of unsupervised techniques is the auto-encoder method proposed by Legarreta \cite{legarreta2021filtering}. This method uses unlabeled whole-brain tractograms to learn a low-dimensional representation of valid streamlines, which can then be used to detect spurious or anatomically invalid streamlines in the test data. Both supervised and unsupervised approaches have reported remarkable success. A generative method based on autoencoders has recently been propsoed to improve the streamline density for tracts that are hard to trace with conventional propagation methods \cite{legarreta2023generative}. Using anatomically valid training streamlines, it learns a latent space representation that can be subsequently used as a ``streamline yard" to synthesize new streamlines for sparsely-populated tracts in a test sample. The synthesized streamlines are accepted or rejected based on a set of criteria related to streamline geometry, connectivity, and agreement with local fiber orientation directions. Experiments with phantom and real data showed that this method significantly improved tract coverage.

\subsection{Delineation of white matter tracts}
\label{sec:TractSeg}

The brain white matter is organized into distinct tracts, which consist of bundles of myelinated axons that connect different brain regions such as the cerebral cortex and the deep gray matter \cite{bullock2022taxonomy, wakana2004fiber, wycoco2013white}. Although they are tightly packed and often cross one another, each tract has a different function and connects different regions of the brain. Accurate segmentation of these tracts is needed in clinical studies and medical research. For example, in surgical planning one needs to know the precise extent of the individual tracts in order to assess the risk of damage to specific neurocognitive functions that may result from surgical removal of brain tissue \cite{kamada2005combined, yang2021diffusion}. Furthermore, changes in the tissue microstructure on specific tracts can be associated with brain development and disorders \cite{glasser2008dti, bubb2018cingulum, zhuang2010white}.

Diffusion MRI is the only non-invasive method that can accurately delineate white matter tracts \cite{wycoco2013white}. Individual tracts may be extracted from whole-brain tractograms by specifying inclusion and exclusion regions of interest (ROIs). This process, which is usually referred to as ``virtual dissection'', is time-consuming, requires substantial expertise, and suffers from high inter-observer variability and low reproducibility \cite{schilling2021tractography, wakana2007reproducibility}. 

Several prior works have aimed at automating the virtual dissection process by learning to compute the inclusion/exclusion ROIs \cite{suarez2012automated, yendiki2011automated, warrington2020xtract}. Moreover, there has been much effort to develop fully automatic methods for tract delineation. These methods can be divided into two categories: (1) methods that start by performing streamline tractography; (2) methods that do not depend on tractography. We have summarized some of the works in these two classes of methods, respectively, in Tables \ref{table:tractseg_table_tractogram} and \ref{table:tractseg_table_direct}. Given the large number of published methods, these tables only list a selection of more recent methods with a focus on deep learning techniques. Below, we describe our main findings from our study of these works.

\begin{table*}[!htb]
\centering
\caption{A summary of selected tractography-based methods for automatic delineation/segmentation of white matter tracts.}
\label{table:tractseg_table_tractogram}
\begin{tabular}{C{25mm} | C{20mm} C{25mm} C{10mm} L{80mm} }
\thickhline
Reference & Model & Input & Number of tracts & Methodology and results \\ \thickhline
Deep white matter analysis (DeepWMA) \cite{zhang2020deep} & CNN & Tractogram & 54 & The method is based on hand-crafted features describing the streamline geometry. A CNN is applied to classify the streamlines based on the feature vectors. The method shows high accuracy across the human lifespan from neonatal to old age and it works well on brains with gross pathologies such as tumors.  \vspace{1mm} \\
DeepBundle \cite{liu2019deepbundle} & Graph CNN & Tractogram & 12 & A graph CNN is used to classify streamlines based on their geometry. Only HCP data is used in this work. The method shows higher accuracy than standard methods, especially for small tracts such as fornix and commissure anterior.  \vspace{1mm} \\
BrainSegNet \cite{gupta2017brainsegnet} & Bidirectional LSTM & Tractogram & 8 & The method works in two steps, in the first step classifying the white matter versus the gray matter and in the second step segmenting the tracts. The method achieves streamline classification accuracy of $>96\%$ and recall of $>73\%$, but it is tested on three subjects only.  \vspace{1mm} \\
FiberNET \cite{gupta2018fibernet} & CNN & Tractogram & 17 & The method uses a harmonic function to compute level-sets based on a brain shape-center, thereby re-parameterizing the streamlines in a manner that is consistent across subjects. A CNN classifies the streamlines based on this representation. The method achieves low false positive rates.   \vspace{1mm}  \\
Xu et al. \cite{xu2019objective} & CNN & Tractogram & 64 & The focus of this work was on detecting eloquent white matter tracts for epilepsy surgery. Different CNN architectures and loss functions were investigated. The models were tested on healthy children and children with epilepsy. A deep CNN with attention mechanism achieved the highest F1 score of 0.95. \vspace{1mm} \\
Spectral clustering \cite{o2007automatic} & Spectral embedding clustering & Tractogram & 10 & The method achieves high tract segmentation accuracy and can be useful for cross-subject white matter studies.  \vspace{1mm} \\
Dayan et al. \cite{dayan2018unsupervised} & Restricted Boltzmann machines (RBM) & Tractogram & 61 & The method offers memory-efficient unsupervised detection of anatomically relevant tracts without the need to explicitly define the tracts. It achieves Dice Similarity Coefficients ranging from under 0.30 to above 0.90 on different tracts. \vspace{1mm} \\
TRACULA \cite{yendiki2011automated} & Bayesian model & Tractogram and anatomical MRI & 18 & The model trained on healthy subjects worked well on schizophrenia patients.  \vspace{1mm} \\
FS2Net \cite{jha2019fs2net} & LSTM & Tractogram & 8 & The authors claim that, because their method relies on streamline geometry, it is particularly useful in situations where the test brains are arbitrarily oriented. They report accuracies above 0.90 but only on three subjects. \vspace{1mm}  \\
Deep Fiber Clustering \cite{chen2023deepfiber} & Siamese Graph convolutional neural network & Tractogram and gray matter parcellation & Unclear & The method produces accurate delineation of tracts across age and gender for healthy as well as diseased brains.  \vspace{1mm} \\
Superficial White Matter Analysis (SupWMA) \cite{xue2023superficial} & MLP & Tractogram & 198 & This method is based on a point cloud-based representation of the streamlines and contrastive supervised learning. Experimental results on six datasets across the human lifespan show that this approach can accurately extract superficial white matter tracts. \vspace{1mm}  \\
Ugurlu et al. \cite{ugurlu2019supervised} & Ensemble of MLPs & Tractogram and FOD map & 9 & This paper presents a way of synthesizing a rich set of invalid streamlines to enable effective training of the model to reject such streamlines in test data. The method is only validated on HCP data. \\
AnatomiCuts \cite{siless2018anatomicuts} & Hierarchical clustering & Tractogram \& cortical and subcortical segmentation & 18 & This method advocates for leveraging the position of the streamline with respect to cortical and sub-cortical landmarks. Experiments on HCP data show that this approach improves the overlap between automatic and manual tracts by 20\%.   \vspace{1mm} \\
TractCloud \cite{xue2023tractcloud} & MLP \& Point-cloud-based networks & Tractogram & 42 & In addition to the streamline to be classified, 20 nearest neighbors and 500 streamlines randomly selected from the tractogram are used as input to the MLP. The latent representation computed by the MLP is used by the point cloud network to classify the streamline. The method is validated on five datasets. \vspace{1mm}  \\
TRACULInA \cite{zollei2019tracts} & Bayesian model & Tractogram and anatomical MRI & 14 & This model addresses the neonatal age range and is in part based on TRACULA \cite{yendiki2011automated}. It is successfully validated on out-of-distribution data such as prematurely-born neonates. \\
\thickhline
\end{tabular}
\end{table*}

\subsubsection{Methods that rely on tractography}

\hspace{3mm}

A large class of automatic tract delineation methods is based on processing of whole-brain tractograms \cite{garyfallidis2018recognition, labra2017fast, clayden2007probabilistic, siless2018anatomicuts, siless2020registration, legarreta2021filtering}. There are two common approaches used by these methods.

\begin{enumerate}

\item One set of methods compare individual streamlines with a predefined set of fibers in an atlas and assign each streamline to a specific tract in the atlas \cite{wassermann2010unsupervised, garyfallidis2018recognition, labra2017fast}.

\item Another set of methods cluster the streamlines based on some measure of pair-wise similarity \cite{garyfallidis2012quickbundles, o2007automatic}. There exist various techniques for implementing the clustering, for example using normalized cuts \cite{brun2004clustering} or k-nearest-neighbors \cite{wang2019fast}.

\end{enumerate}

Both these sub-classes of methods can be computationally expensive, especially for larger number of streamlines and higher image resolution. Consequently, much effort has been directed towards developing fast streamline clustering methods \cite{wang2018gife, garyfallidis2012quickbundles, vazquez2020ffclust}. Some techniques additionally take into account the location of the streamlines relative to anatomical landmarks in the brain \cite{siless2018anatomicuts, siless2020registration, maddah2008mathematical, li2010hybrid}. Most clustering-based methods aim at finding streamlines that have similar geometric shapes and spatial extents. There are also methods that cluster streamlines based on their start/end points \cite{wassermann2016white}. However, methods that use streamline start/end points may not perform as well as streamline clustering techniques \cite{zhang2017comparison}.

Overall, methods that use anatomical information achieve better results than methods that rely on streamline shape alone. Such information can be encoded and incorporated in the tract segmentation computation in various ways. For example, Maddah et al. \cite{maddah2008mathematical} used the anatomical information provided by an atlas of white matter tracts, while Tunc et al. \cite{tuncc2014automated, tuncc2016individualized} used the connectivity signature of the streamlines. Yendiki \cite{yendiki2011automated} has proposed a method that uses the tractography results computed from the dMRI signal and the anatomical tissue segmentation information for automatic tract delineation. Theirs is a probabilistic method, where prior probabilities of the tracts are computed from a set of manually-segmented training subjects. This method, trained with data from healthy subjects, performed well on schizophrenia patients. It has been extended to neonatal brain using a dedicated image processing pipeline to address the challenges of early brain data \cite{zollei2019tracts}. This extended method also showed good generalizability to data from term- and preterm-born neonates imaged with different scanners and acquisition protocols.

Sydnor et al. \cite{sydnor2018comparison} critically compared a manual ROI-based method \cite{catani2008diffusion} and two automatic tractography-based tract segmentation techniques \cite{wassermann2016white, o2007automatic} in terms of various measures of accuracy and reproducibility. The results showed that although clustering methods were overall better, they were not consistently superior and that an optimal method might have to rely on a combination of the two approaches. In a rather similar work, O'Donnell et al. \cite{o2013fiber} compared streamline clustering methods with techniques that clustered streamlines based on their start and end points in the cortex. They highlighted the pros and cons of each class of methods and advocated for hybrid methods that could benefit from the strengths of both classes of techniques. Similarly, Chekir at al. \cite{chekir2014hybrid} and Xu et al. \cite{xu2013gray} point out the limitations of the purely clustering-based methods and methods that are based on anatomical priors. They propose methods that synergistically combine the advantages of the two strategies.

A growing number or works have applied deep learning methods on whole-brain tractograms to extract/cluster individual tracts. These methods have followed very different designs. Given the high flexibility in designing deep learning methods, some of them cannot be categorized under any of the two broad classes mentioned above. A typical recent example is Deep White Matter Analysis (DeepWMA) \cite{zhang2020deep}. DeepWMA first computes a set of feature descriptors to represent the spatial coordinate location of points along each streamline. In order to render the feature descriptor independent of the orientation/order of streamline, the coordinates are also flipped and added to the descriptor via concatenation. The resulting 1D representation is then repeated row-wise to create a 2D feature map that a CNN can efficiently process. A CNN is trained to classify the streamlines based on the feature maps. In order to enable their model to identify false positive streamlines, they include an extra class label to represent those streamlines. In DeepBundle \cite{liu2019deepbundle}, on the other hand, streamlines are represented in terms of the point coordinates and a graph CNN is trained to extract geometric features to classify the streamlines. Hence, DeepBundle sidesteps the manual feature engineering that is used in some tractography-based methods. The graph CNN includes a series of graph convolution and graph pooling layers. TRAFIC \cite{lam2018trafic} computes streamline curvature, torsion, and Euclidean distance to anatomical landmarks for each point on the streamline. Fibernet \cite{gupta2017fibernet,gupta2018fibernet}, on the other hand, represents the streamlines using an iso-surface map of the brain. At least one study has observed that using streamline coordinates as input leads to more accurate tract segmentation than features such as curvature and torsion \cite{xu2019objective}. TractCloud \cite{xue2023tractcloud} uses the streamline to be classified, its nearest neighbors, and a set of randomly selected streamlines from the whole-brain tractogram as input to an MLP to compute a latent representation, which is then used by a point cloud network to classify the streamline.

\subsubsection{Methods that avoid tractography}

\hspace{3mm}

In addition to the streamline similarity measures and anatomical information described above, tractography-based methods have resorted to various other information to improve their accuracy. These include homology between the brain hemispheres, spatial and shape priors, and population averages (atlases). However, all tractography-based methods are inherently limited by the errors in streamline tracing \cite{maier2017challenge, hua2008tract}. Moreover, they typically involve several processing steps such as streamline propagation, white matter segmentation, gray matter parcellation, and clustering or similar computations to extract individual tracts. Each of these computations may introduce its own errors that depend on method settings, and it is difficult to jointly optimize all these processing steps. Moreover, some of these computations, e.g., tractography and streamline clustering, can require long computation times.

\begin{table*}[!htb]
\centering
\caption{A summary of selected non-tractography-based methods for automatic delineation/segmentation of white matter tracts.}
\label{table:tractseg_table_direct}
\begin{tabular}{C{25mm} | C{25mm} C{25mm} C{10mm} L{75mm} }
\thickhline
Reference & Model & Input & Number of tracts & Methodology and results \\ \thickhline
TractSeg \cite{wasserthal2018tractseg} & Set of 2D CNNs & Orientation of the major FOD peaks & 72 & TractSeg was the first work to demonstrably show that deep learning models can segment various white matter tracts without the need for tractography. It achieved an average Dice Similarity Coefficient of 0.84 on HCP data and 0.82 on clinical quality data.   \vspace{1mm} \\
Liu et al. \cite{liu2022volumetric} & CNN & Orientation of the major FOD peaks & 72 & This work showed that modeling the label correlations improved the segmentation accuracy, especially for tracts that were more difficult to segment. Experiments show that this method works better than an atlas-based method and TractSeg \cite{wasserthal2018tractseg}. \vspace{1mm} \\
Lu et al. \cite{lu2021volumetric} & CNN & Orientation of the major FOD peaks & 72 & This work advocates for self-supervised pre-training of the CNN using two pretext tasks: (1) tractography density prediction, (2) segmentation with noisy labels generated with an atlas-based method. It reports improved segmentation accuracy.  \vspace{1mm} \\
Lucena et al. \cite{lucena2022informative, lucena2023assessing} & CNN & dMRI data represented in a spherical harmonics basis & 72 & This work evaluates the accuracy of the dMRI-based segmentation in terms of its overlap with the responses of navigated transcranial magnetic stimulation. \vspace{1mm} \\
Kebiri et al. \cite{kebiri2023direct} & CNN & dMRI data & 72 & This work showed that it was possible to segment all 72 tracts form the TractSeg study using only six dMRI measurements as input.  However, only one independent dataset was used for validation. \vspace{1mm} \\
HAMLET \cite{reisert2018hamlet} & CNNs with rotation equivariance & Raw diffusion data & 12 & The CNN model uses rotationally equivariant convolutional operations, thereby eliminating the need to learn data orientations. Experiments show high test-retest reproducibility and generalizability on low-quality data from an external scanner. \vspace{1mm}  \\
Neuro4Neuro \cite{li2020neuro4neuro} & CNN & Diffusion tensor & 25 & This method produced highly accurate and reproducible tract segmentations and it was generalizable to unseen data from external scanners and from dementia patients.  \vspace{1mm} \\
Xu et al. \cite{xu2023registration} & CNN & Orientation of the major FOD peaks & 72 & This work used a registration-base label propagation method to synthesize noisy labels for unlabeled images, thereby increasing the size of labeled training data. The authors reported that with this approach, only a single manually labeled image was sufficient for training.  \\
\thickhline
\end{tabular}
\end{table*}

To avoid these errors and limitations, several studies have proposed to segment the tracts on diffusion tensor or fiber orientation images, thereby avoiding the tractography. Some of the classical machine learning methods that have been used for this purpose include Markov Random Fields \cite{bazin2011direct}, k-nearest neighbors technique \cite{ratnarajah2014multi}, level-set methods \cite{lenglet2006dti, jonasson2005white, guo2008geometric}, template/atlas-based techniques \cite{eckstein2009active, hua2008tract}.

More recently, deep learning has shown unprecedented accuracy for this application as well \cite{wasserthal2018tractseg, dong2019multimodality}. Wasserthal et al. \cite{wasserthal2018tract} represented overlapping/crossing tracts using a simple but powerful concept that they named tract orientation maps (TOMs). Each tract is represented by a separate TOM, where each voxel contains a vector representing the local orientation of that tract. If a tract does not cross a certain voxel, the value of that voxel in the TOM for that tract will be a vector of zeros. In their framework, a standard method such as CSD is applied to compute the FOD and the FOD peaks are extracted to build the TOMs. These peak orientations are fed into an FCN, which is trained to compute the TOMs. The TOMs can be used as a prior or direct input for bundle-specific tractography. The authors claim that their method is a viable solution to excessive false positive rate of standard tractography techniques. They support this claim by demonstrating that their method leads to superior tract bundle reconstruction compared with several state of the art methods such as TractQuerier \cite{wassermann2016white}, RecoBundles \cite{garyfallidis2018recognition}, and WhiteMatterAnalysis (WMA) \cite{o2007automatic}. In subsequent works, they build upon this method and extend it in a few important directions. In TractSeg \cite{wasserthal2018tractseg}, they develop an FCN-based method to map the TOMs to the segmentation probability maps for each tract. To enable the method to run on limited GPU memory while working on full-resolution images to segment the complete set of 72 tracts, their model is a 2D FCN that works on axial, coronal, and sagittal slices separately. TractSeg achieves a mean Dice Similarity Coefficient (DSC) of 0.84. On all tracts except for commissure anterior and fornix it achieves a mean DSC of higher than 0.75. While the tract segmentation masks generated by TractSeg can be used for accurate bundle-specific tractography \cite{wasserthal2018tractseg}, a more complete method is presented in a subsequent work by the same authors \cite{wasserthal2019combined}, where a separate deep learning model segments the start and end regions of each tract.

A number of studies have built upon the methods proposed by Wasserthal et al. \cite{wasserthal2018tract, wasserthal2019combined}. A representation similar to TOM was proposed in HAMLET \cite{reisert2018hamlet}, where the model output is a tensor field that indicates the presence and local orientation of a tract. Li et al. \cite{li2020neuro4neuro} trained a CNN to segment 25 white matter tracts directly from diffusion tensor images. Their extensive experiments showed high segmentation accuracy, low scan-rescan variability in tract-specific FA analysis, good generalizability to cross-center data, and successful application in a tract-specific population study on dementia patients. Their method was more accurate than tractography-based techniques, while being orders of magnitude faster. They also found that adding anatomical (T1-weighted) images to the diffusion tensor as input did not improve the segmentation accuracy. Moreover, using the directions of the three major peaks led to lower accuracy than diffusion tensor \cite{li2020neuro4neuro}. Liu et al. followed an approach similar to TractSeg but proposed to model the correlation between tract labels \cite{liu2022volumetric}. Specifically, they introduced an auxiliary neural network that mapped the native tract labels into a lower-dimensional label space. The main segmentation network predicted this lower-dimensional label, which could be used to predict the tract segmentation maps. The authors argue that this approach makes the task of tract segmentation simpler to learn and show that it can improve the segmentation of certain tracts such as the fornix that are especially difficult to segment with TractSeg. Similarly, Lu et al. follow the general approach proposed by TractSeg, but propose self-supervised pre-training \cite{lu2021volumetric} and transfer learning \cite{lu2021knowledge} to reduce the required labeled data. These works have shown that as few as five labeled images may be sufficient to train CNNs for segmenting certain tracts. To reduce the required manual labelling, another study proposed a deep learning registration method to align the labeled images to unlabeled images, thereby synthesizing a large training dataset \cite{xu2023registration}. Experiments showed that even a single labeled image was sufficient to train an accurate deep learning model.

In order to determine the orientation of major fascicles, previous works have relied on computation of the diffusion tensor or FOD. However, none of these intermediate computations have an unambiguous biophysical meaning and they entail unavoidable estimation errors. Moreover, the intermediate computations for most existing methods assume availability of dense multi-shell measurements, which are not acquired in many clinical and research applications. Recently, some studies have suggested to perform the segmentation based directly on the dMRI signal \cite{lucena2022informative, lucena2023assessing, kebiri2023direct, mukherjee2020deep}. One study demonstrated the feasibility of segmenting the corticospinal tract (CST) from the dMRI data for neonatal subjects in the dHCP dataset \cite{mukherjee2020deep}. Another work segmented the 72 tracts from the TractSeg study \cite{kebiri2023direct}. It showed that it was possible to achieve a similar level of segmentation accuracy as TractSeg while using only six measurements.

\subsubsection{Superficial white matter analysis methods}

\hspace{3mm}

While most studies have focused on deep white matter, a few studies have addressed the arguably more challenging task of segmenting the superficial white matter tracts, also known as the U-fibers \cite{guevara2012automatic, xue2023superficial, roman2017clustering, vazquez2020automatic, li2025insights}. These tracts are critical for assessing brain connectivity as they account for the majority of the cortico-cortical connections. However, they are difficult to study with standard tractography techniques due to their small size, partial volume effects, and high inter-subject variability \cite{guevara2020superficial, vazquez2020automatic, roman2017clustering}. Xue et al. \cite{xue2023superficial} developed a neural network model to segment 198 superficial white matter tracts via clustering the streamlines using a deep neural network. Streamlines were represented as point clouds and a contrastive learning approach was used to train the network. The authors reported remarkable accuracy (cluster identification rates of 87-99\%, except for a neonatal dataset) for different age groups and health conditions including brain tumor patients. Another study \cite{roman2017clustering} developed a hierarchical clustering method to identify representative groups of short association fibers from a population of subjects. This was accomplished by identifying fibers that were present in the majority of the subjects in a common (Talairach) space. The method was used to develop an atlas of 93 reproducible tracts, which could then be used to segment these tracts from whole-brain tractograms of test subjects. Another study suggested clustering of superficial white matter tracts based on their connectivity patterns using cortical brain parcellations inferred with an atlas \cite{vazquez2020automatic}.

\subsection{dMRI registration}
\label{sec:Registration}

Tract-specific studies, as their name implies, study the organization, development, and tissue microstructure of specific white matter tracts \cite{smith2014cross}. They can be of two types: (i) longitudinal studies consider the same subject(s) over time, while (ii) population studies compare different cohorts of subjects such as diseased versus control groups. They are among the most common dMRI studies because many neurodevelopmental deficits and neurological disorders are linked to damage to the tissue microstructure on specific tracts \cite{pini2016brain, sexton2011meta}. The success of these studies depends on ensuring that the same tracts are compared between scans/subjects, which is typically achieved via precise alignment of different brain scans. Regardless of the approach taken, this is a complex and error-prone computation. Existing computational methods vary greatly in terms of spatial accuracy and the required manual input \cite{smith2014cross}. Voxel-based morphometry methods are simple but incapable of accurate tract-specific comparisons \cite{ashburner2000voxel}, while semi-automatic and automatic tractography-based methods are limited by the tractography errors \cite{pagani2005method, suarez2012automated}. Among the automatic methods, TBSS \cite{smith2006tract} is the most widely used. However, it suffers from important inaccuracies and shortcomings \cite{bach2014methodological, madhyastha2014longitudinal, edden2011spatial}. Machine learning techniques have a unique potential to develop more reliable methods to address these limitations.

A main source of error in many tract-specific analysis methods is poor registration. Some methods, such as TBSS, perform the registration based on FA images, which ignores the orientation-dependent microstructure information. Leveraging the fiber orientation information may significantly improve alignment accuracy \cite{raffelt2011symmetric, zhang2006deformable}. Several deep learning-based dMRI registration methods have been proposed in the past few years and they have shown a tremendous potential for accurate and robust registration \cite{grigorescu2020diffusion, zhang2021deep, bouza2023geometric, grigorescu2022attention}. Overall, these methods follow the recent advancements in deep learning-based registration, which are mainly based on fully convolutional networks, spatial transformer networks, and unsupervised training \cite{balakrishnan2019voxelmorph}. Additionally, they usually use the finite strain technique \cite{alexander2001spatial} to properly reorient the white matter structures based on the computed non-linear deformations. One study has shown that using the diffusion tensor images in addition to anatomical (T1- and T2-weighted) images can lead to higher registration accuracy \cite{grigorescu2020diffusion}. A subsequent work has shown that attention maps can be learned to weight the contribution of the two modalities (i.e., anatomical and microstructural images) to achieve better registration \cite{grigorescu2022attention}. Another study used the tract orientation maps \cite{wasserthal2018tract} for 40 white matter tracts as well as FA maps to compute 41 separate deformation fields \cite{zhang2021deep}. A fusion neural network combined these into a final deformation field. On data from different age groups, scanners, and imaging protocols, this method was more accurate than existing DTI- and FOD-based registration techniques.

Li et al. developed a deep learning–based framework for precise alignment of diffusion tensor images of the fetal brain \cite{li2025fetdtialign}. This is a particularly challenging problem due to the low quality and scarcity of fetal imaging data, the rapid structural changes that occur during gestation, and the limited availability of anatomical landmarks to guide registration. The proposed method, FetDTIAlign, performs both affine and deformable registration. The affine registration module adopts an iterative strategy: the output of each alignment step is recursively fed back into the model to progressively refine accuracy. For deformable registration, FetDTIAlign leverages segmentation masks of 60 white matter tracts, which serve as anatomical constraints to improve robustness and precision. Extensive evaluations demonstrated that FetDTIAlign outperformed both conventional and deep learning–based registration approaches across a wide range of gestational ages (23–36 weeks). Such tools are urgently needed, as diffusion MRI is increasingly applied to the study of fetal and neonatal brain development, yet most existing registration methods have been designed for the adult brain and fail to address the unique challenges of early brain imaging.

\subsection{Tract-specific analysis}
\label{sec:TractSpecific}

A few studies have proposed machine learning methods for cross-subject tract-specific analysis. Most these methods, but not all, are based on tract segmentation and alignment of data from different subjects/scans. Such methods are urgently needed since existing solutions such as TBSS are known to suffer from important limitations. These works are briefly discussed below under two classes of approaches.

\subsubsection{Tractography-based methods}

\hspace{3mm}

Prasad et al. \cite{prasad2014automatic} developed a framework that consisted of streamline clustering, skeletonization based on highest streamline density, and registration based on geodesic curve matching. They showed that this method had a higher sensitivity than TBSS. Jin et al. proposed a similar approach based on multi-atlas clustering of streamlines with label fusion \cite{jin2014automatic}. Zhang et al. \cite{zhang2018whole} used group-wise registration and analysis of whole-brain tractograms to compute an atlas of valid streamlines, which they then used to extract subject-specific fiber tracts. They used this method to analyze the tissue microstructure in terms of FA and MD for a population of autism spectrum disorder children and healthy controls. If the goal is to predict subject-level neurocognitive scores, cross-subject data alignment or spatial correspondence can be side-stepped. TractGeoNet \cite{chen2024tractgeonet}, for example, uses a neural network to directly predict the score from streamline-level tissue microstructure information.

\subsubsection{Deep learning-based methods}

\hspace{3mm}

Segis-Net \cite{li2021longitudinal} is a deep learning method for tract segmentation and registration. The segmentation and registration modules are optimized jointly. The segmentation module uses diffusion tensor as input, while the registration module uses FA images as input. The loss function includes separate terms for segmentation and registration as well as a joint term to encourage tract segmentation overlap between pairs of registered images. The advantage of this approach is the joint learning of the segmentation and registration tasks and reliance on an implicit unbiased reference image \cite{li2021learning}. Segis-Net was applied to analyze six tracts in a longitudinal study including 8045 scans and was shown to lead to higher reproducibility than a non-learning method and a machine learning pipeline that addressed segmentation and registration tasks separately. Segis-Net also reduced the sample size requirements for tract-specific studies by a factor of up to three and it was orders of magnitude faster than methods based on conventional segmentation and registration techniques. In another work, the authors develop an accurate and detailed dMRI atlas that includes 72 tracts \cite{karimi2023tbss++}. For each test subject, the method segments the tracts and registers the subject data onto the atlas. This method offers higher reproducibility and better robustness to measurement noise than TBSS. In a different work, an autoencoder has been proposed to detect abnormal tissue microstructure indices on white matter tracts \cite{chamberland2021detecting}. The autoencoder input consists of microstructural indices on white matter tract profiles. Following the standard autoencoder approach, the model is trained on a population of normal subjects to learn to reconstruct healthy brain data. When applied on data from pathological brains, the autoencoder can detect abnormal microstructure indices via large reconstruction errors. The authors validate this method on data from patients with various neurological disorders including epilepsy and schizophrenia. These experiments show that the new method is more accurate than standard abnormality detection techniques. The authors characterize their method as moving beyond group-wise analysis and enabling assessment of single patients at an individual level. However, one can argue that techniques such as \cite{li2021longitudinal, karimi2023tbss++} can offer the same capability.

\subsection{Segmentation}

Tissue segmentation in the dMRI space is needed for anatomically constrained tractography and structural connectivity analysis \cite{smith2012anatomically, sotiropoulos2019building}. Given the lower spatial resolution and tissue contrast of dMRI, this segmentation is usually obtained via segmenting anatomical MRI data followed by registration to the dMRI space. This registration can introduce significant error because the contrast and image distortions between the modalities can be very different. Therefore, direct tissue segmentation in dMRI is desirable. Machine learning methods, and in particular deep learning techniques, have shown great promise here as well.

Prior to the development of deep learning-based segmentation methods, classical machine learning methods such as fuzzy c-means clustering with spatial constraints \cite{wen2013brain} and sparse representation of the dMRI signal in dictionaries \cite{yap2015brain} were used to segment the brain tissue. Ciritsis et al. determined the optimal set of dMRI measurements and DTI features for tissue classification with support vector machines (SVM) \cite{ciritsis2018automated}. Schnell et al. computed rotation-invariant features from HARDI measurements and used them as input to an SVM to segment the brain tissue into six classes. The class labels included white matter with parallel (single) fibers and white matter with crossing fibers \cite{schnell2009fully}.

More recently, Golkov et al. \cite{golkov2016q} extended their deep learning tissue microstructure estimation technique to develop a segmentation method that mapped the dMRI signal in each voxel to a probability for white matter (WM), gray mater (GM), cerebrospinal fluid (CSF), and MS lesion. Their simple and flexible approach allowed them to include an extra FLAIR channel, which could be especially useful for MS lesion segmentation. Their model accurately segmented the brain tissue, and detected MS lesions with area under the receiver operator characteristic curve (AUC) in the range 0.88-0.94. The same authors also proposed a variational auto-encoder model for novelty detection based on q-space dMRI data and showed that it could accurately segment MS lesions with an AUC of approximately 0.90 \cite{vasilev2020q}. Another study used DTI and DKI features as input to a multi-view CNN to achieve accurate segmentation of WM, GM, and CSF \cite{zhang2021deepseg}. The model was trained on HCP-quality data, where the anatomical and dMRI data could be registered precisely. The trained model achieved high segmentation accuracy on clinical dMRI acquisitions. The same research team trained a deep learning model to parcellate the brain into 101 regions using DTI-derived parameter maps as input \cite{zhang2023ddparcel}. The authors showed the utility of the new method in improving tract-specific streamline bundle extraction from whole-brain tractograms. 

A recent study addressed the challenging task of \emph{fetal} brain segmentation in dMRI \cite{calixto2024anatomically}. This is a particularly difficult application due to the low contrast of fetal brain dMRI. The authors used DTI-derived parameters as the input to their model, which was based on vision transformers. They proposed a novel self-supervised learning approach that enabled them to train the model on a large unlabeled dataset and a smaller labeled dataset. This method achieved a mean DSC of 0.84-0.90 on different tissue types. Experiments showed that this method drastically improved fetal brain tractography. Building upon their improved tractography results, the same research team has developed a multi-task learning method that segments the brain tissue, white matter tracts, and gray matter regions \cite{karimi2024detailed}. For isointense stage of brain development where white matter and gray matter have similar intensity in T1- and T2-weighted images (approximately between 6 and 8 months), one study used an additional FA channel to accurately segment the brain tissue with an FCN \cite{zhang2015deep}. Segmentation of the intracranial volume also know as brain masking or skull stripping, which is typically performed as a pre-processing step, has also been addressed with machine learning methods \cite{reid2018diffusion, wang2021u, karimi2020deep, faghihpirayesh2023fetal}.

\section{Discussion}
\label{sec:discussion}

\begin{figure*}[!htb]
\centering
\includegraphics[width=\textwidth]{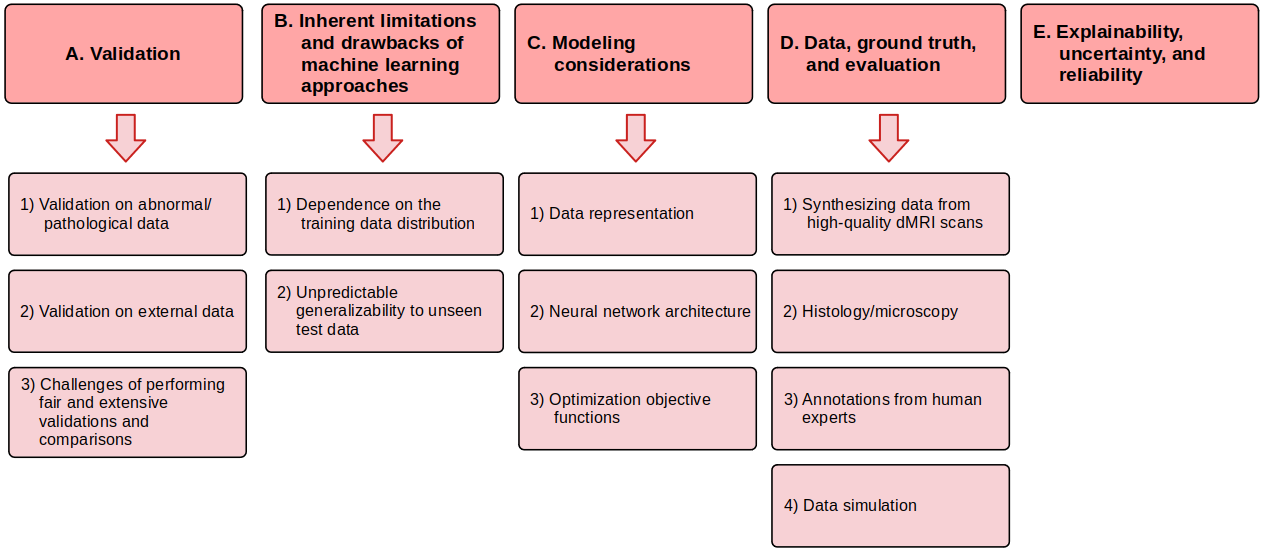}
\caption{Outline of the main aspects of application of machine learning in dMRI that are discussed in Section \ref{sec:discussion}.}
\label{fig:outline_discussion}
\end{figure*}

This section presents some thoughts on the application of machine learning for dMRI analysis. It discusses important factors that need to be taken into considerations when designing, training, and validating new methods. Additionally, it points out some of the limitations of prior works, persistent challenges, and topics that require further investigation. Figure \ref{fig:outline_discussion} shows the outline of the topics discussed in this section.

\subsection{Validation}

A shortcoming of many of the studies discussed above is with regard to validation. This subsection breaks down the main aspects of this limitation.

\subsubsection{Validation on abnormal/pathological data} 

\hspace{3mm}

The methods are usually trained and validated on data from normal brains. This can be a major flaw because brain pathologies may represent vastly different and heterogeneous structure and tissue microstructure that drastically alter the dMRI signal \cite{jelescu2020challenges, szczepankiewicz2016link, reynaud2017time}. Brain pathologies such as tumors can also change the brain anatomy and impact the performance of automatic tract analysis methods \cite{lucena2023assessing, lucena2022informative}. For tract delineation, one study has shown that on pathological brains conventional atlas-based methods work better than the more recent deep learning techniques \cite{young2024fibre}. Another study \cite{peretzke2023attractive} found that the accuracy of automatic tract segmentation methods such as TractSeg dropped significantly (DSC=0.34) on brains with tumors. A new human-in-the-loop method based on active learning achieved a mean DSC of 0.71 on the same dataset. 

For tissue microstructure mapping, some studies have validated the new methods on selected pathologies such as MS \cite{golkov2016q} and stroke \cite{li2021superdti}. One study trained a model on healthy brain data to jointly estimate the DTI, DKI, and multi-compartment models and tested the model on stroke patients \cite{hashemizadehkolowri2022jointly}. Results showed that the model accuracy on pathological tissue suffered more for multi-compartment models (e.g., NODDI) than for DTI. Similarly, a few studies have validated tract-specific analysis methods on clinical data. For example, one study \cite{tallus2023comparison} showed that automatic tract segmentation with TractSeg \cite{wasserthal2018tractseg} was more accurate in detecting microstructural changes due to traumatic brain injury than a method based on manual tract segmentation. Li at al. \cite{li2020neuro4neuro} validated their joint tract segmentation and registration method on a clinical population.

Although most published results on abnormal brains are positive and encouraging, they are not sufficient. Validation of deep learning methods on abnormal brain data is especially important because these methods are highly complex and difficult to interpret, and because the generalizability of these methods to unseen data is hard to predict \cite{arrieta2020explainable}. As an example, for tissue microstructure mapping one study has found that training with a loss function in terms of error in microstructure estimation may lead to better generalizability to pathological brains than loss functions in terms of signal reconstruction error \cite{grussu2021deep}. Other factors may also influence the generalizability of machine learning methods to pathological test data, but there has been little work to investigate this topic.

\subsubsection{Validation on external data} 

\hspace{3mm}

A number of works have reported successful application of trained machine learning models on data from other scanners/centers without any data harmonization or re-training \cite{zheng2023microstructure}. Other studies have suggested that when dMRI intensities or subject demographics deviate from training data, intensity normalization, data harmonization, or model re-training/fine-tuning may be necessary \cite{golkov2016q, lin2023cross}. However, data harmonization in dMRI is a complex problem of its own as mentioned above in Section \ref{sec:harmonization}. Moreover, there may be a tendency in the published works not to report discouraging results when it comes to validation on external data. Given the significant diversity across datasets and applications and strong dependence of machine learning methods on training data distribution, model settings, and design choices, more extensive validations are needed to assess the true advantages of machine learning methods in real-world settings with heterogeneous data.

\subsubsection{Challenges of performing fair and extensive validations and comparisons} 

\hspace{3mm}

Studies often fail to perform careful and extensive comparison with competing methods. This makes it difficult to judge the merits of a new method and contrast its advantages and disadvantages compared with existing technology. This may be due to the inherent difficulties in performing such comparisons. For tractography, as an example, Poulin \cite{poulin2019tractography} have pointed out that factors such as dMRI pre-processing, streamline seeding approach, and training and test data distributions can be major confounding factors that make it difficult to compare machine learning techniques. Lack of standardized dMRI pre-processing pipelines has been a primary challenge for assessing non-machine-learning methods as well \cite{tax2022s}. Recent efforts to develop such pipelines (e.g., \cite{theaud2020tractoflow, cai2021prequal, cieslak2021qsiprep}) may greatly benefit validation and comparison of machine learning methods for dMRI analysis. Nonetheless, as new methods appear in the literature at an accelerated pace, progress in the field can only be achieved if new methods are properly compared with the state of the art.

\subsection{Inherent limitations and drawbacks of machine learning approaches}

Despite the indisputable advantages of machine learning methods, some of which have been discussed in Section \ref{sec:Why_ML}, machine learning methods also suffer from inherent limitations. This section describes some of these, with a focus on applications in dMRI analysis.

\subsubsection{Dependence on the training data distribution} 

\hspace{3mm}

We pointed out that some machine learning models, such as neural networks, have a very high expressive capacity that allows them to learn complex functions from large training datasets. This property, however, can present itself as a downside. For example, the training data may not be sufficiently rich to enable learning of the true target function. In such a scenario, classical methods will likely continue to work as usual, whereas machine learning methods will probably make large and unpredictable errors \cite{gyori2022training}. In other words, machine learning methods can only be expected to work as far as the training data allows. For tractography, for instance, classical methods are likely to work well on brains with different morphology, whereas machine learning methods may fail if the test brain is vastly different from the training data.

Obtaining reliable training data is an important challenge in many dMRI applications discussed above. Many studies have resorted to using simulated data for model training. Realistic simulations, physical phantoms, and histologically validated dMRI data may be useful in many applications \cite{golkov2016q}. However, acquiring these types of data can often be very costly or infeasible.

\subsubsection{Unpredictable generalizability to unseen test data} 

\hspace{3mm}

Generalizability of machine learning methods to new test data can be affected by factors that may not be clear beforehand \cite{gyori2022training}. For instance, it has been suggested that factors such as signal intensity and acquisition parameters such as echo time may impact the performance of a neural network model to the extent that a complete re-training may become necessary \cite{rensonnet2021solving, golkov2016q}. Performance of conventional (i.e., non-learning) methods is likely much less affected by such unforeseen factors.

Closely related to the above point is the issue of out-of-distribution (OOD) data, which refers to data samples that are outside the distribution of the training data. Detecting OOD data in higher-dimensional signal spaces is very challenging, and machine learning methods have no performance guarantee on OOD data \cite{ruff2021unifying, liu2020energy}. A machine learning technique that works well on in-distribution data can experience unexpected and catastrophic failure on OOD test data. Obtaining a training dataset that is sufficiently rich and heterogeneous to represent all data that will be potentially encountered at test time can be difficult if not impossible because many factors may influence the data distribution. In tractography, for example, data noise and artifacts can be important factors and the trained model may produce inferior results if the training and test data are different with regard to these aspects \cite{poulin2019tractography, theberge2021track}. Non-learning methods are less prone to such failures.

It has been shown that machine learning methods can lead to systematic and unpredictable biases in microstructure mapping from dMRI signal \cite{gong2023machine, gyori2022training}. Moreover, this bias can depend on factors such as SNR, partial volume effects, and distribution of the microstructure values in the training and test data \cite{gong2023machine, gyori2022training}. As a result, predictions of these methods may lead to erroneous conclusions when studying the impact of pathology on microstructure \cite{gyori2022training}. While such biases may be present in many machine learning methods, one will require extensive validation and analysis to discover those biases. Most published works lack adequate experiments to address this issue. These suggest that machine learning methods may suffer from shortcomings that are not easy to recognize. In general, most modern machine learning methods are complex and difficult to interpret, a topic that is further discussed below in Section \ref{subsection:XAI}.

\subsection{Modeling considerations}
\label{sec:modeling_considerations}

This section describes some of the technical considerations that commonly arise in developing machine learning models in dMRI.

\subsubsection{Data representation}

\hspace{2mm}

The dMRI data used as input to machine learning models are typically in the so-called q space, where each measurement in a voxel is associated with a gradient strength and a gradient direction. The gradient strength/direction are typically different between subjects/scans. Many different approaches have been used to uniformly represent the data.

\begin{itemize}

\item A simple approach is to interpolate the data unto a fixed spherical grid, where a typical grid size is approximately 100 \cite{neher2015machine}.

\item A common approach is representation in spherical harmonics. There exists several formulations for extending the spherical harmonics representation to multi-shell data \cite{descoteaux2011multiple, cheng2011theoretical}. 

\item Zucchelli et al. \cite{zucchelli2021investigating, zucchelli2021brain} have advocated for signal representation in terms of their rotationally invariant features \cite{zucchelli2020computational}. They have shown that, compared with the two common representations mentioned above, their representation leads to better estimation of multi-compartment models \cite{zucchelli2021investigating} and FOD \cite{zucchelli2020computational}. 

\item Another approach is sparse representation in dictionaries \cite{ye2019deep}. In \cite{ye2020improved}, dictionaries that were separable in spatial and angular domains \cite{schwab2016spatial} were used. Compared with non-separable dictionaries, separable formulations significantly reduce the number of parameters and may improve the analysis results \cite{ye2020improved}.

\item Graphs offer a flexible representation framework that has been successfully used by prior works. In \cite{hong2019longitudinal}, the authors constructed the adjacency matrix in terms of the difference between pairs of measurements. They computed the difference in terms of spatial distance, angular difference between gradient directions, and difference in gradient strengths. A similar representation was used in \cite{chen2017neighborhood} to extend the non-local means denoising algorithm to dMRI.

\item Another approach, proposed in \cite{park2021diffnet}, consists of projecting the q-space data onto the standard (xy, xz, and yz) planes and binning them into fixed grids. For multi-shell data, each shell is binned separately. 

\item For applications that require large spatial fields of view, such as in tract segmentation \cite{wasserthal2018tractseg}, limited GPU memory typically does not allow for using the dMRI data as model input. In such applications, as mentioned above in Sections \ref{sec:Tractography}, \ref{sec:TractSeg}, and \ref{sec:TractSpecific}, DTI or FOD maps may be computed and used as input for machine learning models.

\end{itemize}

\subsubsection{Neural network architecture}

\hspace{2mm}

Most deep learning works discussed in this paper appear to have designed their network architecture in an ad-hoc way as they have provided little or no justification for their choices. One study \cite{chen2022prediction} has reported successful application of network architecture search \cite{zoph2016neural} for dMRI data prediction.

Standard convolutional operations are equivariant to translation, which is one of the key reasons for their success in computer vision. However, they are not equivariant to rotation. Rotation-invariance and rotation-equivariance (i.e., a rotation of the input to result in an identical rotation in the output) are desired in many applications. Incorporating these properties in the network architecture removes the need for the model to learn every possible orientation, which would otherwise require large datasets or massive data augmentation. It can improve the performance in various computer vision tasks \cite{esteves2018learning}. In dMRI, the q-space data in each voxel can be represented as spherical maps. Therefore, in addition to the rotations in the voxel (physical) space, q-space rotations should also be accounted for \cite{elaldi20233}. Most standard neural network models such as CNNs have been designed for signals in Euclidean space such as digital images. An ideal convolutional operation for q-space data should be either rotation-invariant (e.g., for estimating scalar maps of tissue microstructure) or rotation-equivariant (e.g., for FOD estimation). Moreover, it should allow for irregular sampling on the sphere. 

Muller et al. \cite{muller2021rotation} have argued that standard (i.e., non-rotationally-equivariant) networks, when applied in dMRI, are at the risk of both under-fitting and over-fitting. This is because they need to learn the rotation-equivariance relationships from training data, which may be biased towards certain orientations. Incorporating rotation-equivariance will eliminate the need for learning these constraints, enable better weight sharing, improve generalization to orientations not seen in the training data, and allow the model to concentrate on learning the actual underlying function of interest \cite{elaldi2021equivariant, muller2021rotation}.

Various techniques have been proposed to build rotation-equivariance into neural networks and to extend standard convolutions to signals on the sphere. Some of these designs are described below.

\begin{itemize}

\item Some studies have proposed to discretize the surface of the sphere and to apply standard 2D convolutions on the resulting grid \cite{su2017learning, boomsma2017spherical, coors2018spherenet}. A straightforward discretization can use an equiangular spacing in the standard spherical-polar coordinate system. Boomsma and Frellsen \cite{boomsma2017spherical} suggest that a better discretization can be achieved by using the ``cubed sphere'' representation originally developed by \cite{ronchi1996cubed}. In this representation, each point on the sphere is represented by the coordinates in one of the six patches of a circumscribed cube. They define cubed-sphere convolution as applying a standard 2D convolution on a uniformly spaced grid on the six sides of the cube. This method can be extended to concentric shells.

\item A common approach in computer vision is to divide the surface of the sphere into small regions and process each region as a planar signal with standard 2D convolutions \cite{gillet2019deep, fluri2018cosmological}. This processing is usually followed by hierarchical pooling. However, this approach does not result in rotation-equivariance.

\item Another approach is based on representing the data as a graph. Perraudin et al. proposed DeepShere \cite{perraudin2019deepsphere}, which processes spherical signals by representing the sphere as a graph. DeepShere relies on filters that are restricted to being radial. As a result, the convolutional operations are rotation-equivariant and the network predictions become invariant or equivariant to input rotations. Moreover, this formulation has a lower computational cost than performing the convolution with spherical harmonics.

\item Various other techniques have been used to develop rotation-equivariant convolutional operations. The work of Cohen et al. \cite{cohen2018spherical} is based on the spherical Fourier transform, where the convolution is performed in the spectral domain based on the convolution theorem. Banerjee et al. \cite{banerjee2020volterranet} have developed a method based on Volterra's function. Bouza et al. \cite{bouza2021higher} have proposed a manifold-valued convolution based on a generalization of the Volterra Series to manifold-valued functions, whereas Esteves et al. \cite{esteves2018learning} propose a rotation-equivariant 3D convolutional network based on spherical harmonics. Cohen and Welling \cite{cohen2016group} have proposed group equivariant convolutional networks. Group convolutions enable equivariance with regard to rotation, thereby increasing the expressive power of convolutions without increasing the number of parameters.

\item Muller et al. \cite{muller2021rotation} have generalized rotation-equivariant neural networks to the 6D space of dMRI data. They have develop a new linear layer that is equivariant under rotations in the voxel space and q-space and under translations in the voxel space. Their formulation enables translation-equivariance in the coordinate space and rotation-invariance in the coordinate space and q-space.

\end{itemize}

Numerous works have reported that these specialized network architectures can improve the performance of deep learning models in different dMRI analysis tasks \cite{bouza2021higher, goodwin2022can, koppers2021enhancing, kerkela2022microstructural, liu2022group}. Bouza et al. found that their proposed convolution operation improved the state of the art in classification of Parkinson’s Disease patients versus healthy controls and in FOD reconstruction \cite{bouza2021higher}. Sedlar et al. reported improved estimation of NODDI parameters with a rotationally-invariant CNN \cite{sedlar2021spherical}, while Liu et al. showed that their network resulted in a 10\% reduction in the number of trainable weights without a reduction in performance in a tissue segmentation application \cite{liu2022bundle}. Kerkela et al. \cite{kerkela2022microstructural} found that a rotation invariant spherical CNN performed better than a standard MLP in estimating a multi-compartment model. For angular super-resolution of dMRI data, Lyon et al. showed that a parametric continuous convolution network outperformed a standard CNN while reducing the number of parameters by an order of magnitude \cite{lyon2023spatio}. Other dMRI studies have shown that rotationally-equivariant networks can lead to more accurate tissue segmentation and lesion segmentation compared with standard networks \cite{liu2022group, muller2021rotation}. Goodwin-Allcock et al. have shown that including rotation equivariance improves the generalizability of deep learning models to new gradient schemes and reduces the required training data size \cite{goodwin2022can}.

\subsubsection{Optimization objective functions}

\hspace{2mm}

For tissue microstructure mapping, one study compared a training loss function based on the error in microstructure prediction with another in terms of the error in signal prediction \cite{grussu2021deep}. The former has been by far more common in machine learning methods, whereas the latter is the common loss function in standard optimization-based methods. It was shown that although the former led to lower microstructure prediction errors when the noise was modeled accurately, the latter resulted in lower signal reconstruction error and it was more practical because it eliminated the need for reference microstructure and enabled training with synthetic and real data \cite{grussu2021deep}. The authors speculated that the weakness of a loss function based on signal reconstruction error could be due to the fact that neural networks might tend to learn features that originate from the noise floor. In the context of IVIM model fitting, a similar method of training based on the signal prediction error was proposed in \cite{kaandorp2021improved}, where the authors used the terms ``unsupervised'' and ``physics-informed'' to describe their method. Parker et al. argue that the use of root mean square loss function is not consistent with the measurement noise distribution in dMRI \cite{parker2023rician}. They propose a loss based on Rician likelihood and show that, for estimation of apparent diffusion coefficient (ADC) and IVIM parameters, it leads to lower estimation bias at lower SNR.

Another work on tissue microstructure estimation has argued that loss functions based purely on the predicted signal are inadequate \cite{chen2023deep}. Instead, the authors propose two alternative loss functions that additionally incorporate the error in microstructure estimation. One of their loss functions consists of the $\ell_1$ norms of the errors of the dMRI signal and the microstructure indices. The other loss function, which they name the spherical variance loss, is based on the spherical variance of the dMRI signal that is correlated with microstructure \cite{huynh2019probing}. Experiments show that, compared with standard loss functions such as the $\ell_1$, $\ell_2$, and Huber loss of the signal prediction error, these two augmented loss functions result in more accurate prediction for a range of parameters including GFA, DKI, and NODDI. However, they do not compare these new loss functions with a loss function based on the error in the predicted tissue microstructure index.

\subsection{Data, ground truth, and evaluation}

For many dMRI applications, obtaining data with reliable target/label is difficult or impossible. 

\subsubsection{Synthesizing data from high-quality dMRI scans} 

\hspace{3mm}

Most commonly, high-quality dMRI measurements have been used to generate the data needed for developing and validating machine learning methods. For tract segmentation/registration, Li et al. used an automatic method based on probabilistic tractography and atlas-based segmentation to generate their training labels \cite{li2021longitudinal}. For tissue microstructure mapping and for FOD estimation, most studies have used dense q-space data and estimated ground truth parameters using standard methods \cite{zheng2023microstructure, koppers2017reconstruction, yao2023unified, nath2020deep, koppers2017reliable, nath2020deep}.

\subsubsection{Histology/microscopy} 

\hspace{3mm}

In certain applications, histological data can be used to establish a reasonable ground truth, but at considerable cost. For estimation of intra-axonal water exchange time, Hill et al. used histology analysis of an in-vivo mouse model of demyelination using electron microscopy to generate ground truth \cite{hill2021machine}. Specifically, they measured the thickness of the myelin sheath as a measure of exchange time. Surprisingly, they found that the prediction of their random forest regression model based on dMRI signal was strongly correlated with histological measurements, with a Pearson correlation coefficient of 0.98. Nath et al. used confocal histology to generate data for FOD estimation \cite{nath2019deep}. Since the data obtained with this technique was costly and very limited, they performed massive data augmentation to increase the effective size of their data. For data harmonization, a large purely dMRI scan-rescan dataset was used to boost a small dataset of paired histology and dMRI \cite{nath2019inter}. Another work used co-registered dMRI and histology data to develop a machine learning method for estimating several microstructural indices and achieved higher estimation accuracy than standard methods \cite{fick2017assessing}. For validating a dMRI-based tract segmentation methods, Lucena et al. \cite{lucena2023assessing, lucena2022informative} have advocated for using navigated transcranial magnetic stimulation to establish a ground truth.

\subsubsection{Annotations from human experts} 

\hspace{3mm}

In some applications, such as those involving segmenting or analyzing specific white matter tracts, labels can be obtained from human experts. This approach, however, may suffer from difficulty of defining anatomical ground truth \cite{wasserthal2018tract}, high intra- and inter-expert variability \cite{wakana2007reproducibility, schilling2021tractography}, and labeling errors. Such label errors need to be properly accounted for using advanced machine learning techniques \cite{karimi2020deep}. In applications such as tractography and tract analysis, obtaining labels from multiple experts may be useful but costly \cite{yendiki2011automated, poulin2019tractography, theberge2021track}.

\subsubsection{Data simulation} 

\hspace{3mm}

Another approach is to use simulation techniques such as the Monte Carlo method. This approach has been widely used for tissue microstructure mapping \cite{hill2021machine, grussu2021deep, gyori2022training}. Despite its limitations, data simulation is sometimes the only reasonable approach when high-quality data are not available to compute a reliable ground truth. For microstructure estimation, numerical simulations and physical phantoms are considered valid and powerful methods for developing and validating computational methods \cite{jelescu2020challenges, fieremans2018physical}. Data simulation has two potential advantages: (i) it enables exploring the full range of parameters that may influence the signal, and (ii) the ground truth target is known accurately \cite{gyori2022training, masutani2019noise}. Because of the flexibility of data simulation methods, they may also be useful for investigating what data acquisition protocols lead to more accurate reconstruction. This approach was used by Lee et al. to determine the optimal b values for IVIM reconstruction with deep learning \cite{lee2021quantification}. Gyori et al. investigated the impact of training data distribution on the test performance of supervised machine learning methods for microstructure mapping \cite{gyori2022training}. When training on simulated data, they found that in order for the model to achieve reasonable accuracy on atypical brain tissue, the training data should be synthesized using the complete range of plausible parameter values. When the parameter space used to generate the training data matched that of healthy brains, estimation accuracy on atypical brains was low, although the estimations showed a deceptively high precision. 

There have been innovative approaches to dMRI data simulation in prior works. Nedjati et al. used Monte Carlo simulation to synthesize dMRI data based on histologically valid microstructure parameters \cite{nedjati2017machine}. Ye and Li have developed a method for synthesizing q-space data that can potentially be used for any machine learning application in dMRI \cite{ye2019q}. Their signal generator is an MLP that is trained to minimize the difference between the distribution of the synthesized signal and that of the observed signal. They use a continuous representation of the dMRI data in the SHORE basis \cite{ozarslan2009simple}. Their preliminary experiments have shown promising results for estimation of the NODDI model. 

Qin et al. have proposed and validated a knowledge transfer method that relies on a source domain where high-quality data are available \cite{qin2020knowledge}. Using the SHORE basis, this method interpolates the source data in the q-space and/or voxel space to synthesize data that match the quality of the target data. Model training is carried out in the source domain and transferred to the target domain. For Gibbs artifact suppression, one study trained a CNN on more than one million natural images and simulated more than 10 million data instances \cite{muckley2021training}. Experiments showed that the size and richness of the collection of natural images used for training was essential to ensuring generalizability of the model to dMRI test data. Graham et al. developed a simulation method to enable direct quantitative evaluation of artifact-correction methods \cite{graham2016realistic}. They subsequently used this method to develop and validate a deep learning motion artifact detection technique \cite{graham2018supervised}. Their work showed that with proper data simulation only a small real dataset was sufficient to calibrate the method.

Using simulated data is typically much more convenient than real-world data. Some works have relied solely on simulated data to train and or evaluate their methods \cite{parker2023rician, weine2022synthetically, kerkela2022improved, gong2023machine}. However, using in-silico simulations to synthesize training data may be inadequate because some factors such as the noise distribution are difficult to model accurately \cite{grussu2021deep}. For IVIM-DKI estimation, one study observed that increasing the noise level in the synthesized training data improved the model's test accuracy \cite{bertleff2017diffusion}. Masutani \cite{masutani2019noise} analyzed the impact of matching the noise level between training and test data. He found that the best test performance was achieved when the noise level in the training data was close to or slightly higher than the noise level in the test data. Martins et al. \cite{de2021neural}, on the other hand, found that training with a much noisier dataset could still result in superior results on test data with relatively lower noise. For DKI estimation, another work found that varying the noise level in the synthetic training data led to better results on clinical test scans compared with a model that was trained using a single noise level \cite{masutani2021synthetic}. In some applications, such as cardiac DTI \cite{weine2022synthetically} and fetal imaging \cite{karimi2021deep} mentioned above, obtaining reliable in-vivo data faces especial difficulties such that using synthetic training data may be unavoidable.

Another important consideration is the choice of evaluation metrics. It is often unclear what metrics should be reported for a comprehensive assessment of a new method and a fair comparison with competing techniques. As an example, for tissue microstructure mapping most studies report the RMSE as the main or the only metric. However, it has been suggested that correlation analysis and sensitivity analysis are important for revealing certain limitations of machine learning methods that cannot be assessed based solely on RMSE \cite{de2021use}. Moreover, rather than reporting a single global RMSE value for a method, detailed analysis of the estimation error such as its variation in different brain regions may reveal biases that are not captured with global RMSE \cite{karimi2021calibrated, epstein2022choice, gyori2022training, gong2023machine}.

Although for many dMRI computations obtaining the ground truth may be difficult or impossible, test-retest reproducibility assessments may be far easier to perform. In fact, repeatability tests have often been used to assess machine learning methods in dMRI. For example, Tunc et al. used repeated scans of the same subjects to assess the repeatability of their tract segmentation method \cite{tuncc2014automated}, while Barbieri et al. computed inter-subject variability to validate a deep learning method for IVIM estimation \cite{barbieri2020deep}. Performing such tests has become more feasible with the growing availability of public datasets that include repeated scans of the same subjects on the same scanner \cite{chamberland2019penthera} or on different scanners \cite{cai2021} or on different scanners with different protocols \cite{cai2021masivar, ning2020cross}.

Where an objective/quantitative ground truth is difficult to establish, visual assessment by human experts may be a viable alternative. One study quantified the agreement between two readers in their interpretation of IVIM parameter maps to compare a deep learning method with conventional estimation methods \cite{barbieri2020deep}. However, the authors have pointed out that this approach may favor methods that wrongly compute consistent results while failing to account for genuine heterogeneity. Another work has reported successful validation of a deep learning method for fetal DTI estimation based on blind randomized assessment and scoring by neuroanatomists and neuroradiologists \cite{karimi2021deep}.

Lack of reliable and standardized data for developing and validating machine learning methods may also be one of the main barriers to quick and widespread adoption of these methods. Despite repeated claims that machine learning methods can outperform classical techniques, these methods have not seen widespread adoption. For tractography, for example, Poulin et al. point out that despite repeated demonstration of the capabilities of machine learning methods, none of these methods have been adopted for real-world applications \cite{poulin2019tractography}. They attribute this, in part, to the lack of well-defined test benchmarks to allow conclusive performance evaluation and comparison.

An effective approach to assessing the potential of machine learning for dMRI analysis is via open competitions, where research teams are free to develop and apply standard non-learning methods and machine learning techniques. A good example is a recent open challenge where research teams were invited to develop deep learning methods for estimating diffusion tensor parameters from reduced measurements \cite{aja2023validation}. Specifically, focusing on distinguishing between episodic and chronic migraine patients, the goal was to investigate whether deep learning methods were able to achieve the same level of sensitivity and specificity when using 21 measurements, compared with 61 measurements typically used with standard methods for this analysis. With conventional estimation methods, 60\% of the differences detected by TBSS with 61 measurements are missed when using 21 measurements. A total of 14 research teams participated in the challenge. Results showed that deep learning methods improved the global image metrics such as PSNR and SSIM. They also improved the sensitivity of TBSS. However, these improvements came at the cost of higher false positive rates, which increased linearly with the true positive rate. The teams had used different deep learning models and had adopted various approaches, such as mapping the reduced measurements directly to the target DTI parameters or estimating dense dMRI measurements first. The study concluded that the results of deep learning methods should be interpreted with caution even when global image quality metrics are improved.

A very appealing approach to assessing the new methods is to use measures of patient outcome as the evaluation target. This approach has been successfully attempted in a few works, where the predicted microstructure indices have been correlated with post‐stroke outcome \cite{gibbons2019simultaneous} and outcome of pancreatic cancer patients \cite{kaandorp2021improved}. Another study used acute-phase dMRI (as well as anatomical MRI) data as input to a small neural network to directly predict the chronic size of ischemic lesions, where the T2-weighted images at 3 months after stroke were used to establish the ground truth \cite{bagher2011predicting}.

\subsection{Explainability, uncertainty, and reliability}
\label{subsection:XAI}

Advanced machine learning methods such as deep neural networks are increasingly used in safety-critical applications such as medicine. However, they are very complex and hard to interpret \cite{arnez2020comparison, arrieta2020explainable}. It is well documented that deep learning models produce overconfident and poorly-calibrated predictions \cite{guo2017, lakshminarayanan2017}, they can be fooled into making erroneous predictions \cite{goodfellow2014explaining, kurakin2016adversarial}, and they fail silently on OOD data \cite{maartensson2020reliability, bulusu2020anomalous}. There have been much effort to address these issues \cite{loquercio2020general, bulusu2020anomalous}. The majority of these efforts have focused on classification problems. Deep learning-based regression has received much less attention. For example, while for classification there is a widely-accepted definition of uncertainty calibration \cite{guo2017, lakshminarayanan2017}, for regression there is much disagreement \cite{kuleshov2018accurate, levi2019evaluating}.

Even though very few studies have investigated these issues for machine learning-based dMRI analysis, they have reported important observations. For example, it has been reported that machine learning methods produce deceptively confident (i.e., high-precision) predictions on noisy test data, while with classical estimation methods higher measurement noise levels are properly reflected in predictions as higher variance \cite{gyori2022training}. Unfortunately, very little work has been devoted to characterize these issues and to devise effective solutions.

A few studies have incorporated uncertainty estimation in their methods. For microstructure mapping, Ye et al. quantified the uncertainty of their neural network-based tissue microstructure estimation using a residual bootstrap technique \cite{ye2020improved}. They validated their uncertainty computation method by assessing the correlation with estimation error and inspecting the confidence intervals. They concluded that this method could compute practically useful uncertainty estimates. For diffusion tensor estimation, another work computed data-dependent uncertainty via loss attenuation and model uncertainty via Monte Carlo dropout \cite{karimi2021calibrated}. Their extensive experiments showed that estimation uncertainties computed by their proposed methods could highlight the model's biases, detect domain shift, and reflect the measurement noise level. Avci et al. have shown that a simple uncertainty quantification method based on dropout reduces the prediction error and can also serve as an indicator of pathology or measurement artifacts \cite{avci2021quantifying}. For IVIM estimation, one study estimated the full Gaussian posteriors (i.e., mean and variance) for the parameters and used the standard deviation of the posterior as a proxy for estimation uncertainty \cite{zhang2019implicit}. Experiments with simulated and real data showed that the computed uncertainties were qualitatively meaningful. For tract segmentation, one work has used dropout and test-time augmentation to compute epistemic and aleatoric uncertainties \cite{lucena2022informative}. After applying post-hoc uncertainty calibration, the computed uncertainty is shown to be well correlated with segmentation error. The authors argue that such uncertainty computations can be useful in surgical planning applications such as for reliable delineation of eloquent areas. In general, it may be easier to probe the explainability of conventional machine learning models than large neural networks. For example, for microstructure estimation, one study was able to extensively examine the importance of different feature sets and different measurement shells in the q-space for a random forest model \cite{fick2017assessing}. Performing a similar analysis for a deep learning model should be more challenging due to the longer training times for these models.

In the context of dMRI super-resolution, Tanno et al. estimated the uncertainty for CNN and random forest models \cite{tanno2017bayesian, tanno2019uncertainty}. For their CNN estimator, they employed a heteroscedastic noise model to compute the intrinsic uncertainty and variational dropout to compute the model uncertainty \cite{tanno2017bayesian}. However, their evaluation of estimation uncertainty relied on a qualitative visual inspection of the correlation between uncertainty and estimation error. They showed that the uncertainty was higher for brain pathologies not observed in the training data and that the computed estimation uncertainty could be useful for downstream computations such as tractography. In another work \cite{tanno2019uncertainty}, they employed similar techniques to compute the estimation uncertainty for dMRI super-resolution based on the IQT approach discussed above. They also developed methods to propagate the uncertainty estimates to downstream analyses such as computation of microstructure indices. Experiments showed that uncertainty quantification could highlight model's failures, explain prediction performance, and improve prediction accuracy. Also in the context of super-resolution, Qin et al. \cite{qin2021super} used a method based on model ensembles \cite{lakshminarayanan2017simple} to quantify the uncertainty for estimation of NODDI parameters. They showed that the computed uncertainties had moderate correlation with estimation error and could be used to improve estimation accuracy.

Model interpretability has also received little attention. Xu et al. \cite{xu2019objective} used attention maps (similar to \cite{xu2015show}) to interpret the decision mechanism of a tractography-based tract segmentation model. They showed that the attention maps provided useful insights that were consistent with the anatomy of the fiber tracts connecting different brain regions. For subject-level classification based on tract-specific tissue microstructure, one study has shown that class activation maps can identify the white matter tracts that underlie the group differences \cite{zhang2022tractoformer}. Varadarajan and Haldar criticized nonlinear machine learning FOD estimation methods for their black-box nature and difficulty of predicting their performance on unseen measurement schemes \cite{varadarajan2018towards}. They formulated FOD computation as a linear estimation problem and achieved prediction accuracy on par with or better than standard methods. One can argue that works of Ye et al. that involve unfolding the iterative algorithms for microstructure estimation \cite{ye2017tissue, ye2019deep} also represent a step towards improving the explainability of deep learning models. Rensonnet et al. \cite{rensonnet2021solving} have further improved Ye's work in this direction by proposing a hybrid method that involves the use of a physics-based fingerprinting dictionary within a deep learning framework. Their method first projects the dMRI signal into a dictionary containing physics-based responses. Subsequently, an MLP uses the sparse representations computed in the first step to compute the target microstructure indices. Compared with an MLP that is trained end-to-end, this new method showed competitive accuracy while having qualitatively more interpretable features.

Better model explainability may also aid in designing more effective computational methods. For example, it is unclear why machine learning methods often achieve more accurate tissue microstructure estimation than standard optimization-based methods. The shortcomings of optimization methods are due to several factors including sensitivity to measurement noise, poor initialization, and degeneracy (i.e., presence of multiple equally-optimal solutions). For DTI estimation, which consists in a linear or non-linear least squares problem, the advantage of deep learning models are likely only due to their ability to use the measurements in neighboring voxels to reduce the impact of measurement noise. For multi-compartment models with a complex optimization landscape, on the other hand, machine learning models may actually be able to learn to avoid local minima and compute better estimates.

\section{Conclusions}

Machine learning has the potential to usher in a new generation of computational techniques for dMRI data processing and analysis. Some of the capabilities that have been demonstrated by multiple studies include: denoising, artifact correction, data harmonization, estimation of microstructural indices, tractography, and tract analysis tasks such as segmentation and registration. Compared with conventional computational techniques in dMRI, the new machine learning methods may offer improved performance in several aspects including: faster computation, ability to handle imperfect data, flexible modular design, end-to-end optimization, seamless and easy integration of spatial information and other sources of information such as anatomical MRI. However, in order to realize the full potential of these methods, we need to overcome several critical limitations and remaining issues. More rigorous validation on rich and heterogeneous datasets in necessary. Standardized data preprocessing pipelines, validation benchmarks, and evaluation metrics can dramatically help the research community in developing and assessing more effective methods and discovering truly meritorious techniques. Finally, enabling model explainability and proper uncertainty estimation may facilitate and expedite the adoption of these methods in clinical and scientific applications.

\section*{Funding}

This research was supported in part by the National Institute of Neurological Disorders and Stroke and Eunice Kennedy Shriver National Institute of Child Health and Human Development of the National Institutes of Health (NIH) under award numbers R01HD110772 and R01NS128281, and in part by NIH awards S10 OD025111 and R01 LM013608. The content of this publication is solely the responsibility of the authors and does not necessarily represent the official views of the NIH.

\bibliographystyle{ieeetr}
\bibliography{davoodreferences}

\begin{thebibliography}{100}

\bibitem{johansen2013diffusion}
H.~Johansen-Berg and T.~E. Behrens, {\em Diffusion MRI: from quantitative
  measurement to in vivo neuroanatomy}.
\newblock Academic Press, 2013.

\bibitem{le2014diffusion}
D.~Le~Bihan, ``Diffusion mri: what water tells us about the brain,'' {\em EMBO
  molecular medicine}, vol.~6, no.~5, pp.~569--573, 2014.

\bibitem{lerch2017studying}
J.~P. Lerch, A.~J. Van Der~Kouwe, A.~Raznahan, T.~Paus, H.~Johansen-Berg, K.~L.
  Miller, S.~M. Smith, B.~Fischl, and S.~N. Sotiropoulos, ``Studying
  neuroanatomy using mri,'' {\em Nature neuroscience}, vol.~20, no.~3,
  pp.~314--326, 2017.

\bibitem{tournier2019diffusion}
J.-D. Tournier, ``Diffusion mri in the brain--theory and concepts,'' {\em
  Progress in Nuclear Magnetic Resonance Spectroscopy}, vol.~112, pp.~1--16,
  2019.

\bibitem{bodini2009diffusion}
B.~Bodini and O.~Ciccarelli, ``Diffusion mri in neurological disorders,'' in
  {\em Diffusion MRI}, pp.~175--203, Elsevier, 2009.

\bibitem{salat2014diffusion}
D.~H. Salat, ``Diffusion tensor imaging in the study of aging and
  age-associated neural disease,'' in {\em Diffusion MRI}, pp.~257--281,
  Elsevier, 2014.

\bibitem{miller2016multimodal}
K.~L. Miller, F.~Alfaro-Almagro, N.~K. Bangerter, D.~L. Thomas, E.~Yacoub,
  J.~Xu, A.~J. Bartsch, S.~Jbabdi, S.~N. Sotiropoulos, J.~L. Andersson, {\em
  et~al.}, ``Multimodal population brain imaging in the uk biobank prospective
  epidemiological study,'' {\em Nature neuroscience}, vol.~19, no.~11,
  pp.~1523--1536, 2016.

\bibitem{van2013wu}
D.~C. Van~Essen, S.~M. Smith, D.~M. Barch, T.~E. Behrens, E.~Yacoub,
  K.~Ugurbil, W.-M.~H. Consortium, {\em et~al.}, ``The wu-minn human connectome
  project: an overview,'' {\em Neuroimage}, vol.~80, pp.~62--79, 2013.

\bibitem{tax2022s}
C.~M. Tax, M.~Bastiani, J.~Veraart, E.~Garyfallidis, and M.~O. Irfanoglu,
  ``What’s new and what’s next in diffusion mri preprocessing,'' {\em
  NeuroImage}, vol.~249, p.~118830, 2022.

\bibitem{jones2010challenges}
D.~K. Jones, ``Challenges and limitations of quantifying brain connectivity in
  vivo with diffusion mri,'' {\em Imaging in Medicine}, vol.~2, no.~3, p.~341,
  2010.

\bibitem{tournier2019mrtrix3}
J.-D. Tournier, R.~Smith, D.~Raffelt, R.~Tabbara, T.~Dhollander, M.~Pietsch,
  D.~Christiaens, B.~Jeurissen, C.-H. Yeh, and A.~Connelly, ``Mrtrix3: A fast,
  flexible and open software framework for medical image processing and
  visualisation,'' {\em Neuroimage}, vol.~202, p.~116137, 2019.

\bibitem{garyfallidis2014dipy}
E.~Garyfallidis, M.~Brett, B.~Amirbekian, A.~Rokem, S.~Van Der~Walt,
  M.~Descoteaux, I.~Nimmo-Smith, and D.~Contributors, ``Dipy, a library for the
  analysis of diffusion mri data,'' {\em Frontiers in neuroinformatics},
  vol.~8, p.~8, 2014.

\bibitem{theaud2020tractoflow}
G.~Theaud, J.-C. Houde, A.~Bor{\'e}, F.~Rheault, F.~Morency, and M.~Descoteaux,
  ``Tractoflow: A robust, efficient and reproducible diffusion mri pipeline
  leveraging nextflow \& singularity,'' {\em Neuroimage}, vol.~218, p.~116889,
  2020.

\bibitem{bach2014methodological}
M.~Bach {\em et~al.}, ``Methodological considerations on tract-based spatial
  statistics (tbss),'' {\em Neuroimage}, vol.~100, pp.~358--369, 2014.

\bibitem{jones2010twenty}
D.~K. Jones and M.~Cercignani, ``Twenty-five pitfalls in the analysis of
  diffusion mri data,'' {\em NMR in Biomedicine}, vol.~23, no.~7, pp.~803--820,
  2010.

\bibitem{golkov2016q}
V.~Golkov, A.~Dosovitskiy, J.~I. Sperl, M.~I. Menzel, M.~Czisch, P.~S{\"a}mann,
  T.~Brox, and D.~Cremers, ``Q-space deep learning: twelve-fold shorter and
  model-free diffusion mri scans,'' {\em IEEE transactions on medical imaging},
  vol.~35, no.~5, pp.~1344--1351, 2016.

\bibitem{wasserthal2018tractseg}
J.~Wasserthal, P.~Neher, and K.~H. Maier-Hein, ``Tractseg-fast and accurate
  white matter tract segmentation,'' {\em NeuroImage}, vol.~183, pp.~239--253,
  2018.

\bibitem{poulin2019tractography}
P.~Poulin, D.~J{\"o}rgens, P.-M. Jodoin, and M.~Descoteaux, ``Tractography and
  machine learning: Current state and open challenges,'' {\em Magnetic
  resonance imaging}, vol.~64, pp.~37--48, 2019.

\bibitem{hu2020distortion}
Z.~Hu, Y.~Wang, Z.~Zhang, J.~Zhang, H.~Zhang, C.~Guo, Y.~Sun, and H.~Guo,
  ``Distortion correction of single-shot epi enabled by deep-learning,'' {\em
  Neuroimage}, vol.~221, p.~117170, 2020.

\bibitem{moyer2020scanner}
D.~Moyer, G.~Ver~Steeg, C.~M. Tax, and P.~M. Thompson, ``Scanner invariant
  representations for diffusion mri harmonization,'' {\em Magnetic resonance in
  medicine}, vol.~84, no.~4, pp.~2174--2189, 2020.

\bibitem{de2021neural}
J.~P. de~Almeida~Martins, M.~Nilsson, B.~Lampinen, M.~Palombo, P.~T. While,
  C.-F. Westin, and F.~Szczepankiewicz, ``Neural networks for parameter
  estimation in microstructural mri: Application to a diffusion-relaxation
  model of white matter,'' {\em NeuroImage}, vol.~244, p.~118601, 2021.

\bibitem{karimi2021deep}
D.~Karimi, C.~Jaimes, F.~Machado-Rivas, L.~Vasung, S.~Khan, S.~K. Warfield, and
  A.~Gholipour, ``Deep learning-based parameter estimation in fetal
  diffusion-weighted mri,'' {\em Neuroimage}, vol.~243, p.~118482, 2021.

\bibitem{li2021longitudinal}
B.~Li, W.~J. Niessen, S.~Klein, M.~de~Groot, M.~A. Ikram, M.~W. Vernooij, and
  E.~E. Bron, ``Longitudinal diffusion mri analysis using segis-net: a
  single-step deep-learning framework for simultaneous segmentation and
  registration,'' {\em NeuroImage}, vol.~235, p.~118004, 2021.

\bibitem{jones2013white}
D.~K. Jones, T.~R. Kn{\"o}sche, and R.~Turner, ``White matter integrity, fiber
  count, and other fallacies: the do's and don'ts of diffusion mri,'' {\em
  Neuroimage}, vol.~73, pp.~239--254, 2013.

\bibitem{pierpaoli2010artifacts}
C.~Pierpaoli, ``Artifacts in diffusion mri,'' {\em Diffusion MRI: theory,
  methods and applications}, pp.~303--318, 2010.

\bibitem{jones2004squashing}
D.~K. Jones and P.~J. Basser, ``“squashing peanuts and smashing pumpkins”:
  how noise distorts diffusion-weighted mr data,'' {\em Magnetic Resonance in
  Medicine: An Official Journal of the International Society for Magnetic
  Resonance in Medicine}, vol.~52, no.~5, pp.~979--993, 2004.

\bibitem{roalf2016impact}
D.~R. Roalf, M.~Quarmley, M.~A. Elliott, T.~D. Satterthwaite, S.~N. Vandekar,
  K.~Ruparel, E.~D. Gennatas, M.~E. Calkins, T.~M. Moore, R.~Hopson, {\em
  et~al.}, ``The impact of quality assurance assessment on diffusion tensor
  imaging outcomes in a large-scale population-based cohort,'' {\em
  Neuroimage}, vol.~125, pp.~903--919, 2016.

\bibitem{yendiki2014spurious}
A.~Yendiki, K.~Koldewyn, S.~Kakunoori, N.~Kanwisher, and B.~Fischl, ``Spurious
  group differences due to head motion in a diffusion mri study,'' {\em
  Neuroimage}, vol.~88, pp.~79--90, 2014.

\bibitem{oldham2020efficacy}
S.~Oldham, A.~Arnatkevici{\=u}t{\.e}, R.~E. Smith, J.~Tiego, M.~A. Bellgrove,
  and A.~Fornito, ``The efficacy of different preprocessing steps in reducing
  motion-related confounds in diffusion mri connectomics,'' {\em NeuroImage},
  vol.~222, p.~117252, 2020.

\bibitem{baum2018impact}
G.~L. Baum, D.~R. Roalf, P.~A. Cook, R.~Ciric, A.~F. Rosen, C.~Xia, M.~A.
  Elliott, K.~Ruparel, R.~Verma, B.~Tun{\c{c}}, {\em et~al.}, ``The impact of
  in-scanner head motion on structural connectivity derived from diffusion
  mri,'' {\em Neuroimage}, vol.~173, pp.~275--286, 2018.

\bibitem{novikov2018modeling}
D.~S. Novikov, V.~G. Kiselev, and S.~N. Jespersen, ``On modeling,'' {\em
  Magnetic resonance in medicine}, vol.~79, no.~6, pp.~3172--3193, 2018.

\bibitem{novikov2019quantifying}
D.~S. Novikov, E.~Fieremans, S.~N. Jespersen, and V.~G. Kiselev, ``Quantifying
  brain microstructure with diffusion mri: Theory and parameter estimation,''
  {\em NMR in Biomedicine}, vol.~32, no.~4, p.~e3998, 2019.

\bibitem{kerkela2022improved}
L.~Kerkel{\"a}, K.~Seunarine, R.~N. Henriques, J.~D. Clayden, and C.~A. Clark,
  ``Improved reproducibility of diffusion kurtosis imaging using regularized
  non-linear optimization informed by artificial neural networks,'' {\em arXiv
  preprint arXiv:2203.07327}, 2022.

\bibitem{jensen2005diffusional}
J.~H. Jensen, J.~A. Helpern, A.~Ramani, H.~Lu, and K.~Kaczynski, ``Diffusional
  kurtosis imaging: the quantification of non-gaussian water diffusion by means
  of magnetic resonance imaging,'' {\em Magnetic Resonance in Medicine: An
  Official Journal of the International Society for Magnetic Resonance in
  Medicine}, vol.~53, no.~6, pp.~1432--1440, 2005.

\bibitem{tabesh2011estimation}
A.~Tabesh, J.~H. Jensen, B.~A. Ardekani, and J.~A. Helpern, ``Estimation of
  tensors and tensor-derived measures in diffusional kurtosis imaging,'' {\em
  Magnetic resonance in medicine}, vol.~65, no.~3, pp.~823--836, 2011.

\bibitem{neto2018advanced}
R.~Neto~Henriques, {\em Advanced methods for diffusion MRI data analysis and
  their application to the healthy ageing brain}.
\newblock PhD thesis, Ph.D. thesis, University of Cambridge, 2018.

\bibitem{maier2017challenge}
K.~H. Maier-Hein {\em et~al.}, ``The challenge of mapping the human connectome
  based on diffusion tractography,'' {\em Nature communications}, vol.~8,
  no.~1, pp.~1--13, 2017.

\bibitem{thomas2014anatomical}
C.~Thomas, Q.~Y. Frank, M.~O. Irfanoglu, P.~Modi, K.~S. Saleem, D.~A. Leopold,
  and C.~Pierpaoli, ``Anatomical accuracy of brain connections derived from
  diffusion mri tractography is inherently limited,'' {\em Proceedings of the
  National Academy of Sciences}, vol.~111, no.~46, pp.~16574--16579, 2014.

\bibitem{calixto2024white}
C.~Calixto, M.~Soldatelli, B.~Li, L.~Pierotich, A.~Gholipour, S.~K. Warfield,
  and D.~Karimi, ``White matter tract crossing and bottleneck regions in the
  fetal brain,'' {\em bioRxiv}, pp.~2024--07, 2024.

\bibitem{cai2021masivar}
L.~Y. Cai, Q.~Yang, P.~Kanakaraj, V.~Nath, A.~T. Newton, H.~A. Edmonson,
  J.~Luci, B.~N. Conrad, G.~R. Price, C.~B. Hansen, {\em et~al.}, ``Masivar:
  Multisite, multiscanner, and multisubject acquisitions for studying
  variability in diffusion weighted mri,'' {\em Magnetic resonance in
  medicine}, vol.~86, no.~6, pp.~3304--3320, 2021.

\bibitem{ning2020cross}
L.~Ning, E.~Bonet-Carne, F.~Grussu, F.~Sepehrband, E.~Kaden, J.~Veraart, S.~B.
  Blumberg, C.~S. Khoo, M.~Palombo, I.~Kokkinos, {\em et~al.}, ``Cross-scanner
  and cross-protocol multi-shell diffusion mri data harmonization: Algorithms
  and results,'' {\em Neuroimage}, vol.~221, p.~117128, 2020.

\bibitem{fick2017assessing}
R.~H. Fick, N.~Sepasian, M.~Pizzolato, A.~Ianus, and R.~Deriche, ``Assessing
  the feasibility of estimating axon diameter using diffusion models and
  machine learning,'' in {\em 2017 IEEE 14th International Symposium on
  Biomedical Imaging (ISBI 2017)}, pp.~766--769, IEEE, 2017.

\bibitem{hornik1989multilayer}
K.~Hornik, M.~Stinchcombe, and H.~White, ``Multilayer feedforward networks are
  universal approximators,'' {\em Neural networks}, vol.~2, no.~5,
  pp.~359--366, 1989.

\bibitem{nedjati2014machine}
G.~L. Nedjati-Gilani, T.~Schneider, M.~G. Hall, C.~A. Wheeler-Kingshott, and
  D.~C. Alexander, ``Machine learning based compartment models with
  permeability for white matter microstructure imaging,'' in {\em Medical Image
  Computing and Computer-Assisted Intervention--MICCAI 2014: 17th International
  Conference, Boston, MA, USA, September 14-18, 2014, Proceedings, Part III
  17}, pp.~257--264, Springer, 2014.

\bibitem{nedjati2017machine}
G.~L. Nedjati-Gilani, T.~Schneider, M.~G. Hall, N.~Cawley, I.~Hill,
  O.~Ciccarelli, I.~Drobnjak, C.~A.~G. Wheeler-Kingshott, and D.~C. Alexander,
  ``Machine learning based compartment models with permeability for white
  matter microstructure imaging,'' {\em NeuroImage}, vol.~150, pp.~119--135,
  2017.

\bibitem{tian2020deepdti}
Q.~Tian, B.~Bilgic, Q.~Fan, C.~Liao, C.~Ngamsombat, Y.~Hu, T.~Witzel,
  K.~Setsompop, J.~R. Polimeni, and S.~Y. Huang, ``Deepdti: High-fidelity
  six-direction diffusion tensor imaging using deep learning,'' {\em
  NeuroImage}, vol.~219, p.~117017, 2020.

\bibitem{qin2021multimodal}
Y.~Qin, Y.~Li, Z.~Zhuo, Z.~Liu, Y.~Liu, and C.~Ye, ``Multimodal super-resolved
  q-space deep learning,'' {\em Medical Image Analysis}, vol.~71, p.~102085,
  2021.

\bibitem{gruen2023spatially}
J.~Gruen, S.~Groeschel, and T.~Schultz, ``Spatially regularized low-rank tensor
  approximation for accurate and fast tractography,'' {\em NeuroImage},
  vol.~271, p.~120004, 2023.

\bibitem{hong2019longitudinal}
Y.~Hong, J.~Kim, G.~Chen, W.~Lin, P.-T. Yap, and D.~Shen, ``Longitudinal
  prediction of infant diffusion mri data via graph convolutional adversarial
  networks,'' {\em IEEE transactions on medical imaging}, vol.~38, no.~12,
  pp.~2717--2725, 2019.

\bibitem{weine2022synthetically}
J.~Weine, R.~J. van Gorkum, C.~T. Stoeck, V.~Vishnevskiy, and S.~Kozerke,
  ``Synthetically trained convolutional neural networks for improved tensor
  estimation from free-breathing cardiac dti,'' {\em Computerized Medical
  Imaging and Graphics}, vol.~99, p.~102075, 2022.

\bibitem{karimi2022diffusion}
D.~Karimi and A.~Gholipour, ``Diffusion tensor estimation with transformer
  neural networks,'' {\em Artificial Intelligence in Medicine}, vol.~130,
  p.~102330, 2022.

\bibitem{kaandorp2021improved}
M.~P. Kaandorp, S.~Barbieri, R.~Klaassen, H.~W. van Laarhoven, H.~Crezee, P.~T.
  While, A.~J. Nederveen, and O.~J. Gurney-Champion, ``Improved unsupervised
  physics-informed deep learning for intravoxel incoherent motion modeling and
  evaluation in pancreatic cancer patients,'' {\em Magnetic resonance in
  medicine}, vol.~86, no.~4, pp.~2250--2265, 2021.

\bibitem{barbieri2020deep}
S.~Barbieri, O.~J. Gurney-Champion, R.~Klaassen, and H.~C. Thoeny, ``Deep
  learning how to fit an intravoxel incoherent motion model to
  diffusion-weighted mri,'' {\em Magnetic resonance in medicine}, vol.~83,
  no.~1, pp.~312--321, 2020.

\bibitem{wasserthal2018tract}
J.~Wasserthal, P.~F. Neher, and K.~H. Maier-Hein, ``Tract orientation mapping
  for bundle-specific tractography,'' in {\em Medical Image Computing and
  Computer Assisted Intervention--MICCAI 2018: 21st International Conference,
  Granada, Spain, September 16-20, 2018, Proceedings, Part III 11}, pp.~36--44,
  Springer, 2018.

\bibitem{li2021superdti}
H.~Li, Z.~Liang, C.~Zhang, R.~Liu, J.~Li, W.~Zhang, D.~Liang, B.~Shen,
  X.~Zhang, Y.~Ge, {\em et~al.}, ``Superdti: Ultrafast dti and fiber
  tractography with deep learning,'' {\em Magnetic resonance in medicine},
  vol.~86, no.~6, pp.~3334--3347, 2021.

\bibitem{neher2015machine}
P.~F. Neher, M.~G{\"o}tz, T.~Norajitra, C.~Weber, and K.~H. Maier-Hein, ``A
  machine learning based approach to fiber tractography using classifier
  voting,'' in {\em Medical Image Computing and Computer-Assisted
  Intervention--MICCAI 2015: 18th International Conference, Munich, Germany,
  October 5-9, 2015, Proceedings, Part I 18}, pp.~45--52, Springer, 2015.

\bibitem{gong2019robust}
T.~Gong, Q.~Tong, H.~He, Z.~Li, and J.~Zhong, ``Robust diffusion parametric
  mapping of motion-corrupted data with a three-dimensional convolutional
  neural network,'' {\em arXiv preprint arXiv:1905.13075}, 2019.

\bibitem{gong2021deep}
T.~Gong, Q.~Tong, Z.~Li, H.~He, H.~Zhang, and J.~Zhong, ``Deep learning-based
  method for reducing residual motion effects in diffusion parameter
  estimation,'' {\em Magnetic Resonance in Medicine}, vol.~85, no.~4,
  pp.~2278--2293, 2021.

\bibitem{wegmayr2018data}
V.~Wegmayr, G.~Giuliari, S.~Holdener, and J.~Buhmann, ``Data-driven fiber
  tractography with neural networks,'' in {\em 2018 IEEE 15th international
  symposium on biomedical imaging (ISBI 2018)}, pp.~1030--1033, IEEE, 2018.

\bibitem{hill2021machine}
I.~Hill, M.~Palombo, M.~Santin, F.~Branzoli, A.-C. Philippe, D.~Wassermann,
  M.-S. Aigrot, B.~Stankoff, A.~Baron-Van~Evercooren, M.~Felfli, {\em et~al.},
  ``Machine learning based white matter models with permeability: An
  experimental study in cuprizone treated in-vivo mouse model of axonal
  demyelination,'' {\em NeuroImage}, vol.~224, p.~117425, 2021.

\bibitem{cai2023convolutional}
L.~Y. Cai, H.~H. Lee, N.~R. Newlin, C.~I. Kerley, P.~Kanakaraj, Q.~Yang, G.~W.
  Johnson, D.~Moyer, K.~G. Schilling, F.~Rheault, {\em et~al.},
  ``Convolutional-recurrent neural networks approximate diffusion tractography
  from t1-weighted mri and associated anatomical context,'' {\em bioRxiv},
  pp.~2023--02, 2023.

\bibitem{zucchelli2021brain}
M.~Zucchelli, S.~Deslauriers-Gauthier, and R.~Deriche, ``Brain tissue
  microstructure characterization using dmri based autoencoder
  neural-networks,'' in {\em Computational Diffusion MRI: 12th International
  Workshop, CDMRI 2021, Held in Conjunction with MICCAI 2021, Strasbourg,
  France, October 1, 2021, Proceedings 12}, pp.~48--57, Springer, 2021.

\bibitem{nath2021dw}
V.~Nath, K.~Ramadass, K.~G. Schilling, C.~B. Hansen, R.~Fick, S.~K. Pathak,
  A.~W. Anderson, and B.~A. Landman, ``Dw-mri microstructure model of models
  captured via single-shell bottleneck deep learning,'' in {\em Computational
  Diffusion MRI: International MICCAI Workshop, Lima, Peru, October 2020},
  pp.~147--157, Springer, 2021.

\bibitem{dietrich2008influence}
O.~Dietrich, J.~G. Raya, S.~B. Reeder, M.~Ingrisch, M.~F. Reiser, and S.~O.
  Schoenberg, ``Influence of multichannel combination, parallel imaging and
  other reconstruction techniques on mri noise characteristics,'' {\em Magnetic
  resonance imaging}, vol.~26, no.~6, pp.~754--762, 2008.

\bibitem{canales2015spherical}
E.~J. Canales-Rodr{\'\i}guez, A.~Daducci, S.~N. Sotiropoulos, E.~Caruyer,
  S.~Aja-Fern{\'a}ndez, J.~Radua, J.~M.~Y. Mendizabal, Y.~Iturria-Medina,
  L.~Melie-Garc{\'\i}a, Y.~Alem{\'a}n-G{\'o}mez, {\em et~al.}, ``Spherical
  deconvolution of multichannel diffusion mri data with non-gaussian noise
  models and spatial regularization,'' {\em PloS one}, vol.~10, no.~10,
  p.~e0138910, 2015.

\bibitem{jones2004effect}
D.~K. Jones, ``The effect of gradient sampling schemes on measures derived from
  diffusion tensor mri: a monte carlo study,'' {\em Magnetic Resonance in
  Medicine: An Official Journal of the International Society for Magnetic
  Resonance in Medicine}, vol.~51, no.~4, pp.~807--815, 2004.

\bibitem{veraart2016diffusion}
J.~Veraart, E.~Fieremans, and D.~S. Novikov, ``Diffusion mri noise mapping
  using random matrix theory,'' {\em Magnetic resonance in medicine}, vol.~76,
  no.~5, pp.~1582--1593, 2016.

\bibitem{manjon2010adaptive}
J.~V. Manj{\'o}n, P.~Coup{\'e}, L.~Mart{\'\i}-Bonmat{\'\i}, D.~L. Collins, and
  M.~Robles, ``Adaptive non-local means denoising of mr images with spatially
  varying noise levels,'' {\em Journal of Magnetic Resonance Imaging}, vol.~31,
  no.~1, pp.~192--203, 2010.

\bibitem{st2016non}
S.~St-Jean, P.~Coup{\'e}, and M.~Descoteaux, ``Non local spatial and angular
  matching: Enabling higher spatial resolution diffusion mri datasets through
  adaptive denoising,'' {\em Medical image analysis}, vol.~32, pp.~115--130,
  2016.

\bibitem{chen2017neighborhood}
G.~Chen, B.~Dong, Y.~Zhang, D.~Shen, and P.-T. Yap, ``Neighborhood matching for
  curved domains with application to denoising in diffusion mri,'' in {\em
  International Conference on Medical Image Computing and Computer-Assisted
  Intervention}, pp.~629--637, Springer, 2017.

\bibitem{lehtinen2018noise2noise}
J.~Lehtinen, J.~Munkberg, J.~Hasselgren, S.~Laine, T.~Karras, M.~Aittala, and
  T.~Aila, ``Noise2noise: Learning image restoration without clean data,'' {\em
  arXiv preprint arXiv:1803.04189}, 2018.

\bibitem{krull2019noise2void}
A.~Krull, T.-O. Buchholz, and F.~Jug, ``Noise2void-learning denoising from
  single noisy images,'' in {\em Proceedings of the IEEE/CVF conference on
  computer vision and pattern recognition}, pp.~2129--2137, 2019.

\bibitem{batson2019noise2self}
J.~Batson and L.~Royer, ``Noise2self: Blind denoising by self-supervision,'' in
  {\em International Conference on Machine Learning}, pp.~524--533, PMLR, 2019.

\bibitem{fadnavis2020patch2self}
S.~Fadnavis, J.~Batson, and E.~Garyfallidis, ``Patch2self: Denoising diffusion
  mri with self-supervised learning,'' {\em Advances in Neural Information
  Processing Systems}, vol.~33, pp.~16293--16303, 2020.

\bibitem{tian2022sdndti}
Q.~Tian, Z.~Li, Q.~Fan, J.~R. Polimeni, B.~Bilgic, D.~H. Salat, and S.~Y.
  Huang, ``Sdndti: Self-supervised deep learning-based denoising for diffusion
  tensor mri,'' {\em Neuroimage}, vol.~253, p.~119033, 2022.

\bibitem{schilling2021patch2self}
K.~G. Schilling, S.~Fadnavis, M.~Visagie, E.~Garyfallidis, B.~A. Landman, S.~A.
  Smith, and K.~P. O’Grady, ``Patch2self denoising of diffusion mri in the
  cervical spinal cord improves repeatability and feature conspicuity,'' in
  {\em International Society for Magnetic Resonance in Medicine Annual
  Meeting}, 2021.

\bibitem{andersson2021diffusion}
J.~L. Andersson, ``Diffusion mri artifact correction,'' in {\em Advances in
  Magnetic Resonance Technology and Applications}, vol.~4, pp.~123--146,
  Elsevier, 2021.

\bibitem{andersson2003correct}
J.~L. Andersson, S.~Skare, and J.~Ashburner, ``How to correct susceptibility
  distortions in spin-echo echo-planar images: application to diffusion tensor
  imaging,'' {\em Neuroimage}, vol.~20, no.~2, pp.~870--888, 2003.

\bibitem{duong2020unsupervised}
S.~T. Duong, S.~L. Phung, A.~Bouzerdoum, and M.~M. Schira, ``An unsupervised
  deep learning technique for susceptibility artifact correction in reversed
  phase-encoding epi images,'' {\em Magnetic Resonance Imaging}, vol.~71,
  pp.~1--10, 2020.

\bibitem{zahneisen2020deep}
B.~Zahneisen, K.~Baeumler, G.~Zaharchuk, D.~Fleischmann, and M.~Zeineh, ``Deep
  flow-net for epi distortion estimation,'' {\em Neuroimage}, vol.~217,
  p.~116886, 2020.

\bibitem{alkilani2023fd}
A.~Z. Alkilani, T.~{\c{C}}ukur, and E.~U. Saritas, ``Fd-net: An unsupervised
  deep forward-distortion model for susceptibility artifact correction in
  epi,'' {\em arXiv preprint arXiv:2303.10436}, 2023.

\bibitem{legouhy2022correction}
A.~Legouhy, M.~Graham, M.~Guerreri, W.~Stee, T.~Villemonteix, P.~Peigneux, and
  H.~Zhang, ``Correction of susceptibility distortion in epi: a semi-supervised
  approach with deep learning,'' in {\em International Workshop on
  Computational Diffusion MRI}, pp.~38--49, Springer, 2022.

\bibitem{schilling2020distortion}
K.~G. Schilling, J.~Blaber, C.~Hansen, L.~Cai, B.~Rogers, A.~W. Anderson,
  S.~Smith, P.~Kanakaraj, T.~Rex, S.~M. Resnick, {\em et~al.}, ``Distortion
  correction of diffusion weighted mri without reverse phase-encoding scans or
  field-maps,'' {\em PLoS One}, vol.~15, no.~7, p.~e0236418, 2020.

\bibitem{qiao2019fod}
Y.~Qiao, W.~Sun, and Y.~Shi, ``Fod-based registration for susceptibility
  distortion correction in brainstem connectome imaging,'' {\em NeuroImage},
  vol.~202, p.~116164, 2019.

\bibitem{qiao2021unsupervised}
Y.~Qiao and Y.~Shi, ``Unsupervised deep learning for fod-based susceptibility
  distortion correction in diffusion mri,'' {\em IEEE transactions on medical
  imaging}, vol.~41, no.~5, pp.~1165--1175, 2021.

\bibitem{schilling2019synthesized}
K.~G. Schilling, J.~Blaber, Y.~Huo, A.~Newton, C.~Hansen, V.~Nath, A.~T.
  Shafer, O.~Williams, S.~M. Resnick, B.~Rogers, {\em et~al.}, ``Synthesized b0
  for diffusion distortion correction (synb0-disco),'' {\em Magnetic resonance
  imaging}, vol.~64, pp.~62--70, 2019.

\bibitem{goodfellow2020generative}
I.~Goodfellow, J.~Pouget-Abadie, M.~Mirza, B.~Xu, D.~Warde-Farley, S.~Ozair,
  A.~Courville, and Y.~Bengio, ``Generative adversarial networks,'' {\em
  Communications of the ACM}, vol.~63, no.~11, pp.~139--144, 2020.

\bibitem{ayub2020inpainting}
R.~Ayub, Q.~Zhao, M.~Meloy, E.~V. Sullivan, A.~Pfefferbaum, E.~Adeli, and K.~M.
  Pohl, ``Inpainting cropped diffusion mri using deep generative models,'' in
  {\em Predictive Intelligence in Medicine: Third International Workshop, PRIME
  2020, Held in Conjunction with MICCAI 2020, Lima, Peru, October 8, 2020,
  Proceedings 3}, pp.~91--100, Springer, 2020.

\bibitem{zhang2019mri}
Q.~Zhang, G.~Ruan, W.~Yang, Y.~Liu, K.~Zhao, Q.~Feng, W.~Chen, E.~X. Wu, and
  Y.~Feng, ``Mri gibbs-ringing artifact reduction by means of machine learning
  using convolutional neural networks,'' {\em Magnetic resonance in medicine},
  vol.~82, no.~6, pp.~2133--2145, 2019.

\bibitem{muckley2021training}
M.~J. Muckley, B.~Ades-Aron, A.~Papaioannou, G.~Lemberskiy, E.~Solomon, Y.~W.
  Lui, D.~K. Sodickson, E.~Fieremans, D.~S. Novikov, and F.~Knoll, ``Training a
  neural network for gibbs and noise removal in diffusion mri,'' {\em Magnetic
  resonance in medicine}, vol.~85, no.~1, pp.~413--428, 2021.

\bibitem{ahmad20233d}
A.~Ahmad, D.~Parker, S.~Dheer, Z.~R. Samani, and R.~Verma, ``3d-qcnet--a
  pipeline for automated artifact detection in diffusion mri images,'' {\em
  Computerized Medical Imaging and Graphics}, vol.~103, p.~102151, 2023.

\bibitem{graham2018supervised}
M.~S. Graham, I.~Drobnjak, and H.~Zhang, ``A supervised learning approach for
  diffusion mri quality control with minimal training data,'' {\em NeuroImage},
  vol.~178, pp.~668--676, 2018.

\bibitem{kelly2017transfer}
C.~Kelly, M.~Pietsch, S.~Counsell, and J.-D. Tournier, ``Transfer learning and
  convolutional neural net fusion for motion artefact detection,'' in {\em
  Proceedings of the Annual Meeting of the International Society for Magnetic
  Resonance in Medicine, Honolulu, Hawaii}, vol.~3523, 2017.

\bibitem{samani2020qc}
Z.~R. Samani, J.~A. Alappatt, D.~Parker, A.~A.~O. Ismail, and R.~Verma,
  ``Qc-automator: Deep learning-based automated quality control for diffusion
  mr images,'' {\em Frontiers in neuroscience}, vol.~13, p.~1456, 2020.

\bibitem{chen2023deep}
G.~Chen, Y.~Hong, K.~M. Huynh, and P.-T. Yap, ``Deep learning prediction of
  diffusion mri data with microstructure-sensitive loss functions,'' {\em
  Medical image analysis}, vol.~85, p.~102742, 2023.

\bibitem{yin2019fast}
S.~Yin, Z.~Zhang, Q.~Peng, and X.~You, ``Fast and accurate reconstruction of
  hardi using a 1d encoder-decoder convolutional network,'' {\em arXiv preprint
  arXiv:1903.09272}, 2019.

\bibitem{koppers2021enhancing}
S.~Koppers and D.~Merhof, ``Enhancing diffusion signal augmentation using
  spherical convolutions,'' in {\em Computational Diffusion MRI: International
  MICCAI Workshop, Lima, Peru, October 2020}, pp.~189--200, Springer, 2021.

\bibitem{lyon2023spatio}
M.~Lyon, P.~Armitage, and M.~A. {\'A}lvarez, ``Spatio-angular convolutions for
  super-resolution in diffusion mri,'' {\em arXiv preprint arXiv:2306.00854},
  2023.

\bibitem{ren2021q}
M.~Ren, H.~Kim, N.~Dey, and G.~Gerig, ``Q-space conditioned translation
  networks for directional synthesis of diffusion weighted images from
  multi-modal structural mri,'' in {\em Medical Image Computing and Computer
  Assisted Intervention--MICCAI 2021: 24th International Conference,
  Strasbourg, France, September 27--October 1, 2021, Proceedings, Part VII 24},
  pp.~530--540, Springer, 2021.

\bibitem{zhang2012noddi}
H.~Zhang, T.~Schneider, C.~A. Wheeler-Kingshott, and D.~C. Alexander, ``Noddi:
  practical in vivo neurite orientation dispersion and density imaging of the
  human brain,'' {\em Neuroimage}, vol.~61, no.~4, pp.~1000--1016, 2012.

\bibitem{lyon2022angular}
M.~Lyon, P.~Armitage, and M.~A. {\'A}lvarez, ``Angular super-resolution in
  diffusion mri with a 3d recurrent convolutional autoencoder,'' in {\em
  International Conference on Medical Imaging with Deep Learning},
  pp.~834--846, PMLR, 2022.

\bibitem{ewert2024geometric}
C.~Ewert, D.~K{\"u}gler, R.~Stirnberg, A.~Koch, A.~Yendiki, and M.~Reuter,
  ``Geometric deep learning for diffusion mri signal reconstruction with
  continuous samplings (discus),'' {\em Imaging Neuroscience}, vol.~2,
  pp.~1--18, 2024.

\bibitem{elsaid2019super}
N.~M. Elsaid and Y.-C. Wu, ``Super-resolution diffusion tensor imaging using
  srcnn: a feasibility study,'' in {\em 2019 41st Annual International
  Conference of the IEEE Engineering in Medicine and Biology Society (EMBC)},
  pp.~2830--2834, IEEE, 2019.

\bibitem{tian2021srdti}
Q.~Tian, Z.~Li, Q.~Fan, C.~Ngamsombat, Y.~Hu, C.~Liao, F.~Wang, K.~Setsompop,
  J.~R. Polimeni, B.~Bilgic, {\em et~al.}, ``Srdti: Deep learning-based
  super-resolution for diffusion tensor mri,'' {\em arXiv preprint
  arXiv:2102.09069}, 2021.

\bibitem{qin2021super}
Y.~Qin, Z.~Liu, C.~Liu, Y.~Li, X.~Zeng, and C.~Ye, ``Super-resolved q-space
  deep learning with uncertainty quantification,'' {\em Medical Image
  Analysis}, vol.~67, p.~101885, 2021.

\bibitem{spears2023learning}
T.~Spears and P.~T. Fletcher, ``Learning spatially-continuous fiber orientation
  functions,'' {\em arXiv preprint arXiv:2312.05721}, 2023.

\bibitem{zeng2022fod}
R.~Zeng, J.~Lv, H.~Wang, L.~Zhou, M.~Barnett, F.~Calamante, and C.~Wang,
  ``Fod-net: A deep learning method for fiber orientation distribution angular
  super resolution,'' {\em Medical Image Analysis}, vol.~79, p.~102431, 2022.

\bibitem{lucena2021enhancing}
O.~Lucena, S.~B. Vos, V.~Vakharia, J.~Duncan, K.~Ashkan, R.~Sparks, and
  S.~Ourselin, ``Enhancing the estimation of fiber orientation distributions
  using convolutional neural networks,'' {\em Computers in Biology and
  Medicine}, vol.~135, p.~104643, 2021.

\bibitem{ye2019super}
C.~Ye, Y.~Qin, C.~Liu, Y.~Li, X.~Zeng, and Z.~Liu, ``Super-resolved q-space
  deep learning,'' in {\em Medical Image Computing and Computer Assisted
  Intervention--MICCAI 2019: 22nd International Conference, Shenzhen, China,
  October 13--17, 2019, Proceedings, Part III 22}, pp.~582--589, Springer,
  2019.

\bibitem{ye2017estimation}
C.~Ye, ``Estimation of tissue microstructure using a deep network inspired by a
  sparse reconstruction framework,'' in {\em Information Processing in Medical
  Imaging: 25th International Conference, IPMI 2017, Boone, NC, USA, June
  25-30, 2017, Proceedings 25}, pp.~466--477, Springer, 2017.

\bibitem{tanno2017bayesian}
R.~Tanno, D.~E. Worrall, A.~Ghosh, E.~Kaden, S.~N. Sotiropoulos, A.~Criminisi,
  and D.~C. Alexander, ``Bayesian image quality transfer with cnns: exploring
  uncertainty in dmri super-resolution,'' in {\em Medical Image Computing and
  Computer Assisted Intervention- MICCAI 2017: 20th International Conference,
  Quebec City, QC, Canada, September 11-13, 2017, Proceedings, Part I 20},
  pp.~611--619, Springer, 2017.

\bibitem{alexander2017image}
D.~C. Alexander, D.~Zikic, A.~Ghosh, R.~Tanno, V.~Wottschel, J.~Zhang,
  E.~Kaden, T.~B. Dyrby, S.~N. Sotiropoulos, H.~Zhang, {\em et~al.}, ``Image
  quality transfer and applications in diffusion mri,'' {\em NeuroImage},
  vol.~152, pp.~283--298, 2017.

\bibitem{blumberg2018deeper}
S.~B. Blumberg, R.~Tanno, I.~Kokkinos, and D.~C. Alexander, ``Deeper image
  quality transfer: Training low-memory neural networks for 3d images,'' in
  {\em Medical Image Computing and Computer Assisted Intervention--MICCAI 2018:
  21st International Conference, Granada, Spain, September 16-20, 2018,
  Proceedings, Part I}, pp.~118--125, Springer, 2018.

\bibitem{murray2023neural}
C.~Murray, O.~Oladosu, M.~Joshi, S.~Kolind, J.~Oh, and Y.~Zhang, ``Neural
  network algorithms predict new diffusion mri data for multi-compartmental
  analysis of brain microstructure in a clinical setting,'' {\em Magnetic
  Resonance Imaging}, vol.~102, pp.~9--19, 2023.

\bibitem{koppers2017diffusion}
S.~Koppers, C.~Haarburger, and D.~Merhof, ``Diffusion mri signal augmentation:
  from single shell to multi shell with deep learning,'' in {\em Computational
  Diffusion MRI: MICCAI Workshop, Athens, Greece, October 2016 19}, pp.~61--70,
  Springer, 2017.

\bibitem{hong2019multifold}
Y.~Hong, G.~Chen, P.-T. Yap, and D.~Shen, ``Multifold acceleration of diffusion
  mri via deep learning reconstruction from slice-undersampled data,'' in {\em
  International Conference on Information Processing in Medical Imaging},
  pp.~530--541, Springer, 2019.

\bibitem{fortin2017harmonization}
J.-P. Fortin, D.~Parker, B.~Tun{\c{c}}, T.~Watanabe, M.~A. Elliott, K.~Ruparel,
  D.~R. Roalf, T.~D. Satterthwaite, R.~C. Gur, R.~E. Gur, {\em et~al.},
  ``Harmonization of multi-site diffusion tensor imaging data,'' {\em
  Neuroimage}, vol.~161, pp.~149--170, 2017.

\bibitem{mirzaalian2016inter}
H.~Mirzaalian, L.~Ning, P.~Savadjiev, O.~Pasternak, S.~Bouix, O.~Michailovich,
  G.~Grant, C.~E. Marx, R.~A. Morey, L.~A. Flashman, {\em et~al.}, ``Inter-site
  and inter-scanner diffusion mri data harmonization,'' {\em NeuroImage},
  vol.~135, pp.~311--323, 2016.

\bibitem{cetin2020exploring}
S.~Cetin-Karayumak, K.~Stegmayer, S.~Walther, P.~R. Szeszko, T.~Crow, A.~James,
  M.~Keshavan, M.~Kubicki, and Y.~Rathi, ``Exploring the limits of combat
  method for multi-site diffusion mri harmonization,'' {\em bioRxiv},
  pp.~2020--11, 2020.

\bibitem{hansen2022contrastive}
C.~B. Hansen, K.~G. Schilling, F.~Rheault, S.~Resnick, A.~T. Shafer, L.~L.
  Beason-Held, and B.~A. Landman, ``Contrastive semi-supervised harmonization
  of single-shell to multi-shell diffusion mri,'' {\em Magnetic Resonance
  Imaging}, vol.~93, pp.~73--86, 2022.

\bibitem{zhu2017unpaired}
J.-Y. Zhu, T.~Park, P.~Isola, and A.~A. Efros, ``Unpaired image-to-image
  translation using cycle-consistent adversarial networks,'' in {\em
  Proceedings of the IEEE international conference on computer vision},
  pp.~2223--2232, 2017.

\bibitem{ozarslan2009simple}
E.~Ozarslan, C.~Koay, T.~M. Shepherd, S.~J. Blackband, and P.~J. Basser,
  ``Simple harmonic oscillator based reconstruction and estimation for
  three-dimensional q-space mri,'' in {\em Proc. Intl. Soc. Mag. Reson. Med},
  vol.~17, p.~1396, Citeseer, 2009.

\bibitem{nath2019inter}
V.~Nath, P.~Parvathaneni, C.~B. Hansen, A.~E. Hainline, C.~Bermudez,
  S.~Remedios, J.~A. Blaber, K.~G. Schilling, I.~Lyu, V.~Janve, {\em et~al.},
  ``Inter-scanner harmonization of high angular resolution dw-mri using null
  space deep learning,'' in {\em Computational Diffusion MRI: International
  MICCAI Workshop, Granada, Spain, September 2018 22}, pp.~193--201, Springer,
  2019.

\bibitem{mirzaalian2018multi}
H.~Mirzaalian, L.~Ning, P.~Savadjiev, O.~Pasternak, S.~Bouix, O.~Michailovich,
  S.~Karmacharya, G.~Grant, C.~E. Marx, R.~A. Morey, {\em et~al.}, ``Multi-site
  harmonization of diffusion mri data in a registration framework,'' {\em Brain
  imaging and behavior}, vol.~12, pp.~284--295, 2018.

\bibitem{koppers2019spherical}
S.~Koppers, L.~Bloy, J.~I. Berman, C.~M. Tax, J.~C. Edgar, and D.~Merhof,
  ``Spherical harmonic residual network for diffusion signal harmonization,''
  in {\em Computational Diffusion MRI: International MICCAI Workshop, Granada,
  Spain, September 2018 22}, pp.~173--182, Springer, 2019.

\bibitem{tong2020deep}
Q.~Tong, T.~Gong, H.~He, Z.~Wang, W.~Yu, J.~Zhang, L.~Zhai, H.~Cui, X.~Meng,
  C.~W. Tax, {\em et~al.}, ``A deep learning--based method for improving
  reliability of multicenter diffusion kurtosis imaging with varied acquisition
  protocols,'' {\em Magnetic Resonance Imaging}, vol.~73, pp.~31--44, 2020.

\bibitem{yao2023deep}
T.~Yao, F.~Rheault, L.~Y. Cai, V.~Nath, Z.~Asad, N.~Newlin, C.~Cui, R.~Deng,
  K.~Ramadass, K.~Schilling, {\em et~al.}, ``Deep constrained spherical
  deconvolution for robust harmonization,'' in {\em Medical Imaging 2023: Image
  Processing}, vol.~12464, pp.~169--176, SPIE, 2023.

\bibitem{blumberg2019multi}
S.~B. Blumberg, M.~Palombo, C.~S. Khoo, C.~M. Tax, R.~Tanno, and D.~C.
  Alexander, ``Multi-stage prediction networks for data harmonization,'' in
  {\em Medical Image Computing and Computer Assisted Intervention--MICCAI 2019:
  22nd International Conference, Shenzhen, China, October 13--17, 2019,
  Proceedings, Part IV 22}, pp.~411--419, Springer, 2019.

\bibitem{st2020harmonization}
S.~St-Jean, M.~A. Viergever, and A.~Leemans, ``Harmonization of diffusion mri
  data sets with adaptive dictionary learning,'' {\em Human brain mapping},
  vol.~41, no.~16, pp.~4478--4499, 2020.

\bibitem{karayumak2019retrospective}
S.~C. Karayumak, S.~Bouix, L.~Ning, A.~James, T.~Crow, M.~Shenton, M.~Kubicki,
  and Y.~Rathi, ``Retrospective harmonization of multi-site diffusion mri data
  acquired with different acquisition parameters,'' {\em Neuroimage}, vol.~184,
  pp.~180--200, 2019.

\bibitem{tax2019cross}
C.~M. Tax, F.~Grussu, E.~Kaden, L.~Ning, U.~Rudrapatna, C.~J. Evans,
  S.~St-Jean, A.~Leemans, S.~Koppers, D.~Merhof, {\em et~al.}, ``Cross-scanner
  and cross-protocol diffusion mri data harmonisation: A benchmark database and
  evaluation of algorithms,'' {\em NeuroImage}, vol.~195, pp.~285--299, 2019.

\bibitem{newlin2023comparing}
N.~R. Newlin, L.~Y. Cai, T.~Yao, D.~Archer, K.~G. Schilling, T.~J. Hohman,
  K.~R. Pechman, A.~Jefferson, A.~T. Shafer, S.~M. Resnick, {\em et~al.},
  ``Comparing voxel-and feature-wise harmonization of complex graph measures
  from multiple sites for structural brain network investigation of aging,'' in
  {\em Medical Imaging 2023: Image Processing}, vol.~12464, pp.~524--530, SPIE,
  2023.

\bibitem{yao2023robust}
T.~Yao, F.~Rheault, L.~Y. Cai, Z.~Asad, N.~Newlin, C.~Cui, R.~Deng,
  K.~Ramadass, A.~Shafer, S.~Resnick, {\em et~al.}, ``Robust fiber odf
  estimation using deep constrained spherical deconvolution for diffusion
  mri,'' {\em arXiv preprint arXiv:2306.02900}, 2023.

\bibitem{edgar2009white}
J.~M. Edgar and I.~R. Griffiths, ``White matter structure: a microscopist’s
  view,'' in {\em Diffusion Mri}, pp.~74--103, Elsevier, 2009.

\bibitem{kuroiwa2006ex}
T.~Kuroiwa, I.~Yamada, N.~Katsumata, S.~Endo, and K.~Ohno, ``Ex vivo
  measurement of brain tissue viscoelasticity in postischemic brain edema,'' in
  {\em Brain Edema XIII}, pp.~254--257, Springer, 2006.

\bibitem{mellergaard1989time}
P.~Mellerg{\aa}rd, F.~Bengtsson, M.-L. Smith, V.~Riesenfeld, and B.~Siesj{\"o},
  ``Time course of early brain edema following reversible forebrain ischemia in
  rats.,'' {\em Stroke}, vol.~20, no.~11, pp.~1565--1570, 1989.

\bibitem{assaf2014inferring}
Y.~Assaf and Y.~Cohen, ``Inferring microstructural information of white matter
  from diffusion mri,'' in {\em Diffusion MRI}, pp.~185--208, Elsevier, 2014.

\bibitem{bozzali2007diffusion}
M.~Bozzali and A.~Cherubini, ``Diffusion tensor mri to investigate dementias: a
  brief review,'' {\em Magnetic resonance imaging}, vol.~25, no.~6,
  pp.~969--977, 2007.

\bibitem{lakhani2020advanced}
D.~A. Lakhani, K.~Schilling, J.~Xu, and F.~Bagnato, ``Advanced multicompartment
  diffusion mri models and their application in multiple sclerosis,'' {\em
  American Journal of Neuroradiology}, vol.~41, no.~5, pp.~751--757, 2020.

\bibitem{calixto2024detailed}
C.~Calixto, M.~D. Soldatelli, C.~Jaimes, S.~K. Warfield, A.~Gholipour, and
  D.~Karimi, ``A detailed spatio-temporal atlas of the white matter tracts for
  the fetal brain,'' {\em bioRxiv}, 2024.

\bibitem{harms2017robust}
R.~L. Harms, F.~Fritz, A.~Tobisch, R.~Goebel, and A.~Roebroeck, ``Robust and
  fast nonlinear optimization of diffusion mri microstructure models,'' {\em
  Neuroimage}, vol.~155, pp.~82--96, 2017.

\bibitem{jelescu2016degeneracy}
I.~O. Jelescu, J.~Veraart, E.~Fieremans, and D.~S. Novikov, ``Degeneracy in
  model parameter estimation for multi-compartmental diffusion in neuronal
  tissue,'' {\em NMR in Biomedicine}, vol.~29, no.~1, pp.~33--47, 2016.

\bibitem{alexander2010orientationally}
D.~C. Alexander, P.~L. Hubbard, M.~G. Hall, E.~A. Moore, M.~Ptito, G.~J.
  Parker, and T.~B. Dyrby, ``Orientationally invariant indices of axon diameter
  and density from diffusion mri,'' {\em Neuroimage}, vol.~52, no.~4,
  pp.~1374--1389, 2010.

\bibitem{alexander2008general}
D.~C. Alexander, ``A general framework for experiment design in diffusion mri
  and its application in measuring direct tissue-microstructure features,''
  {\em Magnetic Resonance in Medicine: An Official Journal of the International
  Society for Magnetic Resonance in Medicine}, vol.~60, no.~2, pp.~439--448,
  2008.

\bibitem{rensonnet2019towards}
G.~Rensonnet, B.~Scherrer, G.~Girard, A.~Jankovski, S.~K. Warfield, B.~Macq,
  J.-P. Thiran, and M.~Taquet, ``Towards microstructure fingerprinting:
  estimation of tissue properties from a dictionary of monte carlo diffusion
  mri simulations,'' {\em NeuroImage}, vol.~184, pp.~964--980, 2019.

\bibitem{daducci2015accelerated}
A.~Daducci, E.~J. Canales-Rodr{\'\i}guez, H.~Zhang, T.~B. Dyrby, D.~C.
  Alexander, and J.-P. Thiran, ``Accelerated microstructure imaging via convex
  optimization (amico) from diffusion mri data,'' {\em Neuroimage}, vol.~105,
  pp.~32--44, 2015.

\bibitem{gong2023machine}
T.~Gong, F.~Grussu, C.~A. Wheeler-Kingshott, D.~C. Alexander, and H.~Zhang,
  ``Machine-learning-informed parameter estimation improves the reliability of
  spinal cord diffusion mri,'' {\em arXiv preprint arXiv:2301.12294}, 2023.

\bibitem{aliotta2019highly}
E.~Aliotta, H.~Nourzadeh, J.~Sanders, D.~Muller, and D.~B. Ennis, ``Highly
  accelerated, model-free diffusion tensor mri reconstruction using neural
  networks,'' {\em Medical physics}, vol.~46, no.~4, pp.~1581--1591, 2019.

\bibitem{park2021diffnet}
J.~Park, W.~Jung, E.-J. Choi, S.-H. Oh, J.~Jang, D.~Shin, H.~An, and J.~Lee,
  ``Diffnet: diffusion parameter mapping network generalized for input
  diffusion gradient schemes and b-value,'' {\em IEEE Transactions on Medical
  Imaging}, vol.~41, no.~2, pp.~491--499, 2021.

\bibitem{goodwin2023patch}
T.~Goodwin-Allcock, T.~Gong, R.~Gray, P.~Nachev, and H.~Zhang, ``Patch-cnn:
  Training data-efficient deep learning for high-fidelity diffusion tensor
  estimation from minimal diffusion protocols,'' {\em arXiv preprint
  arXiv:2307.01346}, 2023.

\bibitem{liu2023accelerated}
S.~Liu, Y.~Liu, X.~Xu, R.~Chen, D.~Liang, Q.~Jin, H.~Liu, G.~Chen, and Y.~Zhu,
  ``Accelerated cardiac diffusion tensor imaging using deep neural network,''
  {\em Physics in Medicine \& Biology}, vol.~68, no.~2, p.~025008, 2023.

\bibitem{huang2024deep}
J.~Huang, P.~F. Ferreira, L.~Wang, Y.~Wu, A.~I. Aviles-Rivero, C.-B.
  Sch{\"o}nlieb, A.~D. Scott, Z.~Khalique, M.~Dwornik, R.~Rajakulasingam, {\em
  et~al.}, ``Deep learning-based diffusion tensor cardiac magnetic resonance
  reconstruction: a comparison study,'' {\em Scientific Reports}, vol.~14,
  no.~1, p.~5658, 2024.

\bibitem{mozumder2019population}
M.~Mozumder, J.~M. Pozo, S.~Coelho, and A.~F. Frangi, ``Population-based
  bayesian regularization for microstructural diffusion mri with noddida,''
  {\em Magnetic resonance in medicine}, vol.~82, no.~4, pp.~1553--1565, 2019.

\bibitem{reisert2017disentangling}
M.~Reisert, E.~Kellner, B.~Dhital, J.~Hennig, and V.~G. Kiselev,
  ``Disentangling micro from mesostructure by diffusion mri: a bayesian
  approach,'' {\em NeuroImage}, vol.~147, pp.~964--975, 2017.

\bibitem{lathuiliere2019comprehensive}
S.~Lathuili{\`e}re, P.~Mesejo, X.~Alameda-Pineda, and R.~Horaud, ``A
  comprehensive analysis of deep regression,'' {\em IEEE transactions on
  pattern analysis and machine intelligence}, vol.~42, no.~9, pp.~2065--2081,
  2019.

\bibitem{karimi2022atlas}
D.~Karimi and A.~Gholipour, ``Atlas-powered deep learning (adl)-application to
  diffusion weighted mri,'' in {\em International Conference on Medical Image
  Computing and Computer-Assisted Intervention}, pp.~123--132, Springer, 2022.

\bibitem{faiyaz2022single}
A.~Faiyaz, M.~Doyley, G.~Schifitto, J.~Zhong, and M.~N. Uddin, ``Single-shell
  noddi using dictionary-learner-estimated isotropic volume fraction,'' {\em
  NMR in Biomedicine}, vol.~35, no.~2, p.~e4628, 2022.

\bibitem{ye2019deep}
C.~Ye, X.~Li, and J.~Chen, ``A deep network for tissue microstructure
  estimation using modified lstm units,'' {\em Medical image analysis},
  vol.~55, pp.~49--64, 2019.

\bibitem{ye2017tissue}
C.~Ye, ``Tissue microstructure estimation using a deep network inspired by a
  dictionary-based framework,'' {\em Medical image analysis}, vol.~42,
  pp.~288--299, 2017.

\bibitem{ye2020improved}
C.~Ye, Y.~Li, and X.~Zeng, ``An improved deep network for tissue microstructure
  estimation with uncertainty quantification,'' {\em Medical image analysis},
  vol.~61, p.~101650, 2020.

\bibitem{gibbons2019simultaneous}
E.~K. Gibbons, K.~K. Hodgson, A.~S. Chaudhari, L.~G. Richards, J.~J. Majersik,
  G.~Adluru, and E.~V. DiBella, ``Simultaneous noddi and gfa parameter map
  generation from subsampled q-space imaging using deep learning,'' {\em
  Magnetic resonance in medicine}, vol.~81, no.~4, pp.~2399--2411, 2019.

\bibitem{zheng2023microstructure}
T.~Zheng, G.~Yan, H.~Li, W.~Zheng, W.~Shi, Y.~Zhang, C.~Ye, and D.~Wu, ``A
  microstructure estimation transformer inspired by sparse representation for
  diffusion mri,'' {\em Medical Image Analysis}, vol.~86, p.~102788, 2023.

\bibitem{kaandorp2023deep}
M.~P. Kaandorp, F.~Zijlstra, C.~Federau, and P.~T. While, ``Deep learning
  intravoxel incoherent motion modeling: Exploring the impact of training
  features and learning strategies,'' {\em Magnetic Resonance in Medicine},
  vol.~90, no.~1, pp.~312--328, 2023.

\bibitem{epstein2022choice}
S.~C. Epstein, T.~J. Bray, M.~Hall-Craggs, and H.~Zhang, ``Choice of training
  label matters: how to best use deep learning for quantitative mri parameter
  estimation,'' {\em arXiv preprint arXiv:2205.05587}, 2022.

\bibitem{hashemizadehkolowri2022jointly}
S.~HashemizadehKolowri, R.-R. Chen, G.~Adluru, and E.~V. DiBella, ``Jointly
  estimating parametric maps of multiple diffusion models from undersampled
  q-space data: A comparison of three deep learning approaches,'' {\em Magnetic
  Resonance in Medicine}, vol.~87, no.~6, pp.~2957--2971, 2022.

\bibitem{li2019fast}
Z.~Li, T.~Gong, Z.~Lin, H.~He, Q.~Tong, C.~Li, Y.~Sun, F.~Yu, and J.~Zhong,
  ``Fast and robust diffusion kurtosis parametric mapping using a
  three-dimensional convolutional neural network,'' {\em IEEE Access}, vol.~7,
  pp.~71398--71411, 2019.

\bibitem{masutani2019noise}
Y.~Masutani, ``Noise level matching improves robustness of diffusion mri
  parameter inference by synthetic q-space learning,'' in {\em 2019 IEEE 16th
  International Symposium on Biomedical Imaging (ISBI 2019)}, pp.~139--142,
  IEEE, 2019.

\bibitem{gyori2022training}
N.~G. Gyori, M.~Palombo, C.~A. Clark, H.~Zhang, and D.~C. Alexander, ``Training
  data distribution significantly impacts the estimation of tissue
  microstructure with machine learning,'' {\em Magnetic resonance in medicine},
  vol.~87, no.~2, pp.~932--947, 2022.

\bibitem{de2021use}
J.~P. de~Almeida~Martins, M.~Nilsson, B.~Lampinen, M.~Palombo, C.-F. Westin,
  and F.~Szczepankiewicz, ``On the use of neural networks to fit
  high-dimensional microstructure models,'' in {\em Proceedings of the ISMRM},
  vol.~401, 2021.

\bibitem{jung2021artificial}
S.~Jung, H.~Lee, K.~Ryu, J.~E. Song, M.~Park, W.-J. Moon, and D.-H. Kim,
  ``Artificial neural network for multi-echo gradient echo--based myelin water
  fraction estimation,'' {\em Magnetic resonance in medicine}, vol.~85, no.~1,
  pp.~380--389, 2021.

\bibitem{aliotta2021extracting}
E.~Aliotta, H.~Nourzadeh, and S.~H. Patel, ``Extracting diffusion tensor
  fractional anisotropy and mean diffusivity from 3-direction dwi scans using
  deep learning,'' {\em Magnetic Resonance in Medicine}, vol.~85, no.~2,
  pp.~845--854, 2021.

\bibitem{karimi2020robust}
D.~Karimi, O.~Afacan, C.~Velasco-Annis, C.~Jaimes, C.~Rollins, S.~Warfield, and
  A.~Gholipour, ``Robust estimation of the fetal brain architecture from
  in-utero diffusion-weighted imaging,'' in {\em 2020 ISMRM \& SMRT Annual
  Meeting \& Exhibition}, 2020.

\bibitem{aja2023validation}
S.~Aja-Fern{\'a}ndez, C.~Mart{\'\i}n-Mart{\'\i}n, {\'A}.~Planchuelo-G{\'o}mez,
  A.~Faiyaz, M.~N. Uddin, G.~Schifitto, A.~Tiwari, S.~J. Shigwan, R.~K. Singh,
  T.~Zheng, {\em et~al.}, ``Validation of deep learning techniques for quality
  augmentation in diffusion mri for clinical studies,'' {\em NeuroImage:
  Clinical}, vol.~39, p.~103483, 2023.

\bibitem{smith2006tract}
S.~M. Smith {\em et~al.}, ``Tract-based spatial statistics: voxelwise analysis
  of multi-subject diffusion data,'' {\em Neuroimage}, vol.~31, no.~4,
  pp.~1487--1505, 2006.

\bibitem{skare2000condition}
S.~Skare, M.~Hedehus, M.~E. Moseley, and T.-Q. Li, ``Condition number as a
  measure of noise performance of diffusion tensor data acquisition schemes
  with mri,'' {\em Journal of magnetic resonance}, vol.~147, no.~2,
  pp.~340--352, 2000.

\bibitem{hill2018deep}
I.~Hill, M.~Palombo, M.~D. Santin, F.~Branzoli, A.-C. Philippe, D.~Wassermann,
  M.-S. Aigrot, B.~Stankoff, H.~Zhang, S.~Lehericy, {\em et~al.}, ``Deep neural
  network based framework for in-vivo axonal permeability estimation,'' in {\em
  Proceedings of the Joint Annual Meeting ISMRM-ESMRMB 2018}, ISMRM
  (International Society for Magnetic Resonance in Medicine), 2018.

\bibitem{ye2017learning}
C.~Ye, ``Learning-based ensemble average propagator estimation,'' in {\em
  Medical Image Computing and Computer Assisted Intervention- MICCAI 2017: 20th
  International Conference, Quebec City, QC, Canada, September 11-13, 2017,
  Proceedings, Part I 20}, pp.~593--601, Springer, 2017.

\bibitem{grussu2021deep}
F.~Grussu, M.~Battiston, M.~Palombo, T.~Schneider, C.~A.~G. Wheeler-Kingshott,
  and D.~C. Alexander, ``Deep learning model fitting for diffusion-relaxometry:
  a comparative study,'' in {\em Computational Diffusion MRI: International
  MICCAI Workshop, Lima, Peru, October 2020}, pp.~159--172, Springer, 2021.

\bibitem{pirk2020deep}
C.~M. Pirk, P.~A. G{\'o}mez, I.~Lipp, G.~Buonincontri, M.~Molina-Romero,
  A.~Sekuboyina, D.~Waldmannstetter, J.~Dannenberg, S.~Endt, A.~Merola, {\em
  et~al.}, ``Deep learning-based parameter mapping for joint relaxation and
  diffusion tensor mr fingerprinting,'' in {\em Medical Imaging with Deep
  Learning}, pp.~638--654, PMLR, 2020.

\bibitem{bertleff2017diffusion}
M.~Bertleff, S.~Domsch, S.~Weing{\"a}rtner, J.~Zapp, K.~O'Brien, M.~Barth, and
  L.~R. Schad, ``Diffusion parameter mapping with the combined intravoxel
  incoherent motion and kurtosis model using artificial neural networks at 3
  t,'' {\em NMR in Biomedicine}, vol.~30, no.~12, p.~e3833, 2017.

\bibitem{parker2023rician}
C.~S. Parker, A.~Schroder, S.~C. Epstein, J.~Cole, D.~C. Alexander, and
  H.~Zhang, ``Rician likelihood loss for quantitative mri using self-supervised
  deep learning,'' {\em arXiv preprint arXiv:2307.07072}, 2023.

\bibitem{zhang2019implicit}
L.~Zhang, V.~Vishnevskiy, A.~Jakab, and O.~Goksel, ``Implicit modeling with
  uncertainty estimation for intravoxel incoherent motion imaging,'' in {\em
  2019 IEEE 16th International Symposium on Biomedical Imaging (ISBI 2019)},
  pp.~1003--1007, IEEE, 2019.

\bibitem{schilling2018histological}
K.~G. Schilling, V.~Janve, Y.~Gao, I.~Stepniewska, B.~A. Landman, and A.~W.
  Anderson, ``Histological validation of diffusion mri fiber orientation
  distributions and dispersion,'' {\em Neuroimage}, vol.~165, pp.~200--221,
  2018.

\bibitem{schilling2016comparison}
K.~Schilling, V.~Janve, Y.~Gao, I.~Stepniewska, B.~A. Landman, and A.~W.
  Anderson, ``Comparison of 3d orientation distribution functions measured with
  confocal microscopy and diffusion mri,'' {\em Neuroimage}, vol.~129,
  pp.~185--197, 2016.

\bibitem{prvckovska2016reproducibility}
V.~Pr{\v{c}}kovska, P.~Rodrigues, A.~Puigdellivol~Sanchez, M.~Ramos,
  M.~Andorra, E.~Martinez-Heras, C.~Falcon, A.~Prats-Galino, and P.~Villoslada,
  ``Reproducibility of the structural connectome reconstruction across
  diffusion methods,'' {\em Journal of Neuroimaging}, vol.~26, no.~1,
  pp.~46--57, 2016.

\bibitem{bucci2013quantifying}
M.~Bucci, M.~L. Mandelli, J.~I. Berman, B.~Amirbekian, C.~Nguyen, M.~S. Berger,
  and R.~G. Henry, ``Quantifying diffusion mri tractography of the
  corticospinal tract in brain tumors with deterministic and probabilistic
  methods,'' {\em NeuroImage: Clinical}, vol.~3, pp.~361--368, 2013.

\bibitem{peled2006geometrically}
S.~Peled, O.~Friman, F.~Jolesz, and C.-F. Westin, ``Geometrically constrained
  two-tensor model for crossing tracts in dwi,'' {\em Magnetic resonance
  imaging}, vol.~24, no.~9, pp.~1263--1270, 2006.

\bibitem{tuch2002high}
D.~S. Tuch, T.~G. Reese, M.~R. Wiegell, N.~Makris, J.~W. Belliveau, and V.~J.
  Wedeen, ``High angular resolution diffusion imaging reveals intravoxel white
  matter fiber heterogeneity,'' {\em Magnetic Resonance in Medicine: An
  Official Journal of the International Society for Magnetic Resonance in
  Medicine}, vol.~48, no.~4, pp.~577--582, 2002.

\bibitem{scherrer2013reliable}
B.~Scherrer, M.~Taquet, and S.~K. Warfield, ``Reliable selection of the number
  of fascicles in diffusion images by estimation of the generalization error,''
  in {\em International Conference on Information Processing in Medical
  Imaging}, pp.~742--753, Springer, 2013.

\bibitem{schultz2010multi}
T.~Schultz, C.-F. Westin, and G.~Kindlmann, ``Multi-diffusion-tensor fitting
  via spherical deconvolution: a unifying framework,'' in {\em Medical Image
  Computing and Computer-Assisted Intervention--MICCAI 2010: 13th International
  Conference, Beijing, China, September 20-24, 2010, Proceedings, Part I 13},
  pp.~674--681, Springer, 2010.

\bibitem{tournier2007robust}
J.-D. Tournier, F.~Calamante, and A.~Connelly, ``Robust determination of the
  fibre orientation distribution in diffusion mri: non-negativity constrained
  super-resolved spherical deconvolution,'' {\em Neuroimage}, vol.~35, no.~4,
  pp.~1459--1472, 2007.

\bibitem{jeurissen2014multi}
B.~Jeurissen, J.-D. Tournier, T.~Dhollander, A.~Connelly, and J.~Sijbers,
  ``Multi-tissue constrained spherical deconvolution for improved analysis of
  multi-shell diffusion mri data,'' {\em NeuroImage}, vol.~103, pp.~411--426,
  2014.

\bibitem{patel2018better}
K.~Patel, S.~Groeschel, and T.~Schultz, ``Better fiber odfs from suboptimal
  data with autoencoder based regularization,'' in {\em Medical Image Computing
  and Computer Assisted Intervention--MICCAI 2018: 21st International
  Conference, Granada, Spain, September 16-20, 2018, Proceedings, Part III 11},
  pp.~55--62, Springer, 2018.

\bibitem{elaldi2021equivariant}
A.~Elaldi, N.~Dey, H.~Kim, and G.~Gerig, ``Equivariant spherical deconvolution:
  learning sparse orientation distribution functions from spherical data,'' in
  {\em Information Processing in Medical Imaging: 27th International
  Conference, IPMI 2021, Virtual Event, June 28--June 30, 2021, Proceedings
  27}, pp.~267--278, Springer, 2021.

\bibitem{elaldi20233}
A.~Elaldi, G.~Gerig, and N.~Dey, ``$ e (3) \times so (3) $-equivariant networks
  for spherical deconvolution in diffusion mri,'' {\em arXiv preprint
  arXiv:2304.06103}, 2023.

\bibitem{nath2019deep}
V.~Nath, K.~G. Schilling, P.~Parvathaneni, C.~B. Hansen, A.~E. Hainline,
  Y.~Huo, J.~A. Blaber, I.~Lyu, V.~Janve, Y.~Gao, {\em et~al.}, ``Deep learning
  reveals untapped information for local white-matter fiber reconstruction in
  diffusion-weighted mri,'' {\em Magnetic resonance imaging}, vol.~62,
  pp.~220--227, 2019.

\bibitem{kebiri2024deep}
H.~Kebiri, A.~Gholipour, R.~Lin, L.~Vasung, C.~Calixto, {\v{Z}}.~Krsnik,
  D.~Karimi, and M.~B. Cuadra, ``Deep learning microstructure estimation of
  developing brains from diffusion mri: a newborn and fetal study,'' {\em
  Medical Image Analysis}, vol.~95, p.~103186, 2024.

\bibitem{koppers2016direct}
S.~Koppers and D.~Merhof, ``Direct estimation of fiber orientations using deep
  learning in diffusion imaging,'' in {\em Machine Learning in Medical Imaging:
  7th International Workshop, MLMI 2016, Held in Conjunction with MICCAI 2016,
  Athens, Greece, October 17, 2016, Proceedings 7}, pp.~53--60, Springer, 2016.

\bibitem{lin2019fast}
Z.~Lin, T.~Gong, K.~Wang, Z.~Li, H.~He, Q.~Tong, F.~Yu, and J.~Zhong, ``Fast
  learning of fiber orientation distribution function for mr tractography using
  convolutional neural network,'' {\em Medical physics}, vol.~46, no.~7,
  pp.~3101--3116, 2019.

\bibitem{ye2017fiber}
C.~Ye and J.~L. Prince, ``Fiber orientation estimation guided by a deep
  network,'' in {\em Medical Image Computing and Computer Assisted
  Intervention- MICCAI 2017: 20th International Conference, Quebec City, QC,
  Canada, September 11-13, 2017, Proceedings, Part I}, pp.~575--583, Springer,
  2017.

\bibitem{karimi2021learning}
D.~Karimi, L.~Vasung, C.~Jaimes, F.~Machado-Rivas, S.~K. Warfield, and
  A.~Gholipour, ``Learning to estimate the fiber orientation distribution
  function from diffusion-weighted mri,'' {\em NeuroImage}, vol.~239,
  p.~118316, 2021.

\bibitem{koppers2017reliable}
S.~Koppers, C.~Haarburger, J.~C. Edgar, and D.~Merhof, ``Reliable estimation of
  the number of compartments in diffusion mri,'' in {\em Bildverarbeitung
  f{\"u}r die Medizin 2017: Algorithmen-Systeme-Anwendungen. Proceedings des
  Workshops vom 12. bis 14. M{\"a}rz 2017 in Heidelberg}, pp.~203--208,
  Springer, 2017.

\bibitem{koppers2017reconstruction}
S.~Koppers, M.~Friedrichs, and D.~Merhof, ``Reconstruction of diffusion
  anisotropies using 3d deep convolutional neural networks in diffusion
  imaging,'' in {\em Modeling, analysis, and visualization of anisotropy},
  pp.~393--404, Springer, 2017.

\bibitem{yao2023unified}
T.~Yao, N.~Newlin, P.~Kanakaraj, V.~Nath, L.~Y. Cai, K.~Ramadass, K.~Schilling,
  B.~A. Landman, and Y.~Huo, ``A unified learning model for estimating fiber
  orientation distribution functions on heterogeneous multi-shell
  diffusion-weighted mri,'' in {\em International Workshop on Computational
  Diffusion MRI}, pp.~13--22, Springer, 2023.

\bibitem{karimi2021machine}
D.~Karimi, L.~Vasung, C.~Jaimes, F.~Machado-Rivas, S.~Khan, S.~K. Warfield, and
  A.~Gholipour, ``A machine learning-based method for estimating the number and
  orientations of major fascicles in diffusion-weighted magnetic resonance
  imaging,'' {\em Medical image analysis}, vol.~72, p.~102129, 2021.

\bibitem{bartlett2023recovering}
J.~Bartlett, C.~Davey, L.~Johnston, and J.~Duan, ``Recovering high-quality fods
  from a reduced number of diffusion-weighted images using a model-driven deep
  learning architecture,'' {\em arXiv preprint arXiv:2307.15273}, 2023.

\bibitem{schultz2012learning}
T.~Schultz, ``Learning a reliable estimate of the number of fiber directions in
  diffusion mri,'' in {\em Medical Image Computing and Computer-Assisted
  Intervention--MICCAI 2012: 15th International Conference, Nice, France,
  October 1-5, 2012, Proceedings, Part III 15}, pp.~493--500, Springer, 2012.

\bibitem{taquet2013estimation}
M.~Taquet, B.~Scherrer, N.~Boumal, B.~Macq, and S.~K. Warfield, ``Estimation of
  a multi-fascicle model from single b-value data with a population-informed
  prior,'' in {\em Medical Image Computing and Computer-Assisted
  Intervention--MICCAI 2013: 16th International Conference, Nagoya, Japan,
  September 22-26, 2013, Proceedings, Part I 16}, pp.~695--702, Springer, 2013.

\bibitem{nath2020deep}
V.~Nath, S.~K. Pathak, K.~G. Schilling, W.~Schneider, and B.~A. Landman, ``Deep
  learning estimation of multi-tissue constrained spherical deconvolution with
  limited single shell dw-mri,'' in {\em Medical Imaging 2020: Image
  Processing}, vol.~11313, pp.~162--171, SPIE, 2020.

\bibitem{perraudin2019deepsphere}
N.~Perraudin, M.~Defferrard, T.~Kacprzak, and R.~Sgier, ``Deepsphere: Efficient
  spherical convolutional neural network with healpix sampling for cosmological
  applications,'' {\em Astronomy and Computing}, vol.~27, pp.~130--146, 2019.

\bibitem{jelescu2020challenges}
I.~O. Jelescu, M.~Palombo, F.~Bagnato, and K.~G. Schilling, ``Challenges for
  biophysical modeling of microstructure,'' {\em Journal of Neuroscience
  Methods}, vol.~344, p.~108861, 2020.

\bibitem{behrens2014mr}
T.~E. Behrens, S.~N. Sotiropoulos, and S.~Jbabdi, ``Mr diffusion
  tractography,'' in {\em Diffusion MRI}, pp.~429--451, Elsevier, 2014.

\bibitem{sotiropoulos2019building}
S.~N. Sotiropoulos and A.~Zalesky, ``Building connectomes using diffusion mri:
  why, how and but,'' {\em NMR in Biomedicine}, vol.~32, no.~4, p.~e3752, 2019.

\bibitem{liu2019deepbundle}
F.~Liu, J.~Feng, G.~Chen, Y.~Wu, Y.~Hong, P.-T. Yap, and D.~Shen, ``Deepbundle:
  fiber bundle parcellation with graph convolution neural networks,'' in {\em
  Graph Learning in Medical Imaging: First International Workshop, GLMI 2019,
  Held in Conjunction with MICCAI 2019, Shenzhen, China, October 17, 2019,
  Proceedings 1}, pp.~88--95, Springer, 2019.

\bibitem{yeh2021mapping}
C.-H. Yeh, D.~K. Jones, X.~Liang, M.~Descoteaux, and A.~Connelly, ``Mapping
  structural connectivity using diffusion mri: Challenges and opportunities,''
  {\em Journal of Magnetic Resonance Imaging}, vol.~53, no.~6, pp.~1666--1682,
  2021.

\bibitem{zhang2022quantitative}
F.~Zhang, A.~Daducci, Y.~He, S.~Schiavi, C.~Seguin, R.~Smith, C.-H. Yeh,
  T.~Zhao, and L.~J. O’Donnell, ``Quantitative mapping of the brain’s
  structural connectivity using diffusion mri tractography: a review,'' {\em
  NeuroImage}, p.~118870, 2022.

\bibitem{lemkaddem2014global}
A.~Lemkaddem, D.~Ski{\"o}ldebrand, A.~Dal~Pal{\'u}, J.-P. Thiran, and
  A.~Daducci, ``Global tractography with embedded anatomical priors for
  quantitative connectivity analysis,'' {\em Frontiers in neurology}, vol.~5,
  p.~232, 2014.

\bibitem{basser2000vivo}
P.~J. Basser, S.~Pajevic, C.~Pierpaoli, J.~Duda, and A.~Aldroubi, ``In vivo
  fiber tractography using dt-mri data,'' {\em Magnetic resonance in medicine},
  vol.~44, no.~4, pp.~625--632, 2000.

\bibitem{smith2012anatomically}
R.~E. Smith, J.-D. Tournier, F.~Calamante, and A.~Connelly,
  ``Anatomically-constrained tractography: improved diffusion mri streamlines
  tractography through effective use of anatomical information,'' {\em
  Neuroimage}, vol.~62, no.~3, pp.~1924--1938, 2012.

\bibitem{mangin2013toward}
J.-F. Mangin, P.~Fillard, Y.~Cointepas, D.~Le~Bihan, V.~Frouin, and C.~Poupon,
  ``Toward global tractography,'' {\em Neuroimage}, vol.~80, pp.~290--296,
  2013.

\bibitem{takemura2016ensemble}
H.~Takemura, C.~F. Caiafa, B.~A. Wandell, and F.~Pestilli, ``Ensemble
  tractography,'' {\em PLoS computational biology}, vol.~12, no.~2,
  p.~e1004692, 2016.

\bibitem{yeh2020shape}
F.-C. Yeh, ``Shape analysis of the human association pathways,'' {\em
  Neuroimage}, vol.~223, p.~117329, 2020.

\bibitem{smith2020quantitative}
R.~Smith, D.~Raffelt, J.-D. Tournier, and A.~Connelly, ``Quantitative
  streamlines tractography: methods and inter-subject normalisation,'' {\em
  Aperture Neuro}, vol.~2, 2020.

\bibitem{yang2021diffusion}
J.~Y.-M. Yang, C.-H. Yeh, C.~Poupon, and F.~Calamante, ``Diffusion mri
  tractography for neurosurgery: the basics, current state, technical
  reliability and challenges,'' {\em Physics in Medicine \& Biology}, vol.~66,
  no.~15, p.~15TR01, 2021.

\bibitem{rheault2020common}
F.~Rheault, P.~Poulin, A.~V. Caron, E.~St-Onge, and M.~Descoteaux, ``Common
  misconceptions, hidden biases and modern challenges of dmri tractography,''
  {\em Journal of neural engineering}, vol.~17, no.~1, p.~011001, 2020.

\bibitem{schilling2019challenges}
K.~G. Schilling, A.~Daducci, K.~Maier-Hein, C.~Poupon, J.-C. Houde, V.~Nath,
  A.~W. Anderson, B.~A. Landman, and M.~Descoteaux, ``Challenges in diffusion
  mri tractography--lessons learned from international benchmark
  competitions,'' {\em Magnetic resonance imaging}, vol.~57, pp.~194--209,
  2019.

\bibitem{malcolm2010filtered}
J.~G. Malcolm, M.~E. Shenton, and Y.~Rathi, ``Filtered multitensor
  tractography,'' {\em IEEE transactions on medical imaging}, vol.~29, no.~9,
  pp.~1664--1675, 2010.

\bibitem{lazar2003white}
M.~Lazar, D.~M. Weinstein, J.~S. Tsuruda, K.~M. Hasan, K.~Arfanakis, M.~E.
  Meyerand, B.~Badie, H.~A. Rowley, V.~Haughton, A.~Field, {\em et~al.},
  ``White matter tractography using diffusion tensor deflection,'' {\em Human
  brain mapping}, vol.~18, no.~4, pp.~306--321, 2003.

\bibitem{neher2017fiber}
P.~F. Neher, M.-A. C{\^o}t{\'e}, J.-C. Houde, M.~Descoteaux, and K.~H.
  Maier-Hein, ``Fiber tractography using machine learning,'' {\em Neuroimage},
  vol.~158, pp.~417--429, 2017.

\bibitem{sarwar2020towards}
T.~Sarwar, C.~Seguin, K.~Ramamohanarao, and A.~Zalesky, ``Towards deep learning
  for connectome mapping: A block decomposition framework,'' {\em NeuroImage},
  vol.~212, p.~116654, 2020.

\bibitem{duru2013self}
D.~G. Duru and M.~{\"O}zkan, ``Self-organizing maps for brain tractography in
  mri,'' in {\em 2013 6th International IEEE/EMBS Conference on Neural
  Engineering (NER)}, pp.~1509--1512, IEEE, 2013.

\bibitem{dedeep2018}
O.~de~Lucena, {\em Deep learning for brain analysis in MR imaging}.
\newblock PhD thesis, Master’s thesis, Universidade Estadual de Campinas,
  2018.

\bibitem{poulin2018bundle}
P.~Poulin, F.~Rheault, E.~St-Onge, P.-M. Jodoin, and M.~Descoteaux,
  ``Bundle-wise deep tracker: Learning to track bundle-specific streamline
  paths,'' {\em Proc. of the Int. Society for Magnetic Resonance in medicine
  ISMRM-ESMRMB}, 2018.

\bibitem{poulin2017learn}
P.~Poulin, M.-A. C{\^o}t{\'e}, J.-C. Houde, L.~Petit, P.~F. Neher, K.~H.
  Maier-Hein, H.~Larochelle, and M.~Descoteaux, ``Learn to track: deep learning
  for tractography,'' in {\em Medical Image Computing and Computer Assisted
  Intervention- MICCAI 2017: 20th International Conference, Quebec City, QC,
  Canada, September 11-13, 2017, Proceedings, Part I 20}, pp.~540--547,
  Springer, 2017.

\bibitem{benou2019deeptract}
I.~Benou and T.~Riklin~Raviv, ``Deeptract: A probabilistic deep learning
  framework for white matter fiber tractography,'' in {\em Medical Image
  Computing and Computer Assisted Intervention--MICCAI 2019: 22nd International
  Conference, Shenzhen, China, October 13--17, 2019, Proceedings, Part III 22},
  pp.~626--635, Springer, 2019.

\bibitem{jorgens2018learning}
D.~J{\"o}rgens, {\"O}.~Smedby, and R.~Moreno, ``Learning a single step of
  streamline tractography based on neural networks,'' in {\em Computational
  Diffusion MRI: MICCAI Workshop, Qu{\'e}bec, Canada, September 2017},
  pp.~103--116, Springer, 2018.

\bibitem{liu2024streamline}
W.~Liu, C.~Calixto, S.~K. Warfield, and D.~Karimi, ``Streamline tractography of
  the fetal brain in utero with machine learning,'' {\em arXiv preprint
  arXiv:2408.14326}, 2024.

\bibitem{theberge2021track}
A.~Th{\'e}berge, C.~Desrosiers, M.~Descoteaux, and P.-M. Jodoin,
  ``Track-to-learn: A general framework for tractography with deep
  reinforcement learning,'' {\em Medical Image Analysis}, vol.~72, p.~102093,
  2021.

\bibitem{wegmayr2021entrack}
V.~Wegmayr and J.~M. Buhmann, ``Entrack: Probabilistic spherical regression
  with entropy regularization for fiber tractography,'' {\em International
  Journal of Computer Vision}, vol.~129, pp.~656--680, 2021.

\bibitem{cai2023implementation}
L.~Y. Cai, H.~H. Lee, N.~R. Newlin, M.~E. Kim, D.~Moyer, F.~Rheault, K.~G.
  Schilling, and B.~A. Landman, ``Implementation considerations for deep
  learning with diffusion mri streamline tractography,'' {\em bioRxiv}, 2023.

\bibitem{cho2014learning}
K.~Cho, B.~Van~Merri{\"e}nboer, C.~Gulcehre, D.~Bahdanau, F.~Bougares,
  H.~Schwenk, and Y.~Bengio, ``Learning phrase representations using rnn
  encoder-decoder for statistical machine translation,'' {\em arXiv preprint
  arXiv:1406.1078}, 2014.

\bibitem{sinzinger2022reinforcement}
F.~L. Sinzinger and R.~Moreno, ``Reinforcement learning based tractography with
  so (3) equivariant agents,'' in {\em Geometric Deep Learning in Medical Image
  Analysis (Extended abstracts)}, 2022.

\bibitem{wanyan2018tractography}
T.~Wanyan, L.~Liu, and E.~Garyfallidis, ``Tractography using reinforcement
  learning and adaptive-expanding graphs,'' in {\em International symposium on
  biomedical imaging}, 2018.

\bibitem{reisert2018hamlet}
M.~Reisert, V.~A. Coenen, C.~Kaller, K.~Egger, and H.~Skibbe, ``Hamlet:
  hierarchical harmonic filters for learning tracts from diffusion mri,'' {\em
  arXiv preprint arXiv:1807.01068}, 2018.

\bibitem{maier2016tractography}
K.~H. Maier-Hein, P.~Neher, J.-C. Houde, M.-A. C{\^o}t{\'e}, E.~Garyfallidis,
  J.~Zhong, M.~Chamberland, F.-C. Yeh, Y.-C. Lin, Q.~Ji, {\em et~al.},
  ``Tractography-based connectomes are dominated by false-positive
  connections,'' {\em BioRxiv}, p.~084137, 2016.

\bibitem{rheault2020tractostorm}
F.~Rheault, A.~De~Benedictis, A.~Daducci, C.~Maffei, C.~M. Tax, D.~Romascano,
  E.~Caverzasi, F.~C. Morency, F.~Corrivetti, F.~Pestilli, {\em et~al.},
  ``Tractostorm: The what, why, and how of tractography dissection
  reproducibility,'' {\em Human brain mapping}, vol.~41, no.~7, pp.~1859--1874,
  2020.

\bibitem{schilling2021tractography}
K.~G. Schilling {\em et~al.}, ``Tractography dissection variability: What
  happens when 42 groups dissect 14 white matter bundles on the same
  dataset?,'' {\em NeuroImage}, vol.~243, p.~118502, 2021.

\bibitem{wilkins2012development}
B.~Wilkins, N.~Lee, and M.~Singh, ``Development and evaluation of a simulated
  fibercup phantom,'' in {\em International Symposium on Magnetic Resonance in
  Medicine (ISMRM’12)}, p.~1938, 2012.

\bibitem{neher2014fiberfox}
P.~F. Neher, F.~B. Laun, B.~Stieltjes, and K.~H. Maier-Hein, ``Fiberfox:
  facilitating the creation of realistic white matter software phantoms,'' {\em
  Magnetic resonance in medicine}, vol.~72, no.~5, pp.~1460--1470, 2014.

\bibitem{poupon2010diffusion}
C.~Poupon, L.~Laribiere, G.~Tournier, J.~Bernard, D.~Fournier, P.~Fillard,
  M.~Descoteaux, and J.-F. Mangin, ``A diffusion hardware phantom looking like
  a coronal brain slice,'' in {\em Proceedings of the international society for
  magnetic resonance in medicine}, vol.~18, p.~581, 2010.

\bibitem{fillard2011quantitative}
P.~Fillard, M.~Descoteaux, A.~Goh, S.~Gouttard, B.~Jeurissen, J.~Malcolm,
  A.~Ramirez-Manzanares, M.~Reisert, K.~Sakaie, F.~Tensaouti, {\em et~al.},
  ``Quantitative evaluation of 10 tractography algorithms on a realistic
  diffusion mr phantom,'' {\em Neuroimage}, vol.~56, no.~1, pp.~220--234, 2011.

\bibitem{poulin2022tractoinferno}
P.~Poulin, G.~Theaud, F.~Rheault, E.~St-Onge, A.~Bore, E.~Renauld,
  L.~de~Beaumont, S.~Guay, P.-M. Jodoin, and M.~Descoteaux, ``Tractoinferno-a
  large-scale, open-source, multi-site database for machine learning dmri
  tractography,'' {\em Scientific Data}, vol.~9, no.~1, p.~725, 2022.

\bibitem{astolfi2020tractogram}
P.~Astolfi, R.~Verhagen, L.~Petit, E.~Olivetti, J.~Masci, D.~Boscaini, and
  P.~Avesani, ``Tractogram filtering of anatomically non-plausible fibers with
  geometric deep learning,'' in {\em Medical Image Computing and Computer
  Assisted Intervention--MICCAI 2020: 23rd International Conference, Lima,
  Peru, October 4--8, 2020, Proceedings, Part VII 23}, pp.~291--301, Springer,
  2020.

\bibitem{petit2019half}
L.~Petit, F.~Rheault, M.~Descoteaux, and N.~Tzourio-Mazoyer, ``Half of the
  streamlines built in a whole human brain tractogram is anatomically
  uninterpretable,'' {\em Proceedings OHBM, p. Th785}, 2019.

\bibitem{ugurlu2019supervised}
D.~Ugurlu, Z.~Firat, U.~Ture, and G.~Unal, ``Supervised classification of white
  matter fibers based on neighborhood fiber orientation distributions using an
  ensemble of neural networks,'' in {\em Computational Diffusion MRI:
  International MICCAI Workshop, Granada, Spain, September 2018 22},
  pp.~143--154, Springer, 2019.

\bibitem{legarreta2021filtering}
J.~H. Legarreta, L.~Petit, F.~Rheault, G.~Theaud, C.~Lemaire, M.~Descoteaux,
  and P.-M. Jodoin, ``Filtering in tractography using autoencoders (finta),''
  {\em Medical Image Analysis}, vol.~72, p.~102126, 2021.

\bibitem{legarreta2023generative}
J.~H. Legarreta, L.~Petit, P.-M. Jodoin, and M.~Descoteaux, ``Generative
  sampling in bundle tractography using autoencoders (gesta),'' {\em Medical
  Image Analysis}, vol.~85, p.~102761, 2023.

\bibitem{bullock2022taxonomy}
D.~N. Bullock, E.~A. Hayday, M.~D. Grier, W.~Tang, F.~Pestilli, and S.~R.
  Heilbronner, ``A taxonomy of the brain’s white matter: twenty-one major
  tracts for the 21st century,'' {\em Cerebral Cortex}, vol.~32, no.~20,
  pp.~4524--4548, 2022.

\bibitem{wakana2004fiber}
S.~Wakana, H.~Jiang, L.~M. Nagae-Poetscher, P.~C. Van~Zijl, and S.~Mori,
  ``Fiber tract--based atlas of human white matter anatomy,'' {\em Radiology},
  vol.~230, no.~1, pp.~77--87, 2004.

\bibitem{wycoco2013white}
V.~Wycoco, M.~Shroff, S.~Sudhakar, and W.~Lee, ``White matter anatomy: what the
  radiologist needs to know,'' {\em Neuroimaging Clinics}, vol.~23, no.~2,
  pp.~197--216, 2013.

\bibitem{kamada2005combined}
K.~Kamada, T.~Todo, Y.~Masutani, S.~Aoki, K.~Ino, T.~Takano, T.~Kirino,
  N.~Kawahara, and A.~Morita, ``Combined use of tractography-integrated
  functional neuronavigation and direct fiber stimulation,'' {\em Journal of
  neurosurgery}, vol.~102, no.~4, pp.~664--672, 2005.

\bibitem{glasser2008dti}
M.~F. Glasser and J.~K. Rilling, ``Dti tractography of the human brain's
  language pathways,'' {\em Cerebral cortex}, vol.~18, no.~11, pp.~2471--2482,
  2008.

\bibitem{bubb2018cingulum}
E.~J. Bubb, C.~Metzler-Baddeley, and J.~P. Aggleton, ``The cingulum bundle:
  anatomy, function, and dysfunction,'' {\em Neuroscience \& Biobehavioral
  Reviews}, vol.~92, pp.~104--127, 2018.

\bibitem{zhuang2010white}
L.~Zhuang, W.~Wen, W.~Zhu, J.~Trollor, N.~Kochan, J.~Crawford, S.~Reppermund,
  H.~Brodaty, and P.~Sachdev, ``White matter integrity in mild cognitive
  impairment: a tract-based spatial statistics study,'' {\em Neuroimage},
  vol.~53, no.~1, pp.~16--25, 2010.

\bibitem{wakana2007reproducibility}
S.~Wakana, A.~Caprihan, M.~M. Panzenboeck, J.~H. Fallon, M.~Perry, R.~L.
  Gollub, K.~Hua, J.~Zhang, H.~Jiang, P.~Dubey, {\em et~al.}, ``Reproducibility
  of quantitative tractography methods applied to cerebral white matter,'' {\em
  Neuroimage}, vol.~36, no.~3, pp.~630--644, 2007.

\bibitem{suarez2012automated}
R.~O. Suarez, O.~Commowick, S.~P. Prabhu, and S.~K. Warfield, ``Automated
  delineation of white matter fiber tracts with a multiple region-of-interest
  approach,'' {\em Neuroimage}, vol.~59, no.~4, pp.~3690--3700, 2012.

\bibitem{yendiki2011automated}
A.~Yendiki {\em et~al.}, ``Automated probabilistic reconstruction of
  white-matter pathways in health and disease using an atlas of the underlying
  anatomy,'' {\em Frontiers in neuroinformatics}, vol.~5, p.~23, 2011.

\bibitem{warrington2020xtract}
S.~Warrington, K.~L. Bryant, A.~A. Khrapitchev, J.~Sallet,
  M.~Charquero-Ballester, G.~Douaud, S.~Jbabdi, R.~B. Mars, and S.~N.
  Sotiropoulos, ``Xtract-standardised protocols for automated tractography in
  the human and macaque brain,'' {\em Neuroimage}, vol.~217, p.~116923, 2020.

\bibitem{zhang2020deep}
F.~Zhang, S.~C. Karayumak, N.~Hoffmann, Y.~Rathi, A.~J. Golby, and L.~J.
  O’Donnell, ``Deep white matter analysis (deepwma): Fast and consistent
  tractography segmentation,'' {\em Medical Image Analysis}, vol.~65,
  p.~101761, 2020.

\bibitem{gupta2017brainsegnet}
T.~Gupta, S.~M. Patil, M.~Tailor, D.~Thapar, and A.~Nigam, ``Brainsegnet: A
  segmentation network for human brain fiber tractography data into
  anatomically meaningful clusters,'' {\em arXiv preprint arXiv:1710.05158},
  2017.

\bibitem{gupta2018fibernet}
V.~Gupta, S.~I. Thomopoulos, C.~K. Corbin, F.~Rashid, and P.~M. Thompson,
  ``Fibernet 2.0: an automatic neural network based tool for clustering white
  matter fibers in the brain,'' in {\em 2018 IEEE 15th International Symposium
  on Biomedical Imaging (ISBI 2018)}, pp.~708--711, IEEE, 2018.

\bibitem{xu2019objective}
H.~Xu, M.~Dong, M.-H. Lee, N.~O’Hara, E.~Asano, and J.-W. Jeong, ``Objective
  detection of eloquent axonal pathways to minimize postoperative deficits in
  pediatric epilepsy surgery using diffusion tractography and convolutional
  neural networks,'' {\em IEEE transactions on medical imaging}, vol.~38,
  no.~8, pp.~1910--1922, 2019.

\bibitem{o2007automatic}
L.~J. O'Donnell and C.-F. Westin, ``Automatic tractography segmentation using a
  high-dimensional white matter atlas,'' {\em IEEE transactions on medical
  imaging}, vol.~26, no.~11, pp.~1562--1575, 2007.

\bibitem{dayan2018unsupervised}
M.~Dayan, V.-M. Katsageorgiou, L.~Dodero, V.~Murino, and D.~Sona,
  ``Unsupervised detection of white matter fiber bundles with stochastic neural
  networks,'' in {\em 2018 25th IEEE International Conference on Image
  Processing (ICIP)}, pp.~3513--3517, IEEE, 2018.

\bibitem{jha2019fs2net}
R.~R. Jha, S.~Patil, A.~Nigam, and A.~Bhavsar, ``Fs2net: fiber structural
  similarity network (fs2net) for rotation invariant brain tractography
  segmentation using stacked lstm based siamese network,'' in {\em Computer
  Analysis of Images and Patterns: 18th International Conference, CAIP 2019,
  Salerno, Italy, September 3--5, 2019, Proceedings, Part II 18}, pp.~459--469,
  Springer, 2019.

\bibitem{chen2023deepfiber}
Y.~Chen, C.~Zhang, T.~Xue, Y.~Song, N.~Makris, Y.~Rathi, W.~Cai, F.~Zhang, and
  L.~J. O'Donnell, ``Deep fiber clustering: Anatomically informed fiber
  clustering with self-supervised deep learning for fast and effective
  tractography parcellation,'' {\em NeuroImage}, vol.~273, p.~120086, 2023.

\bibitem{xue2023superficial}
T.~Xue, F.~Zhang, C.~Zhang, Y.~Chen, Y.~Song, A.~J. Golby, N.~Makris, Y.~Rathi,
  W.~Cai, and L.~J. O’Donnell, ``Superficial white matter analysis: An
  efficient point-cloud-based deep learning framework with supervised
  contrastive learning for consistent tractography parcellation across
  populations and dmri acquisitions,'' {\em Medical Image Analysis}, vol.~85,
  p.~102759, 2023.

\bibitem{siless2018anatomicuts}
V.~Siless {\em et~al.}, ``Anatomicuts: Hierarchical clustering of tractography
  streamlines based on anatomical similarity,'' {\em NeuroImage}, vol.~166,
  pp.~32--45, 2018.

\bibitem{xue2023tractcloud}
T.~Xue, Y.~Chen, C.~Zhang, A.~J. Golby, N.~Makris, Y.~Rathi, W.~Cai, F.~Zhang,
  and L.~J. O’Donnell, ``Tractcloud: Registration-free tractography
  parcellation with a novel local-global streamline point cloud
  representation,'' in {\em International Conference on Medical Image Computing
  and Computer-Assisted Intervention}, pp.~409--419, Springer, 2023.

\bibitem{zollei2019tracts}
L.~Z{\"o}llei, C.~Jaimes, E.~Saliba, P.~E. Grant, and A.~Yendiki, ``Tracts
  constrained by underlying infant anatomy (traculina): An automated
  probabilistic tractography tool with anatomical priors for use in the newborn
  brain,'' {\em Neuroimage}, vol.~199, pp.~1--17, 2019.

\bibitem{garyfallidis2018recognition}
E.~Garyfallidis {\em et~al.}, ``Recognition of white matter bundles using local
  and global streamline-based registration and clustering,'' {\em NeuroImage},
  vol.~170, pp.~283--295, 2018.

\bibitem{labra2017fast}
N.~Labra {\em et~al.}, ``Fast automatic segmentation of white matter
  streamlines based on a multi-subject bundle atlas,'' {\em Neuroinformatics},
  vol.~15, no.~1, pp.~71--86, 2017.

\bibitem{clayden2007probabilistic}
J.~D. Clayden, A.~J. Storkey, and M.~E. Bastin, ``A probabilistic model-based
  approach to consistent white matter tract segmentation,'' {\em IEEE
  transactions on medical imaging}, vol.~26, no.~11, pp.~1555--1561, 2007.

\bibitem{siless2020registration}
V.~Siless {\em et~al.}, ``Registration-free analysis of diffusion mri
  tractography data across subjects through the human lifespan,'' {\em
  NeuroImage}, vol.~214, p.~116703, 2020.

\bibitem{wassermann2010unsupervised}
D.~Wassermann, L.~Bloy, E.~Kanterakis, R.~Verma, and R.~Deriche, ``Unsupervised
  white matter fiber clustering and tract probability map generation:
  Applications of a gaussian process framework for white matter fibers,'' {\em
  NeuroImage}, vol.~51, no.~1, pp.~228--241, 2010.

\bibitem{garyfallidis2012quickbundles}
E.~Garyfallidis, M.~Brett, M.~M. Correia, G.~B. Williams, and I.~Nimmo-Smith,
  ``Quickbundles, a method for tractography simplification,'' {\em Frontiers in
  neuroscience}, vol.~6, p.~175, 2012.

\bibitem{brun2004clustering}
A.~Brun, H.~Knutsson, H.-J. Park, M.~E. Shenton, and C.-F. Westin, ``Clustering
  fiber traces using normalized cuts,'' in {\em Medical Image Computing and
  Computer-Assisted Intervention--MICCAI 2004: 7th International Conference,
  Saint-Malo, France, September 26-29, 2004. Proceedings, Part I 7},
  pp.~368--375, Springer, 2004.

\bibitem{wang2019fast}
J.~Wang and Y.~Shi, ``A fast fiber k-nearest-neighbor algorithm with
  application to group-wise white matter topography analysis,'' in {\em
  International Conference on Information Processing in Medical Imaging},
  pp.~332--344, Springer, 2019.

\bibitem{wang2018gife}
J.~Wang and Y.~Shi, ``Gife: Efficient and robust group-wise isometric fiber
  embedding,'' in {\em Connectomics in NeuroImaging: Second International
  Workshop, CNI 2018, Held in Conjunction with MICCAI 2018, Granada, Spain,
  September 20, 2018, Proceedings 2}, pp.~20--28, Springer, 2018.

\bibitem{vazquez2020ffclust}
A.~V{\'a}zquez, N.~L{\'o}pez-L{\'o}pez, A.~S{\'a}nchez, J.~Houenou, C.~Poupon,
  J.-F. Mangin, C.~Hern{\'a}ndez, and P.~Guevara, ``Ffclust: Fast fiber
  clustering for large tractography datasets for a detailed study of brain
  connectivity,'' {\em NeuroImage}, vol.~220, p.~117070, 2020.

\bibitem{maddah2008mathematical}
M.~Maddah, L.~Zollei, W.~E.~L. Grimson, C.-F. Westin, and W.~M. Wells, ``A
  mathematical framework for incorporating anatomical knowledge in dt-mri
  analysis,'' in {\em 2008 5th IEEE International Symposium on Biomedical
  Imaging: From Nano to Macro}, pp.~105--108, IEEE, 2008.

\bibitem{li2010hybrid}
H.~Li, Z.~Xue, L.~Guo, T.~Liu, J.~Hunter, and S.~T. Wong, ``A hybrid approach
  to automatic clustering of white matter fibers,'' {\em NeuroImage}, vol.~49,
  no.~2, pp.~1249--1258, 2010.

\bibitem{wassermann2016white}
D.~Wassermann, N.~Makris, Y.~Rathi, M.~Shenton, R.~Kikinis, M.~Kubicki, and
  C.-F. Westin, ``The white matter query language: a novel approach for
  describing human white matter anatomy,'' {\em Brain Structure and Function},
  vol.~221, pp.~4705--4721, 2016.

\bibitem{zhang2017comparison}
F.~Zhang, I.~Norton, W.~Cai, Y.~Song, W.~M. Wells, and L.~J. O'Donnell,
  ``Comparison between two white matter segmentation strategies: an
  investigation into white matter segmentation consistency,'' in {\em 2017 IEEE
  14th International Symposium on Biomedical Imaging (ISBI 2017)},
  pp.~796--799, IEEE, 2017.

\bibitem{tuncc2014automated}
B.~Tun{\c{c}}, W.~A. Parker, M.~Ingalhalikar, and R.~Verma, ``Automated tract
  extraction via atlas based adaptive clustering,'' {\em Neuroimage}, vol.~102,
  pp.~596--607, 2014.

\bibitem{tuncc2016individualized}
B.~Tun{\c{c}}, M.~Ingalhalikar, D.~Parker, J.~Lecoeur, N.~Singh, R.~L. Wolf,
  L.~Macyszyn, S.~Brem, and R.~Verma, ``Individualized map of white matter
  pathways: connectivity-based paradigm for neurosurgical planning,'' {\em
  Neurosurgery}, vol.~79, no.~4, p.~568, 2016.

\bibitem{sydnor2018comparison}
V.~J. Sydnor, A.~M. Rivas-Grajales, A.~E. Lyall, F.~Zhang, S.~Bouix,
  S.~Karmacharya, M.~E. Shenton, C.-F. Westin, N.~Makris, D.~Wassermann, {\em
  et~al.}, ``A comparison of three fiber tract delineation methods and their
  impact on white matter analysis,'' {\em Neuroimage}, vol.~178, pp.~318--331,
  2018.

\bibitem{catani2008diffusion}
M.~Catani and M.~T. De~Schotten, ``A diffusion tensor imaging tractography
  atlas for virtual in vivo dissections,'' {\em cortex}, vol.~44, no.~8,
  pp.~1105--1132, 2008.

\bibitem{o2013fiber}
L.~J. O'Donnell, A.~J. Golby, and C.-F. Westin, ``Fiber clustering versus the
  parcellation-based connectome,'' {\em NeuroImage}, vol.~80, pp.~283--289,
  2013.

\bibitem{chekir2014hybrid}
A.~Chekir, M.~Descoteaux, E.~Garyfallidis, M.-A. C{\^o}t{\'e}, and F.~O.
  Boumghar, ``A hybrid approach for optimal automatic segmentation of white
  matter tracts in hardi,'' in {\em 2014 IEEE Conference on Biomedical
  Engineering and Sciences (IECBES)}, pp.~177--180, IEEE, 2014.

\bibitem{xu2013gray}
Q.~Xu, A.~W. Anderson, J.~C. Gore, and Z.~Ding, ``Gray matter parcellation
  constrained full brain fiber bundling with diffusion tensor imaging,'' {\em
  Medical Physics}, vol.~40, no.~7, p.~072301, 2013.

\bibitem{lam2018trafic}
P.~D.~N. Lam, G.~Belhomme, J.~Ferrall, B.~Patterson, M.~Styner, and J.~C.
  Prieto, ``Trafic: fiber tract classification using deep learning,'' in {\em
  Medical Imaging 2018: Image Processing}, vol.~10574, pp.~257--265, SPIE,
  2018.

\bibitem{gupta2017fibernet}
V.~Gupta, S.~I. Thomopoulos, F.~M. Rashid, and P.~M. Thompson, ``Fibernet: An
  ensemble deep learning framework for clustering white matter fibers,'' in
  {\em Medical Image Computing and Computer Assisted Intervention- MICCAI 2017:
  20th International Conference, Quebec City, QC, Canada, September 11-13,
  2017, Proceedings, Part I 20}, pp.~548--555, Springer, 2017.

\bibitem{hua2008tract}
K.~Hua, J.~Zhang, S.~Wakana, H.~Jiang, X.~Li, D.~S. Reich, P.~A. Calabresi,
  J.~J. Pekar, P.~C. van Zijl, and S.~Mori, ``Tract probability maps in
  stereotaxic spaces: analyses of white matter anatomy and tract-specific
  quantification,'' {\em Neuroimage}, vol.~39, no.~1, pp.~336--347, 2008.

\bibitem{liu2022volumetric}
W.~Liu, Q.~Lu, Z.~Zhuo, Y.~Li, Y.~Duan, P.~Yu, L.~Qu, C.~Ye, and Y.~Liu,
  ``Volumetric segmentation of white matter tracts with label embedding,'' {\em
  Neuroimage}, vol.~250, p.~118934, 2022.

\bibitem{lu2021volumetric}
Q.~Lu, Y.~Li, and C.~Ye, ``Volumetric white matter tract segmentation with
  nested self-supervised learning using sequential pretext tasks,'' {\em
  Medical Image Analysis}, vol.~72, p.~102094, 2021.

\bibitem{lucena2022informative}
O.~Lucena, P.~Borges, J.~Cardoso, K.~Ashkan, R.~Sparks, and S.~Ourselin,
  ``Informative and reliable tract segmentation for preoperative planning,''
  {\em Frontiers in Radiology}, vol.~2, p.~866974, 2022.

\bibitem{lucena2023assessing}
O.~Lucena, J.~P. Lavrador, H.~Irzan, C.~Semedo, P.~Borges, F.~Vergani,
  A.~Granados, R.~Sparks, K.~Ashkan, and S.~Ourselin, ``Assessing informative
  tract segmentation and ntms for pre-operative planning,'' {\em Journal of
  Neuroscience Methods}, vol.~396, p.~109933, 2023.

\bibitem{kebiri2023direct}
H.~Kebiri, A.~Gholipour, M.~B. Cuadra, and D.~Karimi, ``Direct segmentation of
  brain white matter tracts in diffusion mri,'' {\em arXiv preprint
  arXiv:2307.02223}, 2023.

\bibitem{li2020neuro4neuro}
B.~Li, M.~De~Groot, R.~M. Steketee, R.~Meijboom, M.~Smits, M.~W. Vernooij,
  M.~A. Ikram, J.~Liu, W.~J. Niessen, and E.~E. Bron, ``Neuro4neuro: A neural
  network approach for neural tract segmentation using large-scale
  population-based diffusion imaging,'' {\em Neuroimage}, vol.~218, p.~116993,
  2020.

\bibitem{xu2023registration}
H.~Xu, T.~Xue, D.~Liu, F.~Zhang, C.-F. Westin, R.~Kikinis, L.~J. O’Donnell,
  and W.~Cai, ``A registration-and uncertainty-based framework for white matter
  tract segmentation with only one annotated subject,'' in {\em 2023 IEEE 20th
  International Symposium on Biomedical Imaging (ISBI)}, pp.~1--5, IEEE, 2023.

\bibitem{bazin2011direct}
P.-L. Bazin {\em et~al.}, ``Direct segmentation of the major white matter
  tracts in diffusion tensor images,'' {\em Neuroimage}, vol.~58, no.~2,
  pp.~458--468, 2011.

\bibitem{ratnarajah2014multi}
N.~Ratnarajah and A.~Qiu, ``Multi-label segmentation of white matter
  structures: application to neonatal brains,'' {\em NeuroImage}, vol.~102,
  pp.~913--922, 2014.

\bibitem{lenglet2006dti}
C.~Lenglet, M.~Rousson, and R.~Deriche, ``Dti segmentation by statistical
  surface evolution,'' {\em IEEE Transactions on Medical Imaging}, vol.~25,
  no.~6, pp.~685--700, 2006.

\bibitem{jonasson2005white}
L.~Jonasson, X.~Bresson, P.~Hagmann, O.~Cuisenaire, R.~Meuli, and J.-P. Thiran,
  ``White matter fiber tract segmentation in dt-mri using geometric flows,''
  {\em Medical Image Analysis}, vol.~9, no.~3, pp.~223--236, 2005.

\bibitem{guo2008geometric}
W.~Guo, Y.~Chen, and Q.~Zeng, ``A geometric flow-based approach for diffusion
  tensor image segmentation,'' {\em Philosophical Transactions of the Royal
  Society A: Mathematical, Physical and Engineering Sciences}, vol.~366,
  no.~1874, pp.~2279--2292, 2008.

\bibitem{eckstein2009active}
I.~Eckstein {\em et~al.}, ``Active fibers: Matching deformable tract templates
  to diffusion tensor images,'' {\em Neuroimage}, vol.~47, pp.~T82--T89, 2009.

\bibitem{dong2019multimodality}
X.~Dong {\em et~al.}, ``Multimodality white matter tract segmentation using
  cnn,'' in {\em Proceedings of the ACM Turing Celebration Conference-China},
  pp.~1--8, 2019.

\bibitem{wasserthal2019combined}
J.~Wasserthal, P.~F. Neher, D.~Hirjak, and K.~H. Maier-Hein, ``Combined tract
  segmentation and orientation mapping for bundle-specific tractography,'' {\em
  Medical image analysis}, vol.~58, p.~101559, 2019.

\bibitem{lu2021knowledge}
Q.~Lu and C.~Ye, ``Knowledge transfer for few-shot segmentation of novel white
  matter tracts,'' in {\em Information Processing in Medical Imaging: 27th
  International Conference, IPMI 2021, Virtual Event, June 28--June 30, 2021,
  Proceedings 27}, pp.~216--227, Springer, 2021.

\bibitem{mukherjee2020deep}
S.~Mukherjee, N.~Paquette, M.~D. Nelson, Y.~Wang, J.~Wallace, A.~Panigrahy, and
  N.~Lepore, ``Deep-learning based tractography for neonates,'' in {\em 16th
  International Symposium on Medical Information Processing and Analysis},
  vol.~11583, pp.~85--91, SPIE, 2020.

\bibitem{guevara2012automatic}
P.~Guevara, D.~Duclap, C.~Poupon, L.~Marrakchi-Kacem, P.~Fillard, D.~Le~Bihan,
  M.~Leboyer, J.~Houenou, and J.-F. Mangin, ``Automatic fiber bundle
  segmentation in massive tractography datasets using a multi-subject bundle
  atlas,'' {\em Neuroimage}, vol.~61, no.~4, pp.~1083--1099, 2012.

\bibitem{roman2017clustering}
C.~Rom{\'a}n, M.~Guevara, R.~Valenzuela, M.~Figueroa, J.~Houenou, D.~Duclap,
  C.~Poupon, J.-F. Mangin, and P.~Guevara, ``Clustering of whole-brain white
  matter short association bundles using hardi data,'' {\em Frontiers in
  neuroinformatics}, vol.~11, p.~73, 2017.

\bibitem{vazquez2020automatic}
A.~V{\'a}zquez, N.~L{\'o}pez-L{\'o}pez, J.~Houenou, C.~Poupon, J.-F. Mangin,
  S.~Ladra, and P.~Guevara, ``Automatic group-wise whole-brain short
  association fiber bundle labeling based on clustering and cortical surface
  information,'' {\em BioMedical Engineering OnLine}, vol.~19, no.~1,
  pp.~1--24, 2020.

\bibitem{li2025insights}
B.~Li, {\v{Z}}.~Krsnik, L.~Pierotich, I.~Kostovi{\'c}, S.~K. Warfield, P.~E.
  Grant, and D.~Karimi, ``Insights into temporal and spatial dynamics of short
  association fiber formation in the human fetal brain,'' {\em bioRxiv},
  pp.~2025--10, 2025.

\bibitem{guevara2020superficial}
M.~Guevara, P.~Guevara, C.~Rom{\'a}n, and J.-F. Mangin, ``Superficial white
  matter: A review on the dmri analysis methods and applications,'' {\em
  Neuroimage}, vol.~212, p.~116673, 2020.

\bibitem{smith2014cross}
S.~M. Smith, G.~Kindlmann, and S.~Jbabdi, ``Cross-subject comparison of local
  diffusion mri parameters,'' in {\em Diffusion MRI}, pp.~209--239, Elsevier,
  2014.

\bibitem{pini2016brain}
L.~Pini {\em et~al.}, ``Brain atrophy in alzheimer’s disease and aging,''
  {\em Ageing research reviews}, vol.~30, pp.~25--48, 2016.

\bibitem{sexton2011meta}
C.~E. Sexton {\em et~al.}, ``A meta-analysis of diffusion tensor imaging in
  mild cognitive impairment and alzheimer's disease,'' {\em Neurobiology of
  aging}, vol.~32, no.~12, pp.~2322--e5, 2011.

\bibitem{ashburner2000voxel}
J.~Ashburner and K.~J. Friston, ``Voxel-based morphometry—the methods,'' {\em
  Neuroimage}, vol.~11, no.~6, pp.~805--821, 2000.

\bibitem{pagani2005method}
E.~Pagani {\em et~al.}, ``A method for obtaining tract-specific diffusion
  tensor mri measurements in the presence of disease: application to patients
  with clinically isolated syndromes suggestive of multiple sclerosis,'' {\em
  Neuroimage}, vol.~26, no.~1, pp.~258--265, 2005.

\bibitem{madhyastha2014longitudinal}
T.~Madhyastha {\em et~al.}, ``Longitudinal reliability of tract-based spatial
  statistics in diffusion tensor imaging,'' {\em Human brain mapping}, vol.~35,
  no.~9, pp.~4544--4555, 2014.

\bibitem{edden2011spatial}
R.~A. Edden and D.~K. Jones, ``Spatial and orientational heterogeneity in the
  statistical sensitivity of skeleton-based analyses of diffusion tensor mr
  imaging data,'' {\em Journal of neuroscience methods}, vol.~201, no.~1,
  pp.~213--219, 2011.

\bibitem{raffelt2011symmetric}
D.~Raffelt, J.-D. Tournier, J.~Fripp, S.~Crozier, A.~Connelly, and O.~Salvado,
  ``Symmetric diffeomorphic registration of fibre orientation distributions,''
  {\em Neuroimage}, vol.~56, no.~3, pp.~1171--1180, 2011.

\bibitem{zhang2006deformable}
H.~Zhang, P.~A. Yushkevich, D.~C. Alexander, and J.~C. Gee, ``Deformable
  registration of diffusion tensor mr images with explicit orientation
  optimization,'' {\em Medical image analysis}, vol.~10, no.~5, pp.~764--785,
  2006.

\bibitem{grigorescu2020diffusion}
I.~Grigorescu, A.~Uus, D.~Christiaens, L.~Cordero-Grande, J.~Hutter, A.~D.
  Edwards, J.~V. Hajnal, M.~Modat, and M.~Deprez, ``Diffusion tensor driven
  image registration: a deep learning approach,'' in {\em International
  Workshop on Biomedical Image Registration}, pp.~131--140, Springer, 2020.

\bibitem{zhang2021deep}
F.~Zhang, W.~M. Wells, and L.~J. O’Donnell, ``Deep diffusion mri registration
  (ddmreg): a deep learning method for diffusion mri registration,'' {\em IEEE
  Transactions on Medical Imaging}, vol.~41, no.~6, pp.~1454--1467, 2021.

\bibitem{bouza2023geometric}
J.~J. Bouza, C.-H. Yang, and B.~C. Vemuri, ``Geometric deep learning for
  unsupervised registration of diffusion magnetic resonance images,'' in {\em
  International Conference on Information Processing in Medical Imaging},
  pp.~563--575, Springer, 2023.

\bibitem{grigorescu2022attention}
I.~Grigorescu, A.~Uus, D.~Christiaens, L.~Cordero-Grande, J.~Hutter,
  D.~Batalle, A.~David~Edwards, J.~V. Hajnal, M.~Modat, and M.~Deprez,
  ``Attention-driven multi-channel deformable registration of structural and
  microstructural neonatal data,'' in {\em International Workshop on Preterm,
  Perinatal and Paediatric Image Analysis}, pp.~71--81, Springer, 2022.

\bibitem{balakrishnan2019voxelmorph}
G.~Balakrishnan, A.~Zhao, M.~R. Sabuncu, J.~Guttag, and A.~V. Dalca,
  ``Voxelmorph: a learning framework for deformable medical image
  registration,'' {\em IEEE transactions on medical imaging}, vol.~38, no.~8,
  pp.~1788--1800, 2019.

\bibitem{alexander2001spatial}
D.~C. Alexander, C.~Pierpaoli, P.~J. Basser, and J.~C. Gee, ``Spatial
  transformations of diffusion tensor magnetic resonance images,'' {\em IEEE
  transactions on medical imaging}, vol.~20, no.~11, pp.~1131--1139, 2001.

\bibitem{li2025fetdtialign}
B.~Li, Q.~Zeng, S.~K. Warfield, and D.~Karimi, ``Fetdtialign: A deep learning
  framework for affine and deformable registration of fetal brain dmri,'' {\em
  NeuroImage}, vol.~311, p.~121190, 2025.

\bibitem{prasad2014automatic}
G.~Prasad, S.~H. Joshi, N.~Jahanshad, J.~Villalon-Reina, I.~Aganj, C.~Lenglet,
  G.~Sapiro, K.~L. McMahon, G.~I. de~Zubicaray, N.~G. Martin, {\em et~al.},
  ``Automatic clustering and population analysis of white matter tracts using
  maximum density paths,'' {\em Neuroimage}, vol.~97, pp.~284--295, 2014.

\bibitem{jin2014automatic}
Y.~Jin, Y.~Shi, L.~Zhan, B.~A. Gutman, G.~I. de~Zubicaray, K.~L. McMahon, M.~J.
  Wright, A.~W. Toga, and P.~M. Thompson, ``Automatic clustering of white
  matter fibers in brain diffusion mri with an application to genetics,'' {\em
  Neuroimage}, vol.~100, pp.~75--90, 2014.

\bibitem{zhang2018whole}
F.~Zhang, P.~Savadjiev, W.~Cai, Y.~Song, Y.~Rathi, B.~Tun{\c{c}}, D.~Parker,
  T.~Kapur, R.~T. Schultz, N.~Makris, {\em et~al.}, ``Whole brain white matter
  connectivity analysis using machine learning: an application to autism,''
  {\em Neuroimage}, vol.~172, pp.~826--837, 2018.

\bibitem{chen2024tractgeonet}
Y.~Chen, L.~R. Zekelman, C.~Zhang, T.~Xue, Y.~Song, N.~Makris, Y.~Rathi, A.~J.
  Golby, W.~Cai, F.~Zhang, {\em et~al.}, ``Tractgeonet: A geometric deep
  learning framework for pointwise analysis of tract microstructure to predict
  language assessment performance,'' {\em Medical Image Analysis}, vol.~94,
  p.~103120, 2024.

\bibitem{li2021learning}
B.~Li, W.~J. Niessen, S.~Klein, M.~A. Ikram, M.~W. Vernooij, and E.~E. Bron,
  ``Learning unbiased group-wise registration (lugr) and joint segmentation:
  evaluation on longitudinal diffusion mri,'' in {\em Medical Imaging 2021:
  Image Processing}, vol.~11596, pp.~136--144, SPIE, 2021.

\bibitem{karimi2023tbss++}
D.~Karimi, H.~Kebiri, and A.~Gholipour, ``Tbss++: A novel computational method
  for tract-based spatial statistics,'' {\em bioRxiv}, pp.~2023--07, 2023.

\bibitem{chamberland2021detecting}
M.~Chamberland, S.~Genc, C.~M. Tax, D.~Shastin, K.~Koller, E.~P. Raven,
  A.~Cunningham, J.~Doherty, M.~B. van~den Bree, G.~D. Parker, {\em et~al.},
  ``Detecting microstructural deviations in individuals with deep diffusion mri
  tractometry,'' {\em Nature computational science}, vol.~1, no.~9,
  pp.~598--606, 2021.

\bibitem{wen2013brain}
Y.~Wen, L.~He, K.~M. von Deneen, and Y.~Lu, ``Brain tissue classification based
  on dti using an improved fuzzy c-means algorithm with spatial constraints,''
  {\em Magnetic Resonance Imaging}, vol.~31, no.~9, pp.~1623--1630, 2013.

\bibitem{yap2015brain}
P.-T. Yap, Y.~Zhang, and D.~Shen, ``Brain tissue segmentation based on
  diffusion mri using l0 sparse-group representation classification,'' in {\em
  Medical Image Computing and Computer-Assisted Intervention--MICCAI 2015: 18th
  International Conference, Munich, Germany, October 5-9, 2015, Proceedings,
  Part III 18}, pp.~132--139, Springer, 2015.

\bibitem{ciritsis2018automated}
A.~Ciritsis, A.~Boss, and C.~Rossi, ``Automated pixel-wise brain tissue
  segmentation of diffusion-weighted images via machine learning,'' {\em NMR in
  Biomedicine}, vol.~31, no.~7, p.~e3931, 2018.

\bibitem{schnell2009fully}
S.~Schnell, D.~Saur, B.~Kreher, J.~Hennig, H.~Burkhardt, and V.~G. Kiselev,
  ``Fully automated classification of hardi in vivo data using a support vector
  machine,'' {\em NeuroImage}, vol.~46, no.~3, pp.~642--651, 2009.

\bibitem{vasilev2020q}
A.~Vasilev, V.~Golkov, M.~Meissner, I.~Lipp, E.~Sgarlata, V.~Tomassini, D.~K.
  Jones, and D.~Cremers, ``q-space novelty detection with variational
  autoencoders,'' in {\em Computational Diffusion MRI: MICCAI Workshop,
  Shenzhen, China, October 2019}, pp.~113--124, Springer, 2020.

\bibitem{zhang2021deepseg}
F.~Zhang, A.~Breger, K.~I.~K. Cho, L.~Ning, C.-F. Westin, L.~J. O’Donnell,
  and O.~Pasternak, ``Deep learning based segmentation of brain tissue from
  diffusion mri,'' {\em NeuroImage}, vol.~233, p.~117934, 2021.

\bibitem{zhang2023ddparcel}
F.~Zhang, K.~I.~K. Cho, J.~Seitz-Holland, L.~Ning, J.~H. Legarreta, Y.~Rathi,
  C.-F. Westin, L.~J. O’Donnell, and O.~Pasternak, ``Ddparcel: deep learning
  anatomical brain parcellation from diffusion mri,'' {\em IEEE Transactions on
  Medical Imaging}, 2023.

\bibitem{calixto2024anatomically}
C.~Calixto, C.~Jaimes, M.~D. Soldatelli, S.~K. Warfield, A.~Gholipour, and
  D.~Karimi, ``Anatomically constrained tractography of the fetal brain,'' {\em
  NeuroImage}, p.~120723, 2024.

\bibitem{karimi2024detailed}
D.~Karimi, C.~Calixto~Nunez, H.~Snoussi, M.~C. Cortes-Albornoz,
  C.~Velasco-Annis, C.~Rollins, C.~Jaimes, A.~Gholipour, and S.~K. Warfield,
  ``Detailed delineation of the fetal brain in diffusion mri via multi-task
  learning,'' {\em bioRxiv}, pp.~2024--08, 2024.

\bibitem{zhang2015deep}
W.~Zhang, R.~Li, H.~Deng, L.~Wang, W.~Lin, S.~Ji, and D.~Shen, ``Deep
  convolutional neural networks for multi-modality isointense infant brain
  image segmentation,'' {\em NeuroImage}, vol.~108, pp.~214--224, 2015.

\bibitem{reid2018diffusion}
R.~I. Reid, Z.~Nedelska, C.~G. Schwarz, C.~Ward, C.~R. Jack, and A.~D.~N.
  Initiative, ``Diffusion specific segmentation: skull stripping with diffusion
  mri data alone,'' in {\em Computational Diffusion MRI: MICCAI Workshop,
  Qu{\'e}bec, Canada, September 2017}, pp.~67--80, Springer, 2018.

\bibitem{wang2021u}
X.~Wang, X.-H. Li, J.~W. Cho, B.~E. Russ, N.~Rajamani, A.~Omelchenko, L.~Ai,
  A.~Korchmaros, S.~Sawiak, R.~A. Benn, {\em et~al.}, ``U-net model for brain
  extraction: Trained on humans for transfer to non-human primates,'' {\em
  Neuroimage}, vol.~235, p.~118001, 2021.

\bibitem{karimi2020deep}
D.~Karimi, H.~Dou, S.~K. Warfield, and A.~Gholipour, ``Deep learning with noisy
  labels: Exploring techniques and remedies in medical image analysis,'' {\em
  Medical image analysis}, vol.~65, p.~101759, 2020.

\bibitem{faghihpirayesh2023fetal}
R.~Faghihpirayesh, D.~Karimi, D.~Erdo{\u{g}}mu{\c{s}}, and A.~Gholipour,
  ``Fetal-bet: Brain extraction tool for fetal mri,'' {\em arXiv preprint
  arXiv:2310.01523}, 2023.

\bibitem{szczepankiewicz2016link}
F.~Szczepankiewicz, D.~van Westen, E.~Englund, C.-F. Westin, F.~St{\aa}hlberg,
  J.~L{\"a}tt, P.~C. Sundgren, and M.~Nilsson, ``The link between diffusion mri
  and tumor heterogeneity: Mapping cell eccentricity and density by diffusional
  variance decomposition (divide),'' {\em Neuroimage}, vol.~142, pp.~522--532,
  2016.

\bibitem{reynaud2017time}
O.~Reynaud, ``Time-dependent diffusion mri in cancer: tissue modeling and
  applications,'' {\em Frontiers in Physics}, vol.~5, p.~58, 2017.

\bibitem{young2024fibre}
F.~Young, K.~Aquilina, K.~K. Seunarine, L.~Mancini, C.~A. Clark, and J.~D.
  Clayden, ``Fibre orientation atlas guided rapid segmentation of white matter
  tracts,'' tech. rep., Wiley Online Library, 2024.

\bibitem{peretzke2023attractive}
R.~Peretzke, K.~H. Maier-Hein, J.~Bohn, Y.~Kirchhoff, S.~Roy, S.~Oberli-Palma,
  D.~Becker, P.~Lenga, and P.~Neher, ``attractive: semi-automatic white matter
  tract segmentation using active learning,'' in {\em International Conference
  on Medical Image Computing and Computer-Assisted Intervention}, pp.~237--246,
  Springer, 2023.

\bibitem{tallus2023comparison}
J.~Tallus, M.~Mohammadian, T.~Kurki, T.~Roine, J.~P. Posti, and O.~Tenovuo, ``A
  comparison of diffusion tensor imaging tractography and constrained spherical
  deconvolution with automatic segmentation in traumatic brain injury,'' {\em
  NeuroImage: Clinical}, vol.~37, p.~103284, 2023.

\bibitem{arrieta2020explainable}
A.~B. Arrieta, N.~D{\'\i}az-Rodr{\'\i}guez, J.~Del~Ser, A.~Bennetot, S.~Tabik,
  A.~Barbado, S.~Garc{\'\i}a, S.~Gil-L{\'o}pez, D.~Molina, R.~Benjamins, {\em
  et~al.}, ``Explainable artificial intelligence (xai): Concepts, taxonomies,
  opportunities and challenges toward responsible ai,'' {\em Information
  Fusion}, vol.~58, pp.~82--115, 2020.

\bibitem{lin2023cross}
R.~Lin, A.~Gholipour, J.-P. Thiran, D.~Karimi, H.~Kebiri, and M.~B. Cuadra,
  ``Cross-age and cross-site domain shift impacts on deep learning-based white
  matter fiber estimation in newborn and baby brains,'' {\em arXiv preprint
  arXiv:2312.14773}, 2023.

\bibitem{cai2021prequal}
L.~Y. Cai, Q.~Yang, C.~B. Hansen, V.~Nath, K.~Ramadass, G.~W. Johnson, B.~N.
  Conrad, B.~D. Boyd, J.~P. Begnoche, L.~L. Beason-Held, {\em et~al.},
  ``Prequal: An automated pipeline for integrated preprocessing and quality
  assurance of diffusion weighted mri images,'' {\em Magnetic resonance in
  medicine}, vol.~86, no.~1, pp.~456--470, 2021.

\bibitem{cieslak2021qsiprep}
M.~Cieslak, P.~A. Cook, X.~He, F.-C. Yeh, T.~Dhollander, A.~Adebimpe, G.~K.
  Aguirre, D.~S. Bassett, R.~F. Betzel, J.~Bourque, {\em et~al.}, ``Qsiprep: an
  integrative platform for preprocessing and reconstructing diffusion mri
  data,'' {\em Nature methods}, vol.~18, no.~7, pp.~775--778, 2021.

\bibitem{rensonnet2021solving}
G.~Rensonnet, L.~Adam, and B.~Macq, ``Solving inverse problems with deep neural
  networks driven by sparse signal decomposition in a physics-based
  dictionary,'' {\em arXiv preprint arXiv:2107.10657}, 2021.

\bibitem{ruff2021unifying}
L.~Ruff, J.~R. Kauffmann, R.~A. Vandermeulen, G.~Montavon, W.~Samek, M.~Kloft,
  T.~G. Dietterich, and K.-R. M{\"u}ller, ``A unifying review of deep and
  shallow anomaly detection,'' {\em Proceedings of the IEEE}, vol.~109, no.~5,
  pp.~756--795, 2021.

\bibitem{liu2020energy}
W.~Liu, X.~Wang, J.~Owens, and Y.~Li, ``Energy-based out-of-distribution
  detection,'' {\em Advances in neural information processing systems},
  vol.~33, pp.~21464--21475, 2020.

\bibitem{descoteaux2011multiple}
M.~Descoteaux, R.~Deriche, D.~Le~Bihan, J.-F. Mangin, and C.~Poupon, ``Multiple
  q-shell diffusion propagator imaging,'' {\em Medical image analysis},
  vol.~15, no.~4, pp.~603--621, 2011.

\bibitem{cheng2011theoretical}
J.~Cheng, T.~Jiang, and R.~Deriche, ``Theoretical analysis and practical
  insights on eap estimation via a unified hardi framework,'' in {\em MICCAI
  Workshop on Computational Diffusion MRI (CDMRI)}, 2011.

\bibitem{zucchelli2021investigating}
M.~Zucchelli, S.~Deslauriers-Gauthier, and R.~Deriche, ``Investigating the
  effect of dmri signal representation on fully-connected neural networks brain
  tissue microstructure estimation,'' in {\em 2021 IEEE 18th International
  Symposium on Biomedical Imaging (ISBI)}, pp.~725--728, IEEE, 2021.

\bibitem{zucchelli2020computational}
M.~Zucchelli, S.~Deslauriers-Gauthier, and R.~Deriche, ``A computational
  framework for generating rotation invariant features and its application in
  diffusion mri,'' {\em Medical image analysis}, vol.~60, p.~101597, 2020.

\bibitem{schwab2016spatial}
E.~Schwab, R.~Vidal, and N.~Charon, ``Spatial-angular sparse coding for
  hardi,'' in {\em Medical Image Computing and Computer-Assisted
  Intervention-MICCAI 2016: 19th International Conference, Athens, Greece,
  October 17-21, 2016, Proceedings, Part III 19}, pp.~475--483, Springer, 2016.

\bibitem{chen2022prediction}
H.~Chen, Z.~Zhang, M.~Jin, and F.~Wang, ``Prediction of dmri signals with
  neural architecture search,'' {\em Journal of Neuroscience Methods},
  vol.~365, p.~109389, 2022.

\bibitem{zoph2016neural}
B.~Zoph and Q.~V. Le, ``Neural architecture search with reinforcement
  learning,'' {\em arXiv preprint arXiv:1611.01578}, 2016.

\bibitem{esteves2018learning}
C.~Esteves, C.~Allen-Blanchette, A.~Makadia, and K.~Daniilidis, ``Learning so
  (3) equivariant representations with spherical cnns,'' in {\em Proceedings of
  the European Conference on Computer Vision (ECCV)}, pp.~52--68, 2018.

\bibitem{muller2021rotation}
P.~M{\"u}ller, V.~Golkov, V.~Tomassini, and D.~Cremers, ``Rotation-equivariant
  deep learning for diffusion mri,'' {\em arXiv preprint arXiv:2102.06942},
  2021.

\bibitem{su2017learning}
Y.-C. Su and K.~Grauman, ``Learning spherical convolution for fast features
  from 360 imagery,'' {\em Advances in Neural Information Processing Systems},
  vol.~30, 2017.

\bibitem{boomsma2017spherical}
W.~Boomsma and J.~Frellsen, ``Spherical convolutions and their application in
  molecular modelling,'' {\em Advances in neural information processing
  systems}, vol.~30, 2017.

\bibitem{coors2018spherenet}
B.~Coors, A.~P. Condurache, and A.~Geiger, ``Spherenet: Learning spherical
  representations for detection and classification in omnidirectional images,''
  in {\em Proceedings of the European conference on computer vision (ECCV)},
  pp.~518--533, 2018.

\bibitem{ronchi1996cubed}
C.~Ronchi, R.~Iacono, and P.~S. Paolucci, ``The “cubed sphere”: A new
  method for the solution of partial differential equations in spherical
  geometry,'' {\em Journal of computational physics}, vol.~124, no.~1,
  pp.~93--114, 1996.

\bibitem{gillet2019deep}
N.~Gillet, A.~Mesinger, B.~Greig, A.~Liu, and G.~Ucci, ``Deep learning from
  21-cm tomography of the cosmic dawn and reionization,'' {\em Monthly Notices
  of the Royal Astronomical Society}, vol.~484, no.~1, pp.~282--293, 2019.

\bibitem{fluri2018cosmological}
J.~Fluri, T.~Kacprzak, A.~Refregier, A.~Amara, A.~Lucchi, and T.~Hofmann,
  ``Cosmological constraints from noisy convergence maps through deep
  learning,'' {\em Physical Review D}, vol.~98, no.~12, p.~123518, 2018.

\bibitem{cohen2018spherical}
T.~S. Cohen, M.~Geiger, J.~K{\"o}hler, and M.~Welling, ``Spherical cnns,'' {\em
  arXiv preprint arXiv:1801.10130}, 2018.

\bibitem{banerjee2020volterranet}
M.~Banerjee, R.~Chakraborty, J.~Bouza, and B.~C. Vemuri, ``Volterranet: A
  higher order convolutional network with group equivariance for homogeneous
  manifolds,'' {\em IEEE Transactions on Pattern Analysis and Machine
  Intelligence}, vol.~44, no.~2, pp.~823--833, 2020.

\bibitem{bouza2021higher}
J.~J. Bouza, C.-H. Yang, D.~Vaillancourt, and B.~C. Vemuri, ``A higher order
  manifold-valued convolutional neural network with applications to diffusion
  mri processing,'' in {\em Information Processing in Medical Imaging: 27th
  International Conference, IPMI 2021, Virtual Event, June 28--June 30, 2021,
  Proceedings 27}, pp.~304--317, Springer, 2021.

\bibitem{cohen2016group}
T.~Cohen and M.~Welling, ``Group equivariant convolutional networks,'' in {\em
  International conference on machine learning}, pp.~2990--2999, PMLR, 2016.

\bibitem{goodwin2022can}
T.~Goodwin-Allcock, J.~McEwen, R.~Gray, P.~Nachev, and H.~Zhang, ``How can
  spherical cnns benefit ml-based diffusion mri parameter estimation?,'' in
  {\em International Workshop on Computational Diffusion MRI}, pp.~101--112,
  Springer, 2022.

\bibitem{kerkela2022microstructural}
L.~Kerkel{\"a}, K.~Seunarine, F.~Szczepankiewicz, and C.~A. Clark,
  ``Microstructural neuroimaging using spherical convolutional neural
  networks,'' {\em arXiv preprint arXiv:2211.09887}, 2022.

\bibitem{liu2022group}
R.~Liu, F.~Lauze, E.~Bekkers, K.~Erleben, and S.~Darkner, ``Group convolutional
  neural networks for dwi segmentation,'' in {\em Geometric Deep Learning in
  Medical Image Analysis}, pp.~96--106, PMLR, 2022.

\bibitem{sedlar2021spherical}
S.~Sedlar, A.~Alimi, T.~Papadopoulo, R.~Deriche, and S.~Deslauriers-Gauthier,
  ``A spherical convolutional neural network for white matter structure imaging
  via dmri,'' in {\em Medical Image Computing and Computer Assisted
  Intervention--MICCAI 2021: 24th International Conference, Strasbourg, France,
  September 27--October 1, 2021, Proceedings, Part III 24}, pp.~529--539,
  Springer, 2021.

\bibitem{liu2022bundle}
R.~Liu, F.~Lauze, K.~Erleben, R.~W. Berg, and S.~Darkner, ``Bundle geodesic
  convolutional neural network for diffusion-weighted imaging segmentation,''
  {\em Journal of Medical Imaging}, vol.~9, no.~6, pp.~064002--064002, 2022.

\bibitem{huynh2019probing}
K.~M. Huynh, T.~Xu, Y.~Wu, G.~Chen, K.-H. Thung, H.~Wu, W.~Lin, D.~Shen, P.-T.
  Yap, and U.~B. C.~P. Consortium, ``Probing brain micro-architecture by
  orientation distribution invariant identification of diffusion
  compartments,'' in {\em Medical Image Computing and Computer Assisted
  Intervention--MICCAI 2019: 22nd International Conference, Shenzhen, China,
  October 13--17, 2019, Proceedings, Part III 22}, pp.~547--555, Springer,
  2019.

\bibitem{fieremans2018physical}
E.~Fieremans and H.-H. Lee, ``Physical and numerical phantoms for the
  validation of brain microstructural mri: A cookbook,'' {\em Neuroimage},
  vol.~182, pp.~39--61, 2018.

\bibitem{lee2021quantification}
W.~Lee, B.~Kim, and H.~Park, ``Quantification of intravoxel incoherent motion
  with optimized b-values using deep neural network,'' {\em Magnetic Resonance
  in Medicine}, vol.~86, no.~1, pp.~230--244, 2021.

\bibitem{ye2019q}
C.~Ye, Y.~Cui, and X.~Li, ``Q-space learning with synthesized training data,''
  in {\em Computational Diffusion MRI: International MICCAI Workshop, Granada,
  Spain, September 2018 22}, pp.~123--132, Springer, 2019.

\bibitem{qin2020knowledge}
Y.~Qin, Y.~Li, Z.~Liu, and C.~Ye, ``Knowledge transfer between datasets for
  learning-based tissue microstructure estimation,'' in {\em 2020 IEEE 17th
  International Symposium on Biomedical Imaging (ISBI)}, pp.~1530--1533, IEEE,
  2020.

\bibitem{graham2016realistic}
M.~S. Graham, I.~Drobnjak, and H.~Zhang, ``Realistic simulation of artefacts in
  diffusion mri for validating post-processing correction techniques,'' {\em
  NeuroImage}, vol.~125, pp.~1079--1094, 2016.

\bibitem{masutani2021synthetic}
Y.~Masutani, T.~Fujiwara, and K.~Sasaki, ``Synthetic q-space learning with
  mixture distribution noise for robust dki parameter inference,'' in {\em
  International Forum on Medical Imaging in Asia 2021}, vol.~11792,
  pp.~181--185, SPIE, 2021.

\bibitem{karimi2021calibrated}
D.~Karimi, S.~K. Warfield, and A.~Gholipour, ``Calibrated diffusion tensor
  estimation,'' {\em arXiv preprint arXiv:2111.10847}, 2021.

\bibitem{chamberland2019penthera}
M.~Chamberland, M.~Bernier, G.~Girard, D.~Fortin, M.~Descoteaux, and
  K.~Whittingstall, ``Penthera 1.5 t,'' {\em URL: https://doi.
  org/10.5281/zenodo}, vol.~2602022, 2019.

\bibitem{cai2021}
L.~Y. Cai, Q.~Yang, P.~Kanakaraj, V.~Nath, A.~T. Newton, H.~A. Edmonson,
  J.~Luci, B.~N. Conrad, G.~R. Price, C.~B. Hansen, C.~I. Kerley, K.~Ramadass,
  F.-C. Yeh, H.~Kang, E.~Garyfallidis, M.~Descoteaux, F.~Rheault, K.~G.
  Schilling, and B.~A. Landman, ``"masivar: Multisite, multiscanner, and
  multisubject acquisitions for studying variability in diffusion weighted
  magnetic resonance imaging",'' {\em OpenNeuro}, 2021.

\bibitem{bagher2011predicting}
H.~Bagher-Ebadian, K.~Jafari-Khouzani, P.~D. Mitsias, M.~Lu,
  H.~Soltanian-Zadeh, M.~Chopp, and J.~R. Ewing, ``Predicting final extent of
  ischemic infarction using artificial neural network analysis of
  multi-parametric mri in patients with stroke,'' {\em PloS one}, vol.~6,
  no.~8, p.~e22626, 2011.

\bibitem{arnez2020comparison}
F.~Arnez, H.~Espinoza, A.~Radermacher, and F.~Terrier, ``A comparison of
  uncertainty estimation approaches in deep learning components for autonomous
  vehicle applications,'' {\em arXiv preprint arXiv:2006.15172}, 2020.

\bibitem{guo2017}
C.~Guo, G.~Pleiss, Y.~Sun, and K.~Q. Weinberger, ``On calibration of modern
  neural networks,'' {\em arXiv preprint arXiv:1706.04599}, 2017.

\bibitem{lakshminarayanan2017}
B.~Lakshminarayanan, A.~Pritzel, and C.~Blundell, ``Simple and scalable
  predictive uncertainty estimation using deep ensembles,'' in {\em Advances in
  Neural Information Processing Systems}, pp.~6402--6413, 2017.

\bibitem{goodfellow2014explaining}
I.~J. Goodfellow, J.~Shlens, and C.~Szegedy, ``Explaining and harnessing
  adversarial examples,'' {\em arXiv preprint arXiv:1412.6572}, 2014.

\bibitem{kurakin2016adversarial}
A.~Kurakin, I.~Goodfellow, S.~Bengio, {\em et~al.}, ``Adversarial examples in
  the physical world,'' 2016.

\bibitem{maartensson2020reliability}
G.~M{\aa}rtensson, D.~Ferreira, T.~Granberg, L.~Cavallin, K.~Oppedal,
  A.~Padovani, I.~Rektorova, L.~Bonanni, M.~Pardini, M.~G. Kramberger, {\em
  et~al.}, ``The reliability of a deep learning model in clinical
  out-of-distribution mri data: a multicohort study,'' {\em Medical Image
  Analysis}, vol.~66, p.~101714, 2020.

\bibitem{bulusu2020anomalous}
S.~Bulusu, B.~Kailkhura, B.~Li, P.~K. Varshney, and D.~Song, ``Anomalous
  example detection in deep learning: A survey,'' {\em IEEE Access}, vol.~8,
  pp.~132330--132347, 2020.

\bibitem{loquercio2020general}
A.~Loquercio, M.~Segu, and D.~Scaramuzza, ``A general framework for uncertainty
  estimation in deep learning,'' {\em IEEE Robotics and Automation Letters},
  vol.~5, no.~2, pp.~3153--3160, 2020.

\bibitem{kuleshov2018accurate}
V.~Kuleshov, N.~Fenner, and S.~Ermon, ``Accurate uncertainties for deep
  learning using calibrated regression,'' in {\em International Conference on
  Machine Learning}, pp.~2796--2804, PMLR, 2018.

\bibitem{levi2019evaluating}
D.~Levi, L.~Gispan, N.~Giladi, and E.~Fetaya, ``Evaluating and calibrating
  uncertainty prediction in regression tasks,'' {\em arXiv preprint
  arXiv:1905.11659}, 2019.

\bibitem{avci2021quantifying}
M.~Y. Avci, Z.~Li, Q.~Fan, S.~Huang, B.~Bilgic, and Q.~Tian, ``Quantifying the
  uncertainty of neural networks using monte carlo dropout for deep learning
  based quantitative mri,'' {\em arXiv preprint arXiv:2112.01587}, 2021.

\bibitem{tanno2019uncertainty}
R.~Tanno, D.~Worrall, E.~Kaden, A.~Ghosh, F.~Grussu, A.~Bizzi, S.~N.
  Sotiropoulos, A.~Criminisi, and D.~C. Alexander, ``Uncertainty quantification
  in deep learning for safer neuroimage enhancement,'' {\em arXiv preprint
  arXiv:1907.13418}, 2019.

\bibitem{lakshminarayanan2017simple}
B.~Lakshminarayanan, A.~Pritzel, and C.~Blundell, ``Simple and scalable
  predictive uncertainty estimation using deep ensembles,'' {\em Advances in
  neural information processing systems}, vol.~30, 2017.

\bibitem{xu2015show}
K.~Xu, J.~Ba, R.~Kiros, K.~Cho, A.~Courville, R.~Salakhudinov, R.~Zemel, and
  Y.~Bengio, ``Show, attend and tell: Neural image caption generation with
  visual attention,'' in {\em International conference on machine learning},
  pp.~2048--2057, PMLR, 2015.

\bibitem{zhang2022tractoformer}
F.~Zhang, T.~Xue, W.~Cai, Y.~Rathi, C.-F. Westin, and L.~J. O’Donnell,
  ``Tractoformer: a novel fiber-level whole brain tractography analysis
  framework using spectral embedding and vision transformers,'' in {\em
  International Conference on Medical Image Computing and Computer-Assisted
  Intervention}, pp.~196--206, Springer, 2022.

\bibitem{varadarajan2018towards}
D.~Varadarajan and J.~P. Haldar, ``Towards optimal linear estimation of
  orientation distribution functions with arbitrarily sampled diffusion mri
  data,'' in {\em 2018 IEEE 15th International Symposium on Biomedical Imaging
  (ISBI 2018)}, pp.~743--746, IEEE, 2018.

\end{thebibliography}

\end{document}